\documentclass[aip, reprint]{revtex4-1}
\usepackage[english]{babel}
\usepackage{array}
\usepackage{tabularx}
\usepackage{multirow}
\usepackage{graphicx}
\usepackage{dcolumn}
\usepackage{bm}
\usepackage{titletoc}

\usepackage{hyperref}
\hypersetup{
 bookmarks=true,		
 unicode=false,			
 pdftoolbar=true,		
 pdffitwindow=false,		
 pdfstartview={FitH},		
 pdfcreator={pdflatex},		
 pdfnewwindow=true,		
 colorlinks=true,		
 linktoc=page,			
 linkcolor=blue,		
 citecolor=blue,		
 filecolor=blue,		
 urlcolor=blue			
}
\usepackage{amsfonts,amsmath,amssymb,bm}
\usepackage{cancel}

\usepackage[utf8]{inputenc}
\usepackage[T1]{fontenc}
\usepackage{mathptmx}
\usepackage{etoolbox}
\usepackage{multirow}
\usepackage{tabularray}
\usepackage{siunitx}
\usepackage{tikz}
\usepackage{ulem}
\usepackage{color, colortbl,xcolor}
\usepackage{xparse}
\usepackage{ifthen}
\usepackage{braket}
\usepackage{wrapfig,lipsum,booktabs}
\usepackage{titlesec}

\titleformat{\section}      {\normalfont\large\bfseries}{\thesection}{1em}{}
\titleformat{\subsection}   {\normalfont\bfseries}{\thesubsection}{1em}{}
\titleformat{\subsubsection}{\normalfont\bfseries}{\thesubsubsection}{1em}{}

\titlespacing\section{0pt}{8pt plus 2pt minus 2pt}{0pt}
\titlespacing\subsection{0pt}{8pt plus 2pt minus 2pt}{0pt}
\titlespacing\subsubsection{0pt}{4pt plus 1pt minus 1pt}{2pt}

\makeatletter
\def\@email#1#2{%
 \endgroup
 \patchcmd{\titleblock@produce}
  {\frontmatter@RRAPformat}
  {\frontmatter@RRAPformat{\produce@RRAP{*#1\href{mailto:#2}{#2}}}\frontmatter@RRAPformat}
  {}{}
}%
\makeatother

\newboolean{draft}
\setboolean{draft}{True}
\setboolean{draft}{False}
\NewDocumentCommand\change{om}{%
  \ifthenelse{\boolean{draft}}
  {\IfNoValueTF{#1}{{\color{orange}#2}}{{\color{lightgray}#1}/{\color{orange}#2}}}
  {#2}%
}

\def\vsi{$\text{V}_{\text{Si}}$}
\def\vsim{$\mathrm{V_{Si}^{-}}$}
\def\gsq{\ensuremath{{^4}\!A_2}}
\def\esqA{\ensuremath{{^4}\!A_2^{'}}}
\def\esqE{\ensuremath{{^{4}}\!E}}
\def\dTT{$\Delta$T/T}

\def \vsih{$\text{V}_{h}$}
\def \vsik{$\text{V}_{k}$}
\def\D#1#2{\ensuremath{\text{D}^{#1}_{#2}}}
\def\C3v{\ensuremath{C_{3v}}}
\def\dT#1{\ensuremath{^2\!T_{#1}}}
\def\dA#1{\ensuremath{^2\!A_{#1}}}
\def\dE{\ensuremath{^2\!E}}

\newcommand{\nocontentsline}[3]{}
\let\origcontentsline\addcontentsline
\newcommand\stoptoc{\let\addcontentsline\nocontentsline}
\newcommand\resumetoc{\let\addcontentsline\origcontentsline}

\begin{document}

\title{Unraveling the electronic structure of silicon vacancy centers in 4H-SiC}

\author{Ali Tayefeh Younesi}
\affiliation{Max-Planck Institut f{\"u}r Polymerforschung$,$ Ackermannweg 10$,$ 55128 Mainz$,$ Germany}
\author{Minh Tuan Luu}
\affiliation{Max-Planck Institut f{\"u}r Polymerforschung$,$ Ackermannweg 10$,$ 55128 Mainz$,$ Germany}
\author{Christopher Linder\"alv}
\affiliation{Department of Physics / Centre for Material Science and Nanotechnology, University of Oslo, Norway}
\author{Vytautas Žalandauskas}
\affiliation{Department of Physics / Centre for Material Science and Nanotechnology, University of Oslo, Norway}
\author{Marianne Etzelm{\"u}ller Bathen}
\affiliation{Department of Physics / Centre for Material Science and Nanotechnology, University of Oslo, Norway}
\author{Nguyen Tien Son}
\affiliation{Department of Physics, Chemistry and Biology, Link\"oping University, SE-58183, Linköping, Sweden}
\author{Takeshi Ohshima}
\affiliation{Takasaki Institute for Advanced
	Quantum Science, National Institutes for Quantum Science
	and Technology: QST, Gunma 370-1292, Japan;
	Department of Materials Science, Tohoku University, Miyagi
	980-8579, Japan}
\author{Gerg\H{o} Thiering}
\affiliation{Wigner Research Centre for Physics, Hungarian Academy of Sciences, PO Box 49, Budapest 1525, Hungary}
\author{Lukas Razinkovas}
\affiliation{Department of Fundamental Research, Center for Physical Sciences and Technology (FTMC), Vilnius LT--10257, Lithuania}
\author{Ronald Ulbricht}
\affiliation{Max-Planck Institut f{\"u}r Polymerforschung$,$ Ackermannweg 10$,$ 55128 Mainz$,$ Germany}
\email{ulbricht@mpip-mainz.mpg.de}

\date{\today}

\frenchspacing

\begin{abstract}

  Point defects in silicon carbide (SiC), particularly the negatively-charged silicon vacancy (\vsim) in 4H-SiC, are leading candidates for scalable
  quantum technologies due to their favorable spin–optical properties and compatibility with industrial semiconductor fabrication processes. Comprehensive knowledge of a defect's electronic structure is essential for interpreting spin--optical dynamics and for the reliable design and optimization of defect-based quantum devices. Despite extensive study, our knowledge of the electronic structure of \vsim\ is limited since key excited-state manifolds have remained inaccessible to conventional steady-state spectroscopy. In this study, transient absorption spectroscopy is utilized to probe non-equilibrium electronic transitions of \vsim\ and to uncover previously unobserved excited states. The first direct observation of the elusive $\text{V2}'$ quartet transition is presented, with its broad spectral signature attributed to nonadiabatic vibronic coupling. Within the spin-doublet manifold, which is central to optically detected magnetic resonance (ODMR) but has remained unresolved spectroscopically, multiple optical transitions are identified. The complete electronic level structure in the relevant energy range is elucidated by combining polarization-resolved spectroscopy, group-theoretical analysis, quantum embedding calculations and first-principles optical lineshape modeling. Collectively, these results provide a microscopic understanding of the \vsim\ electronic structure. Our approach also establishes a general framework for resolving and understanding complex excited-state manifolds in wide-bandgap color centers.
\end{abstract}

\maketitle

\stoptoc
\section*{Introduction}

Point defects in wide-bandgap semiconductors have emerged as a promising
platform for quantum information and sensing applications, primarily due to
their favorable optical and spin properties. In light of the advancements in
utilizing nitrogen vacancy (NV) centers in diamond as a quantum platform, there
has been a growing demand for alternative host materials with properties
comparable to or superior to diamond, albeit with a more straightforward
manufacturing process. Silicon carbide (SiC) shows favorable characteristics in
this context, enabling the creation of defects with electronic levels deep
within the band gap that give rise to optically bright emission and long-lived,
controllable spin states required for quantum applications.\cite{anderson_2019}
Furthermore, the mature fabrication and processing of SiC\cite{Kimoto_2014}
ensure CMOS (complementary metal--oxide--semiconductor) compatibility, enabling
on-chip integration of optically addressable single spins with electronic and
photonic devices.\cite{Castelletto_2020,Norman_2020,Bathen_QuTe_2021}

The negatively-charged silicon vacancy (\vsim) in 4H-SiC shows particular promise for a variety of applications, exhibiting single-photon emission in
the near-infrared\cite{Janzen2009,Widmann_2014} and millisecond spin coherence
times\cite{Widmann_2014} at room temperature. \vsim\ is one of the few color
centers, and the only one in SiC, where indistinguishable photon emission has
been demonstrated.\cite{Morioka_2020} In short, the intriguing spin-photon
interface,\cite{Soykal_2016} demonstrated charge-state
control,\cite{wolfowicz2017optical,bathen2019electrical,widmann_2019} Stark and
strain tuning,\cite{Ruhl_2020,Vasquez_2020} Purcell enhancement \cite{Bracher_2017}, and
photonic integration \cite{Lukin_2020b} of the \vsim\ all mark it as highly exploitable in
next-generation scalable quantum devices for sensing, communication, and
computing. While these properties highlight the technological potential of \vsim, a complete understanding of its electronic structure, central to both its fundamental description and future device concepts, remains lacking.
\begin{figure*}[t]
    \centering
    \includegraphics[width=0.95\linewidth]{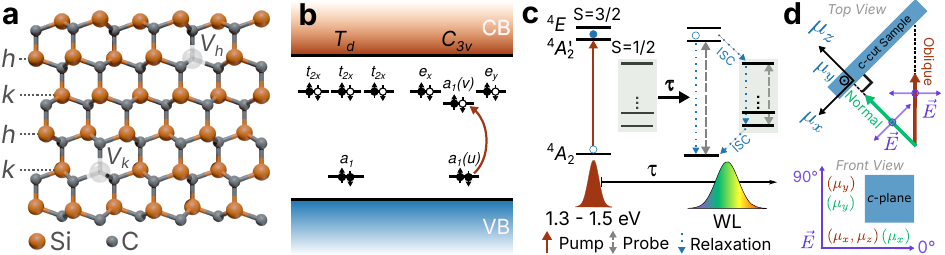}
    \caption{\textbf{Overview.} 
    \textbf{a} Crystal structure of the \vsi\ center in 4H-SiC, with the basal-plane
    hexagonal (\textit{h}) and quasi-cubic (\textit{k}) lattice sites indicated.
    \textbf{b} Single-particle energy levels of $\mathrm{V_{Si}}^-$. Left: Idealized $T_d$
    symmetry; Right: crystal-field splitting resulting in $\C3v$ symmetry. Red arrow: excitation to $^4A_2'$ excited state.
    \textbf{c} Electronic states in the quartet ($S=3/2$) and doublet ($S=1/2$) spin channels and schematic of the transient absorption (TA)
measurement scheme. 
    \textbf{d} Top: Schematic of the probing
    geometry in both the oblique and normal incidence cases of \vsim\
    electronic transitions relative to the \textit{c}-plane, with a coordinate 
    system for the transition dipole moments defined by $\mu_x$, $\mu_y$ and $\mu_z$. 
    Purple arrows indicate the electric field vectors of the probe beam. 
    Bottom: Front view of sample ($c$-plane), i.e., in the laser beam direction, with summary of the transition dipole moments (or "dipoles") excited by the 0$^\circ$ or 90$^\circ$ polarization of the electric field vector.\label{fig1}}
    \vspace{-1em}
\end{figure*}

In this article, we comprehensively characterize the electronic structure of both
spin channels of \vsim\ using a combination of experimental and theoretical
methods, including transient absorption (TA) spectroscopy, group theory, density
functional theory (DFT) and quantum embedding calculations. For the
spin-quartet transitions, we identify the so far elusive V2$^{\prime}$
zero-phonon line (ZPL), which appears strongly broadened as compared to the
other ZPLs. We attribute this behavior to non-adiabatic coupling to vibronic
modes. In the spin doublet channel, we for the first time measure the
spectroscopic signatures of the relevant optical transitions, from which we
can infer the electronic structure using polarization-resolved TA spectra and group theory. The holistic methodology presented herein can be applied to other color centers and host materials, making it an interesting
tool for characterizing their electronic structure to aid the broader
development of color center-based quantum technologies.
\goodbreak

\section*{Results}
\subsection*{The \vsi \ in 4H-SiC}

We begin by describing the crystallographic environments and symmetry
characteristics of the \vsi\ in 4H–SiC, which serve as the foundation for
understanding its electronic structure. As illustrated in Fig.~\ref{fig1}a, the
Si--C bilayers in 4H–SiC stack along the crystallographic $c$-axis ([0001]
direction), giving rise to two types of lattice sites: hexagonal ($h$)
and quasi-cubic ($k$). Consequently, a silicon vacancy can occur in two
inequivalent configurations, denoted $V_{h}$ and $V_{k}$. Both share the same
first-neighbor tetrahedral coordination, with local symmetry close to $T_{d}$,
but they differ in the arrangement of their second-nearest neighbors. The actual
point-group symmetry of both configurations is $\C3v$, with the threefold axis
oriented along the crystallographic $c$-axis. The electronic structure of both
the \textit{h} and \textit{k} vacancies can be described in terms of four
molecular orbitals formed from the tetrahedrally oriented carbon dangling bonds
surrounding the vacancy. Because the local symmetry is close to $T_{d}$, these
orbitals can initially be classified according to the irreducible
representations of this point group (left side of Fig.~\ref{fig1}b): one $a_{1}$
orbital, given by the equal-weight linear superposition of the four carbon
dangling bonds, and three degenerate orbitals of $t_{2}$ symmetry. Owing to its
bonding character, the $a_{1}$ orbital lies lowest in energy, while the $t_{2}$
set resides at higher energies. When the symmetry is reduced from $T_{d}$ to
$\C3v$, the $t_{2}$ triplet splits into an orbital doublet of $e$ symmetry and
a singlet of $a_{1}$ symmetry (right side of Fig.~\ref{fig1}b).

In the negatively charged state, the \vsim \ defect accommodates five electrons,
giving rise to a rich electronic structure. The ground-state configuration,
shown in Fig.~\ref{fig1}b, is a spin quartet ($S=3/2$) of $A_{2}$ orbital symmetry
($\gsq$). Optical excitation corresponds to promoting an electron from the
lower-lying $a_{1}(u)$ orbital to either the $a_{1}(v)$ or $e$ orbitals (as
depicted by brown arrows in Fig.~\ref{fig1}b), resulting in the
$^{4}\!A_{2}^{\prime}$ and $^{4}\!E$ excited states, separated by a small
crystal-field splitting, see Fig.~\ref{fig1}c. The electronic structure further
includes a spin-doublet ($S=1/2$) channel that can be accessed through
intersystem crossing (ISC) between both spin channels.
\goodbreak

\begin{figure*}[ht]
	\centering
	\includegraphics[width=0.95\linewidth]{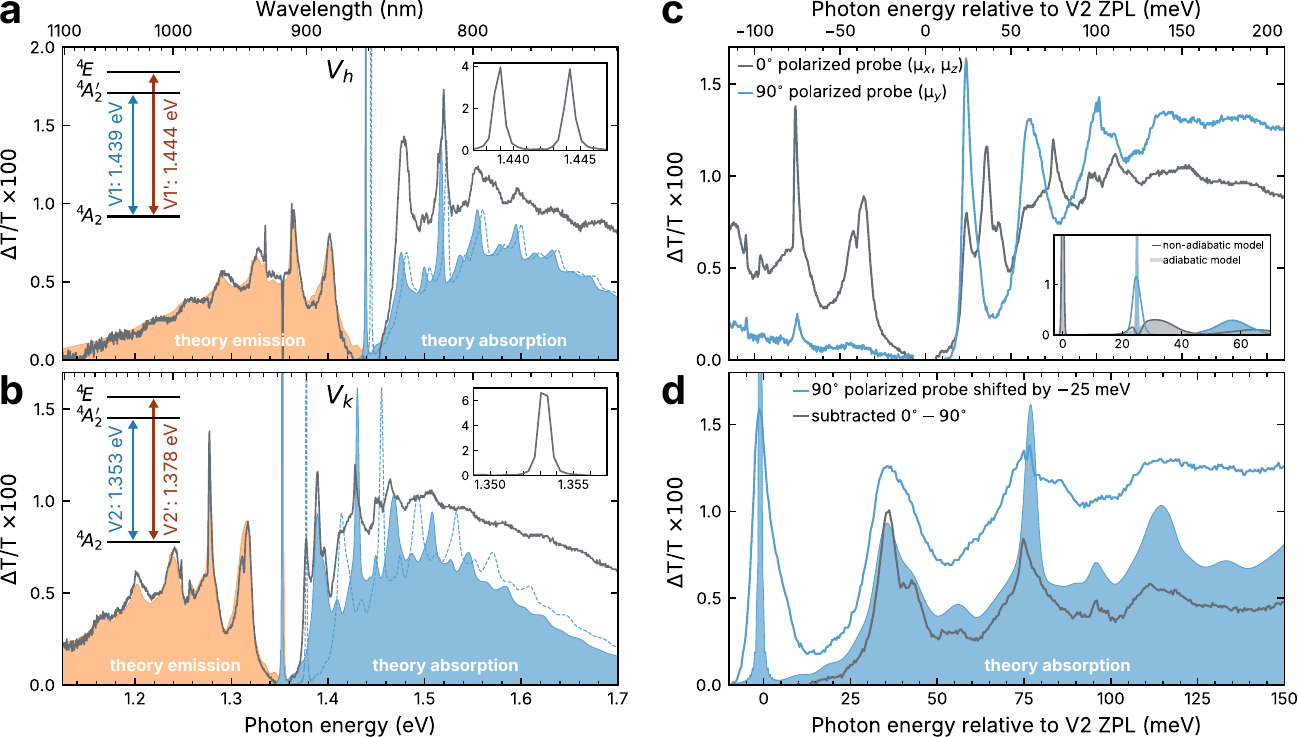}
	\caption{\textbf{Spin-quartet electronic states and optical transitions.} 
    \textbf{a} $V_{h}$ and \textbf{b} $V_{k}$: TA spectra at pump-probe delay of $t = 4$~ns (black curves),
		calculated PSBs of V1 and V2 in emission (orange shaded) and absorption
		(blue shaded); insets: enlarged version of the ZPL spectral regions
		(right) and electronic level scheme with optical transitions and transition energies
		in eV (left); 
    \textbf{c} TA spectra of \vsik\ for 0$^\circ$ and 90$^\circ$ probe
		polarizations with the energy axis referenced to the V2 ZPL; heuristic model
		simulating the broadening of the V2$^{\prime}$ ZPL in the inset; 
    \textbf{d} TA spectra of \vsik, focusing on the GSB by comparing the V2 PSB with absorption
		lineshape simulation (shaded) and V2$^{\prime}$ PSB red-shifted by 25~meV.\label{fig2}}
\end{figure*}

The sharp zero-phonon
lines (ZPLs) commonly observed in photoluminescence (PL) spectra of 4H-SiC have
been identified with different configurations and excited states, namely V1
	[1.439~eV, $V_{h}$], V1$^{\prime}$ [1.444~eV, $V_{h}$], and V2 [1.353~eV, $V_{k}$],\cite{Janzen2009} see also Figs.~\ref{fig2}a~and~\ref{fig2}b. These ZPLs are accompanied
by broad phonon sidebands (PSBs) arising from the coupling of the electronic
transitions to lattice vibrations (see
Fig.~\ref{fig2}).\cite{udvarhelyi2020vibronic} Interestingly, the V2$^{\prime}$
ZPL, which should arise from the transition to $^{4}\!E$ of $V_{k}$ and is
predicted to lie 22~meV above \esqA,\cite{udvarhelyi2020vibronic} has not yet been
observed.

While the quartet states and their
associated transitions have been extensively characterized, considerably less is
known about the spin-doublet channel of \vsim. This spin manifold can be
populated from and de-populated to the quartet manifolds via ISC, which plays
the key role in the spin polarization cycle that enables optically-detected
magnetic resonance (ODMR) used in quantum applications. Despite this importance,
ISC rates have only recently been measured\cite{liu2024silicon} and so far, the
electronic structure of the $S=1/2$ manifold has been established primarily
through group-theoretical many-body analyses combined with physically motivated
level ordering.\cite{Soykal_2016,dong2019spin} Its role has only been inferred
indirectly from spin-polarization dynamics rather than from direct spectroscopic
observations.

\subsection*{Experimental conditions}

To experimentally investigate the optical transitions and relaxation dynamics of
\vsim\ ensembles, we employ TA spectroscopy, a pump--probe technique that has
been successfully applied to NV centers in
diamond.\cite{Younesi2022,luu2024nitrogen,Luu2025} In this approach, a pump
pulse excites the system, while a time-delayed probe monitors the resulting
pump-induced change in sample transmission \dTT. To this end, the photon energy
of the pump pulse is tuned in the range of 1.3--1.5~eV to selectively excite
ZPLs or the PSB associated with the \vsim\ quartet transitions, followed by broadband
whitelight (WL) that transiently probes \dTT\ from 0.6--1.8 eV (see
Fig.~\ref{fig1}c), thereby monitoring both the quartet and doublet relaxation
channels. Positive values of \dTT\ originate from ground-state bleaching (GSB)
and stimulated emission (SE) of the quartet transitions, whereas negative values
arise from absorption of the doublet transitions.

The sample under study is a $c$-cut 4H-SiC wafer housed in a cryostat for measurements down to 4~K temperature. For the polarization components of the probe beam to access all
possible transition dipole moments, the sample is held at an angle of either
0$^\circ$ (`normal incidence') or 45$^\circ$ (`oblique incidence') between the
laser beams and the $c$-axis, see Fig.~\ref{fig1}d. We define a coordinate
system for the transition dipole moments $\mu$, where $\mu_x$ and $\mu_y$ are
within the $c$-plane and $\mu_z$ is perpendicular to it, i.e. along the $c$-axis
(see Fig.~\ref{fig1}d). Next to the normal or oblique incidence beam
configurations, utilizing two orthogonal probe polarizations (0$^\circ$ and
90$^\circ$) provides additional selectivity in picking certain transitions based
on the optical selection rules. Where normal incidence only accesses
polarization components within the $c$-plane ($\mu_x$ for 0$^\circ$ and $\mu_y$
for 90$^\circ$), oblique incidence additionally also probes $\mu_z$, however
only for 0$^\circ$ (`p-polarization'). According to the optical selection rules for the
quartet transitions, V1 and V2 only have active $\mu_z$ transition dipoles,
whereas V1$^{\prime}$ and V2$^{\prime}$ have both $\mu_x$ and $\mu_y$. Thus, in oblique incidence, 0$^\circ$ will probe all transitions ($\mu_x$ and $\mu_z$), while 90$^\circ$ will probe only V1$^{\prime}$/V2$^{\prime}$ ($\mu_y$). In normal incidence, only
V1$^{\prime}$/V2$^{\prime}$ is probed for all polarizations ($\mu_x$ and $\mu_y$).

\subsection*{Electronic structure of spin quartets}

At early (< a few ns) pump-probe delays, the transient response is dominated by
quartet optical transitions. Figs.~\ref{fig2}a~and~\ref{fig2}b show \dTT\ (gray curves)
for both the \vsih\ and \vsik\ configurations of \vsi\ under resonant ZPL
excitation, with the ZPL region excluded due to the pump laser. The spectra
exhibit broad PSBs, each comprising distinct vibronic peaks on the blue (GSB)
and red (SE) sides, reflecting absorption and emission contributions,
respectively. The insets in Fig.~\ref{fig2}a~and~\ref{fig2}b show the ZPLs measured under
excitation into the PSBs. In the inset of Fig.~\ref{fig2}a, both the V1 and
V1$^{\prime}$ ZPLs associated with the \vsih\ configuration are clearly resolved
and are separated by 5~meV, while in the inset of Fig.~\ref{fig2}b, only one ZPL
related to the V2 transition is visible in the displayed energy region.

To gain further insight into the vibronic structure behind the observed PSBs,
Figs.~\ref{fig2}a~and~\ref{fig2}b also show the overlaid theoretical absorption (blue
shaded areas) and emission (orange shaded areas) lineshapes calculated for the
$\gsq\leftrightarrow \esqA$ optical transitions within the adiabatic Huang--Rhys
(HR) model, see also Methods and Supplementary Section \ref{SM_HR}. For both the V1 and V2 transitions, the calculated emission
lineshapes show strong agreement with the experimental data by capturing the overall
redistribution of spectral weight and resolving most of the observed PSB fine
structure. While at low temperatures the emission peaks can be explained by
transitions from the lower-lying \esqA\ state, the vibronic peaks on the blue
side (GSB) consist of two quartet absorption channels corresponding to the
$\gsq\to \esqA$ and $\gsq\to \esqE$ transitions. For the \vsih\ configuration,
the absorption lineshape calculated for the $\gsq\to\esqA$ (V1) transition
reproduces almost all observed vibronic features (blue shaded area in Fig.~\ref{fig2}a), while the additional experimental broadening can be attributed to
a second ZPL located 5~meV higher in energy. To illustrate this, we include an
additional absorption lineshape, obtained by rigidly shifting the calculated
$\gsq\to\esqA$ lineshape by 5~meV to the V1$^{\prime}$ ZPL energy (dashed
blue line in Fig.~\ref{fig2}a). The ability of the adiabatic HR model to
reproduce nearly all observed sideband features implies that non-adiabatic
Jahn--Teller coupling within the $\esqE$ manifold and between the $\esqA$ and $\esqE$ states is not relevant.

Turning to the \vsik\ configuration, the situation is qualitatively different.
In contrast to \vsih, only a single ZPL associated with the V2 transition is
clearly observed in the experiment (inset of Fig.~\ref{fig2}b). When overlaying
the calculated absorption lineshape for the $\gsq \to \esqA$ (V2) transition,
it becomes evident that the observed GSB on the blue side cannot
be explained by this channel alone. In particular, an additional absorption
feature appears at approximately 25~meV above the V2 ZPL, which cannot be
associated with a characteristic vibrational resonance of the V2 transition
($\approx$35~meV above ZPL). Notably, this energy separation is close to the
theoretically predicted splitting between the V2 and V2$^{\prime}$ ZPLs
(22~meV), as reported in Ref.~\onlinecite{udvarhelyi2020vibronic}. Motivated by this
proximity, we tentatively include an additional absorption contribution obtained
by rigidly shifting the calculated V2 absorption lineshape by 25~meV (dashed
blue line in Fig~\ref{fig2}b). The resulting superposition reproduces several
additional features in the experimental PSB, suggesting that overlapping
absorption from two transitions may contribute to the observed spectrum.

Assuming that the absorption feature observed 25~meV above the V2 ZPL
corresponds to the $\gsq\to\esqE$ transition, we next turn to
polarization-resolved experiments designed to test this hypothesis. From
symmetry considerations, the $\gsq\to\esqA$ transition is expected to have a
transition dipole moment that is oriented perpendicular to the $c$-plane ($\mu_z$), whereas the
$\gsq\to\esqE$ transition should exhibit an in-plane polarization ($\mu_x$ and $\mu_y$).
Fig.~\ref{fig2}c shows the corresponding TA spectra measured at oblique
incidence using two orthogonal probe polarizations. For the $0^\circ$-polarized
probe, which contains both out-of-plane and in-plane polarization components,
contributions from both quartet transitions are expected and indeed observed,
resulting in pronounced GSB and SE features. In contrast, the
$90^\circ$-polarized probe predominantly samples the in-plane response and
therefore probes mainly the $\gsq\leftrightarrow\esqE$ channel, exhibiting
almost no SE and yielding a clear GSB signal that closely resembles an
absorption lineshape, with an apparent ZPL located approximately 25~meV above
the V2 transition and an associated PSB.

To further support this assignment, we analyze the polarization-resolved GSB
line shapes in more detail. Fig.~\ref{fig2}d shows the $90^\circ$-polarized
GSB spectrum, where the first peak is shifted to zero energy and treated as an
effective ZPL (blue line), together with the calculated absorption lineshape
(blue shaded area). The agreement between features in both lines confirms that
this feature indeed corresponds to a ZPL accompanied by its PSB. In addition,
Fig.~\ref{fig2}d also displays the $0^\circ$-polarized spectrum (dark line)
after subtracting the $90^\circ$-polarized contribution to effectively
remove the absorption feature located 25~meV above the V2 ZPL. Notably, the
resulting sideband shows the same set of vibronic peaks as the shifted
$90^\circ$ spectrum and the calculated absorption lineshape, indicating that
all three spectra essentially share the same phonon sideband structure.

The only remaining feature that cannot be explained within the adiabatic
framework is the anomalous linewidth of the V2$^{\prime}$ ZPL, which is broader
by a factor of about eight compared to the other quartet ZPLs (see Supplementary Table \ref{table:zpl_linewidth}). To rationalize this behavior, we introduce an effective many-mode vibronic model
that captures non-adiabatic Jahn--Teller-type interactions between the $\esqA$
and $\esqE$ quartet manifolds. The main idea is that the nominal zero-phonon
state of the $\esqE$ manifold becomes resonantly coupled to a dense set of
vibronic states originating from the $\esqA$ configuration. This coupling occurs
because the V2$^{\prime}$ ZPL lies within an energy range (around 25~meV) that
overlaps with the frequency range of the first vacancy-related vibrational
resonance (corresponding to the first PSB peak at about 35~meV). As a result, the oscillator
strength of a single transition is redistributed among many closely spaced
vibronic states, effectively broadening the observed ZPL. The result is
shown in the inset of Fig.~\ref{fig2}c, where lineshapes for the two orthogonal
polarizations are modeled both without vibronic coupling (shaded areas) and with
vibronic coupling included (solid lines); see Supplementary Section \ref{SM_nonadiabatic}. Our model captures the core physics and reproduces the broadening of the V2$^{\prime}$ ZPL, resulting in a lineshape that resembles the experimental ZPL.

\begin{figure*}[ht]
    \centering
    \includegraphics[width=0.8\linewidth]{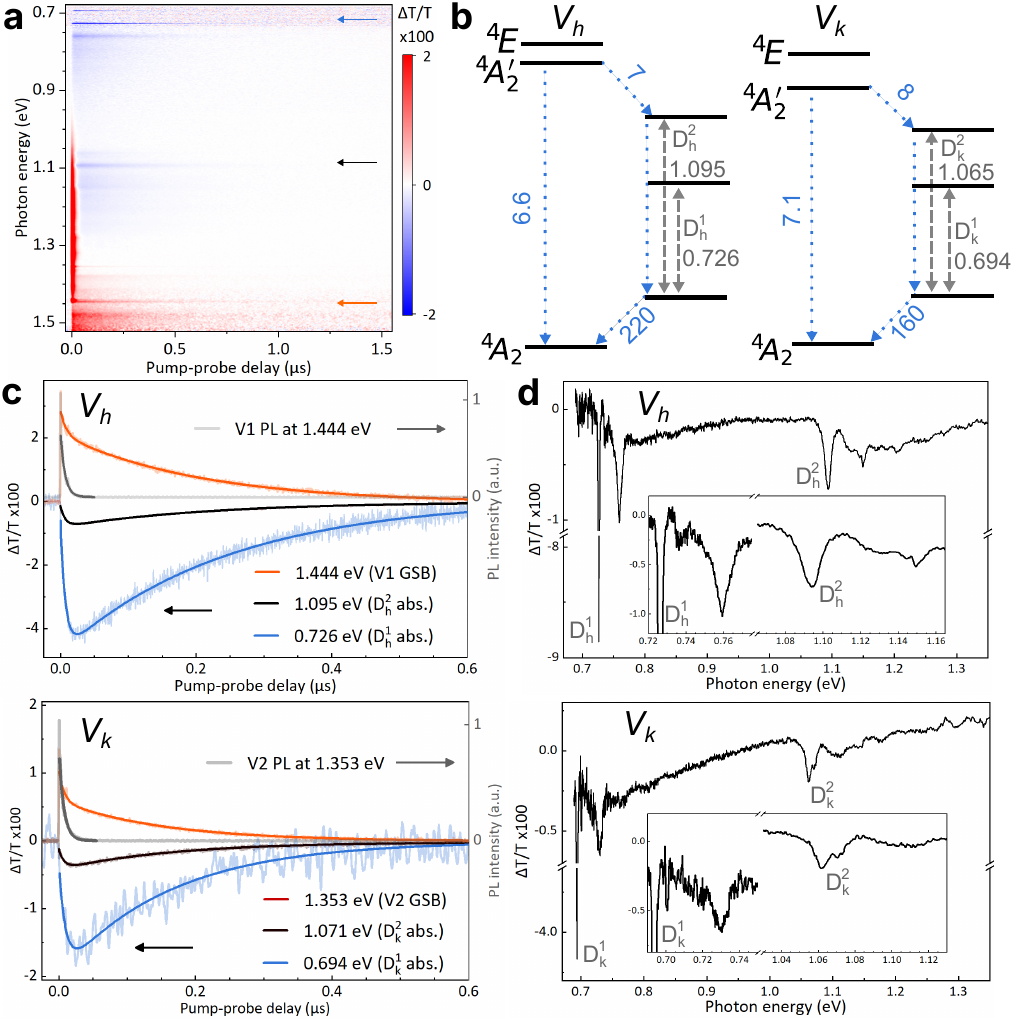}
    \caption{\textbf{Spin-doublet electronic states and optical transitions.} 
    \textbf{a} Spectrally-resolved TA dynamics for $V_{h}$ and $V_{k}$; colored arrows indicate transitions whose dynamics are shown in panel c. 
    \textbf{b} Extracted electronic level scheme and relaxation dynamics of $V_{h}$ and $V_{k}$. Gray dashed arrows denote optical transition energies (in eV), while blue dashed arrows indicate relaxation pathways with time constants (in ns).
    \textbf{c} TA dynamics of spin-quartet and spin-doublet transitions and PL decay (right axis) of spin-quartet transitions of $V_{h}$ and $V_{k}$; 
    \textbf{d} TA spectra of spin-doublet transitions of $V_{h}$ and $V_{k}$ at a pump–probe delay of 90~ns. The insets show magnified views of selected doublet features.\label{fig3}}
\end{figure*}

The five times larger energy difference between \esqA\ and \esqE\ for \vsik\ as
compared to \vsih\ (25~meV vs. 5~meV) explains why previous PL measurements did
not detect V2$^{\prime}$: because these measurements are usually done at
cryogenic temperatures, the population of \esqE\ after photoexcitation, as
described by the Boltzmann distribution, is simply too low to be observed in PL.
Absorption measurements (or here: GSB in TA) capture both transitions regardless
of temperature. However, the non-adiabatic broadening of the V2$^{\prime}$ ZPL
makes it easy to be mistaken for a vibronic peak.

\subsection*{Optical transitions in the spin doublet channel}

By analyzing the TA dynamics over extended timescales, we can track the entire
spin-polarization cycle by monitoring relevant electronic transitions. This is
demonstrated in Fig.~\ref{fig3}a, which displays spectro-temporal TA results
obtained under broadband excitation into the PSB of the $V_h$ and $V_k$ quartet
transitions ($\approx 1.55$~eV). The data show positive differential
transmission signals (red) linked to the quartet transition, as well as new
absorptive features (blue) corresponding to the expected doublet transitions.
The measurement was conducted at oblique incidence using unpolarized probe light,
ensuring that all components of the transition dipole moment ($\mu_x$, $\mu_y$
and $\mu_z$) are detected.

Fig.~\ref{fig3}b schematically shows the electronic energy levels and
relaxation time constants for the $V_{h}$ and $V_{k}$ configurations, as
determined from this and complementary measurements. The blue-dashed arrows are relaxation pathways with time constants in ns, whereas gray-dashed arrows denote the observed doublet transitions with photon energies in eV. For each
defect site ($h$ and $k$), two doublet transitions are observed: the first
around 0.7~eV ($\D1h$ and $\D1k$) and the second around 1.1~eV ($\D2h$ and
$\D2k$), which appear as blue features in Fig.~\ref{fig3}a.

Fig.~\ref{fig3}c shows the dynamics of the relevant transitions for both $V_h$
(top) and $V_k$ (bottom) sites (semi-transparent lines), along with fits to the
appropriate kinetic models (solid lines). The gray traces represent
time-resolved PL measurements of the V1 and V2 ZPLs, which were fitted using an
exponential decay model to determine the \esqA\ to \gsq\ radiative lifetimes
shown in Fig.~\ref{fig3}b. The orange trace shows the TA GSB dynamics of the V1
and V2 ZPLs, exhibiting a bi-exponential decay with a fast component associated
with the upper ISC (7~ns and 8~ns for $V_h$ and $V_k$, respectively), and a slow
component governed by the lower ISC (220~ns for $V_h$ and 160~ns for $V_k$). The
numbers for the lower ISC time are in very good agreement with previous results
\cite{liu2024silicon}, whereas the values for the upper ISC time are in good
agreement with the radiative lifetime of the quartet transition as measured with
time-resolved PL (Fig.~\ref{fig3}c).

The blue and black plots in Fig.~\ref{fig3}c show the TA dynamics of absorption
(i.e., negative \dTT\ values) from the two doublet transitions $\text{D}^1$ and $\text{D}^2$,
as illustrated in Fig.~\ref{fig3}b. Both signals initially display an increasing
negative \dTT\ amplitude, followed by a decrease at longer pump–probe delays.
The initial rise is attributed to population transfer from the quartet to the
doublet spin channel via the upper ISC, while the subsequent decline indicates
doublet-to-quartet relaxation via the lower ISC. The observed timescales align
with those derived from the quartet GSB dynamics (see Supplementary Sections
\ref{SM_vsih} and \ref{SM_vsik}). The lifetime of the
upper doublet state cannot be directly resolved in our TA measurements. Such a
state would give rise to stimulated emission at early pump–probe delays, as
observed for spin-singlet states in NV centers \cite{ulbricht2018excited}. While
our sub-picosecond time resolution is adequate, the expected weak doublet SE
signal is masked by the dominant SE of the quartet transition.

To better isolate the doublet spectral signatures of the $V_h$ and $V_k$
configurations, we resonantly excited the V1 and V2 ZPLs to enable a clear
separation of their respective TA responses. The TA spectra at a pump--probe
delay of 90~ns are shown in Fig.~\ref{fig3}d, with the $V_h$ and $V_k$
configurations displayed in the top and bottom panels, respectively. The
$\text{D}^1$ ZPLs appear as sharp resonances with approximately 1.5~meV
linewidth, similar to most of the quartet ZPLs, at transition energies of
0.726~eV ($\D1h$) and 0.694~eV ($\D1k$). Both ZPLs are accompanied by distinct
vibronic peaks about 35~meV to the blue side, like for the quartet transitions.
A second set of transitions are found at around 1.095~eV for $V_{h}$ ($\D2h$)
and about about 1.07~eV for $V_{k}$ ($\D2k$). The insets of Fig.~\ref{fig3}d
show an enlarged version of these peaks, revealing a splitting of the transition
in two components, most pronounced for $\D2k$. This double-peak feature becomes
clearer in the probe polarization dependence that we will discuss further down,
while the data shown here were recorded with an unpolarized probe.

\begin{figure*}
    \centering
    \includegraphics[width=0.8\linewidth]{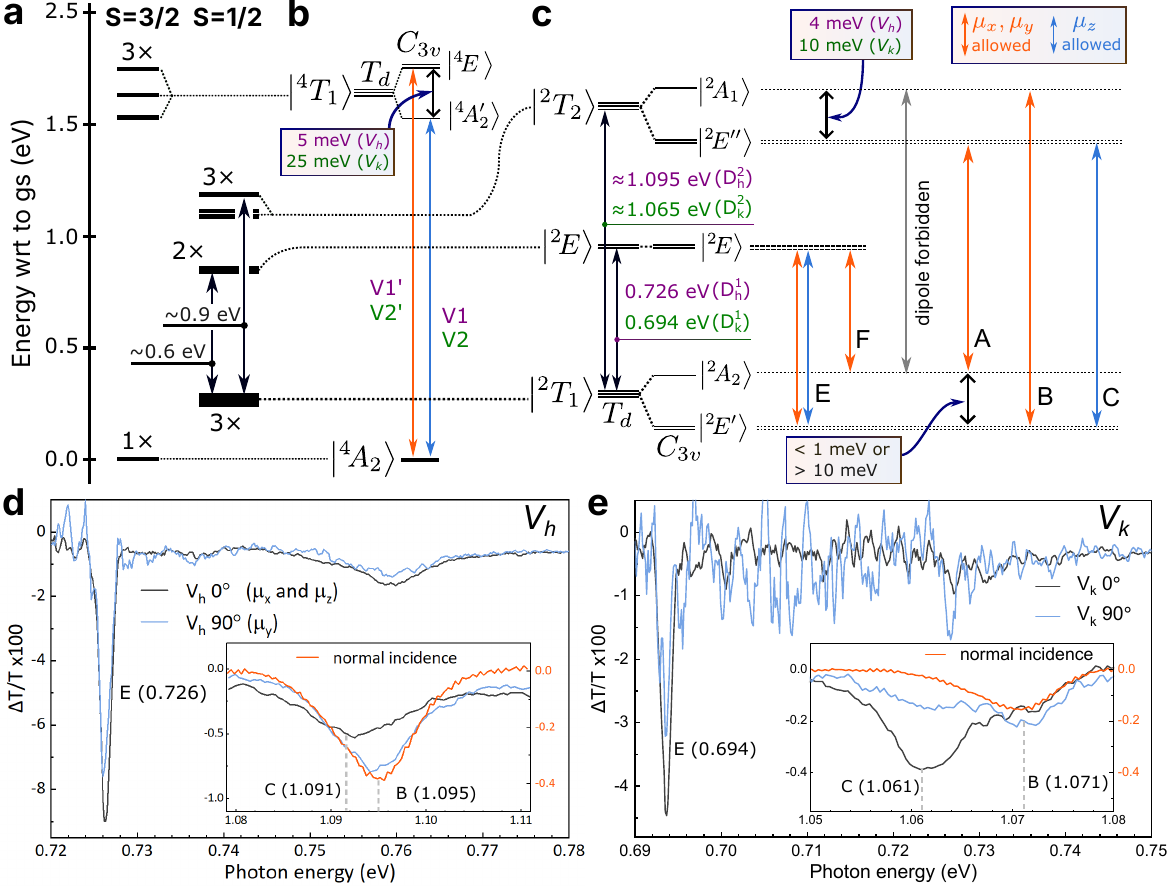}
    \caption{\textbf{Electronic structure and optical selection rules of spin-doublet and spin-quartet channels.} 
    \textbf{a} Electronic levels for \vsik\ calculated using quantum embedding. \textbf{b,c} Electronic structure and optical selection rules of the spin-quartet and spin-doublet channels, respectively. \textbf{d,e} TA probe polarization dependence of spin-doublet transitions for the $V_{h}$ and $V_{k}$ configurations, measured under oblique incidence (blue and black traces) and normal incidence (orange traces).\label{fig4}}
\end{figure*}

\subsection*{Electronic structure of spin doublets}

Having established the energies and kinetics of the spin-doublet transitions, we
now turn to their microscopic interpretation. In the following, we analyze the
symmetry structure of the doublet manifold, resolve the ordering of the states
using group theory and quantum embedding calculations, examine crystal-field
splittings, and derive the resulting optical selection rules and polarization
dependence.

As discussed earlier, the silicon vacancy center is conveniently described
within a near-$T_d$ local environment. The $\C3v$ symmetry results from a weak
trigonal distortion, as shown by the small splittings observed in the excited
quartet manifold (5~meV and 25~meV for \vsih\ and \vsik, respectively). This indicates that the
$T_d \rightarrow \C3v$ crystal field acts only as a minor perturbation.
Therefore, the dominant energy separations correspond to the term splittings
emerging from many-body electron--electron interactions already present in ideal
$T_d$ symmetry. 
This type of energy scale supports the construction of
many-body electronic states within the parent $T_d$ symmetry for simplicity, accompanied by the treatment
of the $\C3v$ crystal field as a small perturbation.

Within $T_d$ symmetry, the low-energy electronic structure of a $t_2^3$ hole
configuration is well described by classical multiplet theory.\cite{sugano1970}
The spin-doublet manifold consists of three distinct symmetry terms: $\dT1$,
$^2\!E$, and $\dT2$. This framework places the $\dT2$ term as the
highest energy term of the $t_2^3$ manifold, whereas the relative ordering of
the remaining $\dT1$ and $^{2}\!E$ terms cannot be determined from symmetry
considerations alone (see Supplementary Section \ref{SM_group_theory}).
As a consequence, the detailed ordering of the spin-doublet manifold must be
established beyond symmetry arguments. Moreover, the multi-determinant character of the
doublet states prevents 
a description within standard single-determinant DFT,
motivating the use of quantum embedding
methodologies.\cite{Bockstedte_2018, Muechler-Dreyer-embedding-2022, ma2021, chen2025} This approach allows for an explicit treatment of electronic
correlations on top of a mean-field DFT description by separating a localized
active space of defect electronic degrees of freedom from the surrounding bulk
environment and solving the many-body problem within this space beyond the
single-determinant level. 

Quantum embedding calculations for $V_k$ were performed using an active space
with four Wannierized spatial orbitals, as detailed in the Methods and
Supplementary Section \ref{subsection:QuantumEmbedding}. The resulting energy
level structure for the maximum spin projections ($m_s = 3/2$ and $m_s = 1/2$)
is shown in Fig.~\ref{fig4}a. Within the quartet ($S = 3/2$) manifold, the
second excited state is located 1.52~eV above the ground state, which aligns
well with the experimental ZPL ($\approx 1.4$~eV). However, contrary to the
experimental data, an artificial splitting between three excited quartet levels
is observed. This splitting should be regarded as a methodological artifact
stemming from the approximate nature of the Wannier construction and the
inability to preserve point-group symmetry in the embedded system.

Within the spin-doublet manifold depicted in Fig.~\ref{fig4}a, the calculated
energy levels form three distinct groups, centered at approximately 0.26~eV
(three states), 0.84~eV (two states), and 1.12~eV (three states) above the
ground state. This grouping, together with the $T_d$ symmetry analysis described
above, allows for identification of the lowest spin-doublet as possessing
$^2T_1$ character, while the next higher doublet exhibits $^2E$ character. The
calculated transition energies between these multiplets, 0.6~eV and 0.9~eV, are
in favorable agreement with the experimentally observed values of approximately
0.7~eV and 1.1~eV for the $\text{D}^1$ and $\text{D}^2$ transitions,
respectively (Fig.~\ref{fig3}b).

Following the identification of the relevant electronic multiplets within the
parent $T_d$ symmetry, we analyze the term structure and polarization dependence
in the actual $\C3v$ symmetry (for group-theoretical details, see Supplementary
Sections \ref{SM_group_theory} and \ref{SM_selection}). Fig.~\ref{fig4}b shows
the already discussed quartet channel, where orange arrows indicate polarization
within the basal $c$-plane ($\mu_x$ and $\mu_y$), while blue arrows correspond
to polarization along the crystallographic $c$-axis ($\mu_z$). With the
exception of the $^4\!T_1$ crystal-field splitting of $V_k$ (25~meV), the
description is consistent with previous work\cite{Janzen2009}.

Fig.~\ref{fig4}c extends this discussion to the spin-doublet channel. Within
the $\C3v$ crystal field, both spin-doublet $T$ terms split into an orbital
singlet and a doublet, i.e., $\dT1 \to {\dE'} \oplus {\dA2}$ and
$\dT2 \to {\dE''}\oplus {\dA1}$. Although symmetry considerations alone do not
determine the relative ordering of the $\dE'$ and $\dA2$ sublevels from $\dT1$,
group-theoretical analysis predicts an inverted ordering for the $\dT2$
splitting compared to the quartet case, resulting in an $\dE''$, $\dA1$ energy
sequence (see Supplementary Section \ref{section:s2f}). In contrast to the quartet manifold, symmetry
arguments predict that linear crystal-field splitting should vanish for the
spin-doublet $T$ levels. The finite splittings observed experimentally, as
discussed below, could therefore be attributed to higher-order crystal-field
effects or coupling to other $T$ states that are not considered in our analysis
(e.g., resulting from $a_1t_2^2$ configuration).

Deriving the multiplet structure from the parent $T_d$ symmetry imposes
additional constraints on the optical selection rules for sublevels of doublet
$T$ states beyond those obtained from $\C3v$ symmetry alone (see Supplementary
Section \ref{SM_selection}). Specifically, transitions between the $E'$ and $E''$ levels are allowed
only for $z$-polarized light (transition C in Fig.~\ref{fig4}c), while in-plane
polarization induces transitions from $A$ to $E$ states (transitions A and B in
Fig.~\ref{fig4}c). In contrast, a description within $\C3v$ symmetry would allow
$E$ to $E$ transitions for all polarizations, as is the case for transitions
between $\dE'$ and $\dE$ (transition E in Fig.~\ref{fig4}c). Finally, the transition F from $\dA2$ to $\dE$ is restricted to in-plane polarization.

  Having established the group-theoretical framework, we turn to
  polarization-resolved TA measurements to examine the optical selection rules
  of the doublet transitions. Figs.~\ref{fig4}d~and~\ref{fig4}e show the
  polarization-resolved TA spectra of the $\text{D}^1$ transition for $V_h$ and
  $V_k$, respectively. The spectra were measured at 5~K at oblique incidence
  with $0^\circ$ and $90^\circ$ probe polarizations probing $(\mu_x,\mu_z)$ and
  $\mu_y$ dipoles. The insets display the corresponding $\text{D}^2$ spectra
  measured for $0^\circ$, $90^\circ$, and normal incidence, the latter being
  sensitive only to the in-plane dipole components $(\mu_x,\mu_y)$ irrespective
  of the probe polarization. From the group-theoretical analysis it follows that
  the $\text{D}^1$ transition should be composed of the E and F sub-transitions.
However, we can only observe a single resonance that is of Gaussian shape with
a linewidth of 1.5~meV, in contrast to the Lorentzian-shaped V1, V1$\prime$ and
V2 ZPLs that have about 0.8~meV linewidth. This larger linewidth with Gaussian
lineshape is most likely because for $\text{D}^1$ we are limited by the spectral
measurement resolution in this energy range (see Supplementary
Section~\ref{SM_setup}). 

Observing only one resonance instead of two (E and F, see Fig.~\ref{fig4}c)
could have two reasons. First, a single resonance would be observed if the
splitting of $^2T_1$ is much less than 1~meV, effectively merging E and F into a
single transition in our measurements. Another possibility is that the splitting
is large enough for the $^2A_2$ level not to be thermally populated at 5~K. We
therefore measured D$^1$ at 60~K, anticipating the appearance of the F
transition as a new peak to the red side of the E transition, which however was
not detected (see Supplementary Section~\ref{SM_temp}). The absence of the F
transition at both 5~K and 60~K sample temperature leads us to estimate the
$^2T_1$ splitting to be either larger than 10~meV or smaller than 1~meV, see
Supplementary Section \ref{SM_temp}.

  The probe-polarization dependence for $\text{D}^1$ transitions (insets in
  Figs.~\ref{fig4}d~and~\ref{fig4}e) reveals a slight reduction of the $\text{D}^1$ amplitude
  when changing from $0^\circ$ to $90^\circ$ polarization. This behavior is
  consistent with the expected electronic structure if $^2A_2$ lies above the
  $E$ level, since the probe sensitivity changes from $(\mu_x,\mu_z)$ to
  $\mu_y$, i.e., the polarization component along the $c$ direction is suppressed.

  Finally, we turn to the $\text{D}^2$ transitions, shown in the insets of
  Figs.~\ref{fig4}d~and~\ref{fig4}e. In contrast to $\text{D}^1$, the polarization
  dependence is markedly different. Here, the relevant sub-transitions are A, B,
  and C, and the insets reveal a characteristic double-peak structure with
  central energies of 1.091~eV and 1.095~eV for $\text{D}^2_h$, and 1.061~eV and
  1.071~eV for $\text{D}^2_k$ (see Supplementary Section~\ref{SM_pol} for
  quantitative peak fits).
  
We can thus conclude that the $^2T_2$ splitting is 4~meV (\vsih) and 10~meV
(\vsik). In normal incidence (orange plots), which probes $\mu_x$ and $\mu_y$ but
not $\mu_z$ (and thus only the B transition), we only observe the high-energy
transitions (1.095~eV and 1.071~eV), which must be the B transition. It follows
that $^2A_1 > E''$ in $^2T_2$ and that the low-energy transition (1.091~eV and
1.061~eV) corresponds to C. We performed further TA measurements of the D$^2$ transitions at
normal incidence, where we varied the sample temperature from 5~K to 120~K (see
Supplementary Section \ref{SM_temp}). Only very subtle changes in the A, B and C
transitions were observed, which is consistent with our hypothesis of the
$^2T_1$ splitting being $< 1$~meV or $> 10$~meV. The assignment of C to the
low-energy transition is consistent with the fact that in oblique incidence, its
peak amplitude is reduced when going from 0$^\circ$ to 90$^\circ$ (black and
blue plots in Fig.~\ref{fig4}d~and~\ref{fig4}e), i.e. from probing $\mu_x$ and $\mu_z$ to
$\mu_y$, since this is the only transition containing $\mu_z$ components.

Having determined the electronic structure of the doublet channel, we want to conclude this section with a short discussion on the linewidths of the doublet transitions. Whereas the linewidth of D$^1$ is similar to V1, V1$^\prime$ and V2 (around 1~meV), the linewidth of D$^2$ is broader and more akin to V2$^\prime$ (6-8~meV). For V2$^\prime$, we explained the broadening with non-adiabatic coupling to vibronic modes of V2 that redistributes its oscillator strength, resulting in the observed broadening. This approach seems less justified for D$^2$ due to the paucity of vibronic modes in the relevant energy range, which is 25~meV for V2$^\prime$ vs. 4~meV (D$^2_h$) and 10~meV (D$^2_k$), and the fact that V1$^\prime$ is not broadened with a distance to V1 of 5~meV, i.e. very similar to D$^2_h$. We are currently not able to explain this broadening. It could be related to Jahn--Teller-induced effects in $E''$ that couple it to $^2A_1$, resulting in additional dephasing processes that broaden the linewidth.\cite{ulbricht2016jahn} This would not occur in the quartet excited state since the splitting there is inverted and $^4E$ > $^4A_2^\prime$.

\section*{Discussion}
In conclusion, we extended the known electronic structure landscape of \vsim \
in 4H-SiC via a combination of experimental and theoretical techniques,
including transient absorption spectroscopy, density functional theory, quantum embedding calculations and group theory. We spectroscopically characterized and analyzed the spin quartet and spin doublet transitions by taking into account the optical polarization dependencies with respect to the crystal orientation
and the resulting optical selection rules. For the quartet transitions, we
identify the so far elusive V2$^{\prime}$ ZPL at a photon energy that is 25~meV
higher than the V2 ZPL. This V2$^{\prime}$ ZPL appears strongly broadened as
compared to the V1, V1$^{\prime}$ and V2 ZPLs, which we attribute to
non-adiabatic coupling to vibronic modes of the V2 transition. A heuristic model
is shown to qualitatively reproduce the broadening effect. Future work could address the underlying mechanism in more detail, both experimentally by, e.g., employing multidimensional coherent spectroscopy, as well as theoretically using non-adiabatic \textit{ab initio} simulations.
We also calculate the full absorption and emission lineshapes, including phonon sidebands, of the spin quartet transitions of \vsim \ in excellent agreement with the measured spectra.

In the spin doublet channel, we for the first time measure the spectroscopic
signatures of the \vsim \ electronic transitions that reveal important details about the doublet channel electronic structure, which is expanded upon using quantum embedding methods and group theory. Each defect configuration (\vsih\ and \vsik) shows two distinct doublet absorption transitions from the lowest doublet level $^2T_1$: D$^1_h$ and D$^1_k$ in the 0.7~eV range and D$^2_h$ and D$^2_k$ in the 1.1~eV range.
Furthermore, the ZPLs of the D$^2$ transitions show a double peak structure that
we attribute to a crystal-field splitting of the $^2T_2$ excited state into
$^2A_1$ and $^2E$ levels (4~meV for \vsih\ and 10~meV for \vsik), similar to the quartet transitions, where it is 5~meV for \vsih\ and 25~meV for \vsik.
The identified doublet transitions suggest a possible alternative ODMR readout signal to the commonly measured quartet PL. Similar to the singlet transition of NV centers in diamond\cite{Acosta2010, TayefehYounesi2025}, the absorption or PL of the doublet transitions could be exploited as a signal for ODMR. This approach could be beneficial for quantum technologies with fiber-based access, since the doublet transitions are situated in the important telecom wavelength range. It is conceivable that employing the doublet transitions for ODMR could improve the spin contrast, as is the case for the singlet transitions of NV centers in diamond\cite{Meirzada2019, TayefehYounesi2025}, which would be interesting to investigate in future work. Our determination of the doublet level energies could potentially also aid in improving state-selective photoionization that forms the basis of photoelectric detection of magnetic resonance (PDMR).\cite{Niethammer2019, Nishikawa2025}

Our combined experimental and theoretical approach described here can be applied to other color centers and host materials, marking it as a versatile toolbox for characterizing their electronic structure to aid the broader development of color center-based quantum technologies. TA spectroscopy can be applied to other `quantum defects', where metastable transitions (and excited states in general) have not been identified yet, such as the boron vacancy in hexagonal boron nitride (h-BN).\cite{Clua-Provost2024, Stern2022} Depending on the crystal structure and defect symmetries, polarization-resolved TA spectra combined with optical selection rules derived by group theory and \textit{ab initio} methods can then be used to infer details about the defect's electronic structure.

\section*{Methods}
\subsection*{Spectroscopy and Materials}

We employed two TA spectroscopy setups, one for long timescale dynamics (ns-$\mu$s)\cite{Younesi2022} and another setup for short timescales (fs-ns)\cite{luu2024nitrogen}. As a pump, we either use an optical parametric oscillator (OPO, Ekspla NT242) that generates wavelength-tunable narrowband pulses with few-ns pulse duration or a non-collinear optical parametric amplifier (NOPA, AG Riedle) that generates broadband pulses with sub-ps pulse duration. The SiC sample was housed in a closed-cycle Helium-cooled cryostat. We used a high-purity semi-insulating (HPSI) 4H-SiC sample that was irradiated by 2-MeV electrons at room temperature to a fluence of 8×10$^{18}$~cm$^{-2}$ at room temperature. More details can be found in Supplementary Section \ref{SM_setup}.

\subsection*{Lineshape calculations}


Lineshape calculations were performed utilizing spin-polarized density functional theory with the
meta-GGA r$^2$SCAN functional\cite{Furness_2020} and a force-constant embedding
approach, which facilitates the description of defect electron–phonon coupling
in the dilute limit.\cite{Razinkovas_2021,Zalandauskas2025} The methodology
considerably reduces finite-size effects that arise from restricted supercell
geometry, such as the limited number of vibrational modes and insufficient
representation of coupling to long-wavelength phonons, thereby enabling the
calculation of high-resolution optical lineshapes.

Optical lineshapes are computed for transitions between \gsq\ and \esqA\ states
using the adiabatic Huang–Rhys framework.\cite{Huang_1950,Alkauskas_2014}
Mode-resolved Huang–Rhys factors are determined by calculating the displacement
between the equilibrium geometries of the ground- and excited-state
potential-energy surfaces in a large supercell. These displacements are then
projected onto the phonon eigenmodes of the embedded system. The resulting
absorption and emission spectra are evaluated using the generating-function
formalism.\cite{Lax_1952,Kubo_1955}

Non-adiabatic vibronic coupling between near-degenerate electronic states,
including Jahn–Teller effects, is not considered in this DFT based study.
Additional technical details about the density functional theory parameters,
embedding procedure, phonon calculations, and lineshape evaluation are available
in the Supplementary Information.

\subsection*{Group-theory framework}

Since the trigonal crystal field in 4H-SiC represents only a minor perturbation
to the local tetrahedral environment, the primary energy separations of the
silicon vacancy center are dictated by the intrinsic multiplet structure of the
underlying many-body configurations in ideal $T_d$ symmetry. Specifically, the
low-energy electronic structure is determined by the multiplet splittings of the
$a_1 t_2^2$ and $t_2^3$ hole configurations, which result from two-body Coulomb
and exchange interactions and are well described by classical multiplet
theory.\cite{sugano1970} Conversely, starting directly from $C_{3v}$ symmetry
and constructing many-body states based on the occupation of slightly split
single-particle levels ($t_2 \rightarrow a_1(v) \oplus e$) can obscure the
parent multiplet structure and potentially result in an incorrect ordering of
the many-body energy levels.

Therefore, a $T_d$-based description is adopted as the starting point. Within
$T_d$ symmetry, the lowest-lying spin-doublet manifold originates from the
$t_2^3$ hole (or equivalently, electron) configuration, whose term structure is
well established in classical multiplet theory.\cite{sugano1970} The
corresponding many-body wavefunctions are detailed in Supplementary
Note~\ref{SM_sym}.

To analyze the polarization dependence of the optical transitions, the resulting
electronic terms are reformulated in a symmetry-adapted basis whose components
transform as irreducible representations of the $C_{3v}$ point group. This is
accomplished by transforming the standard representation matrices of the $T_2$
irreducible representation of $T_d$ so that its components correspond to the
$A_1$ and $E$ representations of the $C_{3v}$ subgroup. This procedure is
equivalent to rotating the coordinate system so that the $z$-axis is aligned
with the crystallographic $c$-axis and coincides with the $C_{3v}$ rotational
axis, while the $x$- and $y$-axes lie within the basal plane of 4H-SiC. This
method preserves the underlying $T_d$ multiplet structure and at the same time
allows the electronic states to be treated as eigenstates of $C_{3v}$ symmetry.
Additional technical details regarding the basis construction and symmetry
reduction are provided in Supplementary Section~\ref{SM_selection}.

\subsection*{cRPA+FCI calculations}

The cRPA calculations were performed in the \texttt{VASP} software
package\cite{Kresse1996a} to extract screened two-body Coulomb matrix elements.
The minimal active space consisted of 4 spatial orbitals localized on the
neighboring atoms. Localization was performed using the Wannier90 package as
interfaced with \texttt{vasp}. Eigenvalues from DFT calculations were used as
approximations to the one-body matrix elements. The mean-field electron-electron
interaction was removed but the exchange-correlation energy contribution from
the active space to the one-body matrix elements remained. The resulting
Hamiltonian was diagonalized using full configuration interaction in an in-house
implementation. A 128 atom supercell was used in conjunction with $\Gamma$-point
sampling of the Brillouin zone. A plane wave cutoff of 520~eV was used. Further
details are given in Supplementary Sections~\ref{subsection:QuantumEmbedding}
and~\ref{SM_QE}.

\section*{Data availability}

The data that support the findings of this study are available from the
corresponding author upon request. Source data are provided with this paper.


\bibliographystyle{unsrt}
\bibliography{ref.bib}

\section*{Acknowledgments}
RU acknowledges funding by the Max-Planck Society. MEB, CL and LR acknowledge financial support that was kindly provided by the Research Council of Norway and the University of Oslo through the frontier research projects QuTe (no. 325573, FriPro ToppForsk-program) and TASQ (no. 354419, FriPro). NTS acknowledges fundings from the European Union under Horizon Europe for the projects QuSPARC (no. 101186889), QRC-4-ESP (no.  101129663) and QUEST (no. 101156088), the Knut and Alice Wallenberg Foundation (KAW 2018.0071), the Vinnova (no. 2024-00461 and no. 2025-03848). Some of the computations were performed on resources provided by UNINETT Sigma2 --- the National Infrastructure for High Performance Computing and Data Storage in Norway, supercomputer GALAX of the Center for Physical Sciences and Technology, Lithuania, and the High Performance Computing Center “HPC Saulėtekis” in the Faculty of Physics, Vilnius University.
GT was supported by the J\'anos Bolyai Research Scholarship of the Hungarian Academy of Sciences and acknowledges the funding from the Hungarian National Research, Development and Innovation Office (NKFIH) under the NKKP STARTING grant (no. 150113).

\section*{Author contributions}
RU initiated and supervised the project. ATY and MTL performed the experiments and analyzed the data. VŽ, LR and MEB performed the lineshape calculations. CL performed the quantum embedding calculations. GT and LR worked on the group theory. NTS provided the SiC sample. TO performed the electron irradiation. All authors participated in writing the manuscript.
\section*{Competing interests}
The authors declare no competing interests.


\clearpage
\onecolumngrid


\titleformat{\section}
  {\centering\normalfont\large\bfseries}
  {}{0pt}{}

\setcounter{subsection}{0}
\counterwithout{subsection}{section}

\renewcommand{\thesubsection}{S\arabic{subsection}}
\renewcommand{\thesubsubsection}{\thesubsection.\arabic{subsubsection}}

\titleformat{\subsection}
  {\normalfont\bfseries}
  {\thesubsection}
  {0.75em}
  {}

\titlespacing{\subsection}
  {0pt}{10pt plus 3pt minus 2pt}{6pt}

\setcounter{tocdepth}{2}

\setcounter{figure}{0}
\setcounter{table}{0}
\setcounter{equation}{0}
\renewcommand{\thefigure}{S\arabic{figure}}
\renewcommand{\thetable}{S\arabic{table}}
\renewcommand{\theequation}{S\arabic{equation}}


\section*{Supplementary Information}
\addcontentsline{toc}{section}{Supplementary Information}

\resumetoc
\tableofcontents
\newpage


\def \vsih{$V_{h}$ }
\def \vsik{$V_{k}$ }
\section*{Experimental methods}
\addcontentsline{toc}{section}{Experimental Methods}

\subsection{Measurement setup}
\label{SM_setup}

A more detailed description of the employed transient absorption (TA) spectroscopy setups has been previously published \cite{luu2024nitrogen,Younesi2022}. Here, we present the essential information explaining the setups, including the specific parameters used in this study. 

Two TA setups are employed: one spanning the femto- to nano-second (fs-ns) range, and the other spanning the nano- to micro-second (ns-$\mu$s) range. In both methods, a fiber laser (Amplitude Tangerine SP) with a repetition rate at typically 2\,kHz is employed as the main output source. Although the repetition rate can reach up to 40\,MHz for this laser, the chosen value is appropriate to allow the measured defects to return to the initial state after pumping, as well as compatibility with the optical parametric oscillator (OPO) laser system, whose repetition rate is locked to 1\,kHz. The output of the fiber laser is about 175\,$\mu$J with a pulse duration of $\sim$150\,fs, which is then distributed to a non-collinear optical amplifier (NOPA, AG Riedle NOPA Rainbow) and a home-built OPA. The spectral output of the NOPA spans in the 390--1300\,nm range, although it is mostly operating in the 620--900\,nm range for the measurements presented in this paper. In this spectral range, the output pulse energy is about 2-4\,$\mu$J depending on the wavelength. The output is employed as the pump in several experiments. As for the OPA, the output spectral range is 660--2100\,nm. The extension to the near IR is due to the collinear geometry allowing the idler to be used as an output.

In order to perform the TA measurements, either the output the NOPA or the output of a tunable OPO laser (Ekspla NT242) was utilized as the pump. The NOPA output has a pulse duration of about 150\,fs (inheriting from the fiber laser) and thus suitable for studying fast population dynamics. However, the broad linewidth (at least $\sim$5\,nm/10\,meV) of the output means that resonant excitation of the ZPLs may not be efficient. The main use of the NOPA output here is to cover the excitation around 700--750\,nm (1.55--1.65\,eV) in non-resonant pumping experiments, since the OPO laser power output is inefficient in this spectral range. In contrast, the OPO laser has a pulse duration of approximately 4\,ns and a linewidth of sub-0.5\,nm (<0.4\,meV) in the interested spectral range. The narrow optical bandwidth of the ns-OPO enables superior spectrally selective pumping, enabling precise resonant excitation of specific ZPLs and hence the straightforward distinction of the features associated with h- and k-site vacancies (\vsih and \vsik, respectively).

For the fs-ns TA setup, the probe beam is the idler output of the OPA at about 1170\,nm focused onto a 4\,mm-thick c-cut sapphire crystal to generate a broadband pulse spanning between the 500-1300\,nm spectral region. The probe beam is then recollimated and subsequently focused on to the sample with a diameter of about 60\,$\mu$m to be completely filled by the pump beam. Exiting the sample, the probe is then imaged onto a line camera (Teledyne e2v Octoplus for < 1100 nm and Xenics Manx for 950 - 1600 nm). The spectral resolution is determined by the groove density of the grating used to image the probe onto each pixel of the line camera. In our case, this resolution is about 0.25\,nm per pixel, which is roughly 0.3-0.6\,meV in the spectral range where this setup is employed. Due to the detection limit of this setup towards the IR, the D$^1$ transitions can not be observed.

In the case of the ns-$\mu$s TA setup, a continuous wave (CW) incoherent broadband WL source (XWS-30, ISTEQ BV) is used as the probe instead. This method allows observation of TA spectra in the $\mu$s range as it is only limited by the repetition rate of the pump pulses. This WL source produces a spectrum in the range of 400--2000\,nm (0.6--3.1\,eV) and thus covers most electronic transitions relevant to the discussion. After interacting with the sample, this probe beam passes through a monochromator and is then detected with a photodiode (PD). The process is repeated for each wavelength until the desired spectral range is covered. The spectral resolution is limited primarily by the slit width of the monochromator, and thus is sometimes a trade-off between spectral resolution and signal-to-noise ratio. Generally, this resolution is sub-2\,meV in most of the interested spectral regions, although further into the IR region ($>$1700\,nm -- $<$0.73\,eV), where the D$^1$ transitions are observed, the slit width is increased and thus the resolution may be higher than the aforementioned value. The setup can also be used to measure time-resolved PL by blocking the WL and simply recording the emission.

All measurements performed in this paper are conducted at 5\,K in a closed-cycle helium cryostat (Advanced Research Systems ARS-10HW), unless stated otherwise. 

\clearpage

\subsection{Oblique Incidence Probe Configuration}
\label{SM_oblique}

 \begin{figure}[h]
    \centering
\includegraphics[width=0.4\columnwidth]{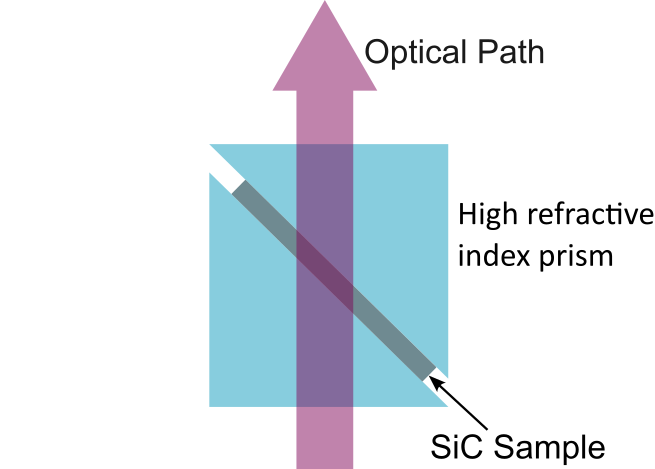}
    \caption{Schematic of the sample sandwiched between two high refractive index rutile prisms and optical path. }
    \label{fig:chap6_Prism_config}
\end{figure}

As illustrated in Figure~\ref{fig1}d, we can infer the coupling of the polarization vector of the probe with the transition dipoles as follows:

\begin{itemize}
    \item Oblique incidence case: 0°-polarization couples to the $\mu_x$ and $\mu_z$ dipoles; 90°-polarization couples to the $\mu_y$ dipole.
    \item Normal incidence case: 0°-polarization couples to the $\mu_x$ dipole; 90°-polarization couples to the $\mu_y$ dipole.
\end{itemize}

Under the $C_{3v}$ point group, the transition dipole moments transform as either $(x,y)$ ($E$) or $z$ ($A_1$) in Cartesian coordinates. For example, optical transitions between electronic states $A_2\leftrightarrow\,A_2$ (whose direct product transforms as $A_1$) are allowed by z dipoles, while the $A_2\leftrightarrow\,E$ transition (direct product transforms as $E$) is allowed with x,y dipoles \cite{udvarhelyi2020vibronic}. A more complete description of selection rules is detailed in \ref{SM_selection} below.

For the oblique incidence case, we can summarize this experimental scheme with a few useful statements:
\begin{itemize}
    \item 0°-polarization probe can observe all transitions in either $V_h$ or $V_k$ configuration.
    \item 90°-polarization probe only couples to the $\mu_y$ dipole, thus any transition allowed by $\mu_z$ cannot be observed by the probe pulse. 
\end{itemize}


To achieve oblique incidence, a c-cut 4H-SiC sample was placed at an angle of 45$^\circ$ with the incident beam. However, due to the considerable refractive index mismatch between the SiC sample and vacuum (approximately 2.7 and 1, respectively), the beam enters the sample at a much smaller angle (\textasciitilde\,15$^\circ$), therefore barely probing the $\mu_z$ component. To resolve this issue, the sample was positioned between two high refractive index rutile right-angle prisms, thereby ensuring the beam passed with minimal refraction (see Fig.\,\ref{fig:chap6_Prism_config} for a schematic representation of the sample configuration). Although passing the laser pulses through the prisms chirps the pulses to a few ps and limits the temporal resolution, all the population dynamics observed in this study are in the ns range.

\subsection{Scanning Excitation Transient Absorption}
\label{SM_SETA}

The narrow linewidth of the EKSPLA NT242 OPO is utilized to scan the pump excitation wavelength on the SiC samples at every 1\,nm, then a TA measurement is carried out at each excitation wavelength. We call this method scanning excitation TA (SETA).

\begin{figure}[ht]
    \centering
    \includegraphics[width=0.7\linewidth]{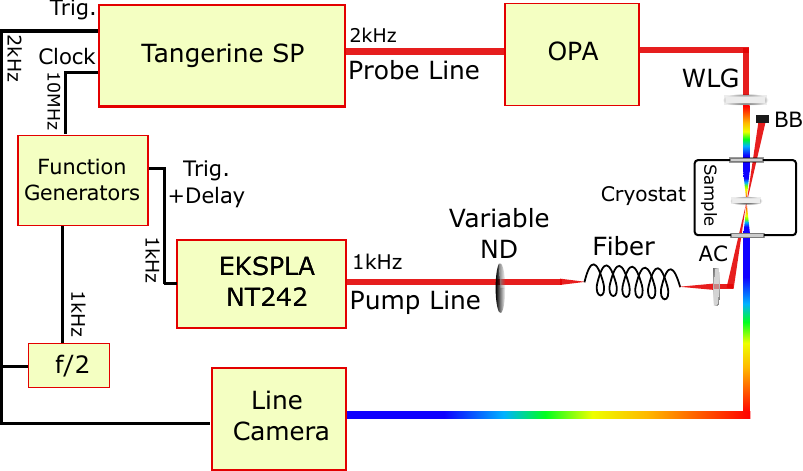}
    \caption{Schematics of the SETA setup including the interlock scheme between the fiber laser (Tangerine SP) and the OPO laser (EKSPLA NT242).}
    \label{fig:SM_SETA}
\end{figure}

The timing of the OPO laser (pump) pulse and the fiber laser (probe) pulse needs to be synchronized and thus requires an interlock scheme that can also generate an electronically delayed signal. This is achieved by using a frequency-divided signal of the internal clock of the fiber laser at 10\,MHz, before using this signal to serve as an external clock for a set of function generators. The function generators produce a pulse train at 1\,kHz, whose delay is adjustable with minimum increments of 5\,ns, relative to the fiber laser pulse. This pulse train is then used to trigger the OPO laser, producing output at 1\,kHz repetition rate and synchronized to the fiber laser. The pulse sequence and synchronization schemes are illustrated in Fig.~\ref{fig:SM_SETA}. 

The OPO laser output beam passes through a variable neutral density filter to allow control of the pulse energy of the pump pulse. It is then coupled to a 5\,m-long multimode fiber by a spherical mirror with a focal length of 200\,mm. The output from the fiber is then collimated before being focused onto the sample using an achromatic lens with anti-reflective coating in the VIS-NIR region (400-1100\,nm) with a focal length of +75\,mm. The spot size of the pump pulse at the sample is approximately 200\,$\mu$m in diameter. The energy output of the OPO laser at each wavelength is measured with a power meter, and is later used to normalize the obtained TA spectra to the photon flux arriving at the sample. 

This method outputs a two-dimensional dataset of the TA spectrum at each excitation wavelength, as shown in Fig.~\ref{fig:chap6_2D_SiC}. The spectra obtained were measured at about 4\,ns pump-probe delay. This delay time was chosen such that the probe pulse would arrive after the temporally-broad pump pulse from the OPO laser, yet early enough to observe population in the excited quartet states of the SiC samples as SE. 

\subsection{Materials}
We used a high-purity semi-insulating (HPSI) 4H-SiC sample that was irradiated by 2-MeV electrons at room temperature to a fluence of 8 × 10$^{18}$~cm$^{-2}$ at room temperature. The samples were cut from a top of a boule and the top layer is curved with original as-grown surface (the $(0001)$ plane).

\clearpage

\section*{Theoretical methods}
\addcontentsline{toc}{section}{Theoretical methods}
\label{section:s2}

\subsection{DFT calculations} 
\label{SM_DFT}

Vibrational properties of the $\mathrm{V_{Si}^{-}}$ defects in 4H-SiC were
computed using density functional theory (DFT) within the Kohn--Sham formalism.
All calculations were performed using the Vienna \textit{ab initio} Simulation
Package (VASP)~\cite{Kresse1993,Kresse1994,Kresse1996a}. Core and valence
electrons were treated using the projector augmented-wave (PAW)
method~\cite{Blochl_1994,Kresse1996} and a plane-wave basis set, respectively.
The meta-GGA functional r$^{2}$SCAN~\cite{Furness_2020} was employed, owing to
its demonstrated accuracy in describing the structural and electronic properties
of point defects in 4H-SiC~\cite{Abbas2025,Zalandauskas2025}.

Hexagonal 4H-SiC supercells containing 576 atoms were constructed by repeating
the primitive cell $6\times6\times2$ times along the crystallographic axes. The
Brillouin zone was sampled at the $\Gamma$ point, and a plane-wave cutoff energy
of $600$~eV was employed. The electronic self-consistent field (SCF) cycle and
atomic forces were converged to within $10^{-8}$~eV and 0.005~eV~\r{A}$^{-1}$,
respectively. Point defects were introduced by removing a silicon atom from the
lattice to create a silicon vacancy ($\mathrm{V_{Si}}$). All calculations were
performed for the negative charge state of the vacancy, $\mathrm{V_{Si}^{-}}$,
which corresponds to the only known optically bright states of this defect center.

Phonon modes of 4H-SiC supercells containing $\mathrm{V_{Si}^{-}}$ defects were
computed using the finite-displacement method, with the displaced atomic
configurations generated by the \textsc{PHONOPY}
package~\cite{Togo-phonopy-2015}. A displacement amplitude of $0.01$~\r{A} from
the equilibrium geometry was used.

\subsection{Electron-phonon coupling}
\label{SM_ep}

Within the Franck--Condon approximation, the normalized luminescence and
absorption lineshapes at absolute zero temperature ($T = 0$~K) are given
by~\cite{Razinkovas_2021}:
\begin{equation}
  \label{eq:lineshape}
  L(\hbar\omega) = C \omega^{\kappa} A(\hbar\omega),
\end{equation}
where $C$ denotes a normalization constant, $A(\hbar\omega)$ denotes the
spectral function of electron--phonon coupling, and the exponent $\kappa$ is
equal to 3 for luminescence and 1 for absorption. The spectral function is then
defined by the equation
\begin{equation}
  \label{eq:non_degen_spectral_function}
  A(\hbar\omega) = \sum_{m}
    \left| \braket{\chi_{i;0}|\chi_{f;m}} \right|^2
    \delta \left( E_{\mathrm{ZPL}} \mp (\varepsilon_{fm} - \varepsilon_{f0}) - \hbar\omega \right),
\end{equation}
where $E_{\mathrm{ZPL}}$ denotes the zero-phonon line (ZPL) energy, and
$\chi_{i;0}$ and $\chi_{f;m}$ are the vibrational wave functions of the initial
and final electronic states, respectively. The quantity $\varepsilon_{f m}$
corresponds to the energy of the $m$-th vibrational level in the final
electronic manifold, measured relative to the minimum of the potential energy
surface. The sign in the argument of the $\delta$ function distinguishes
luminescence (minus sign) from absorption (plus sign) processes. The optical
spectral function $A(\hbar\omega)$ describes the vibrational transition
amplitudes and plays a crucial role in determining the lineshape.

Owing to the intrinsic differences between the vibrational structures of the
ground and excited electronic states, the overlap integrals
$\braket{\chi_{i;0}|\chi_{f;m}}$ appearing in
Eq.~\eqref{eq:non_degen_spectral_function} are inherently multidimensional. A
direct evaluation of these integrals in defect systems with numerous vibrational
modes becomes computationally challenging. To overcome this limitation we employ
equal-mode approximation, which assumes identical vibrational mode shapes and
frequencies in the initial and final electronic
states~\cite{Razinkovas_2021,Markham_1959}. The spectral function
$A(\hbar\omega)$ can be obtained in the time domain using the generating
function approach introduced by Kubo and Lax~\cite{Kubo_1955,Lax_1952}. Within
this approach, $A(\hbar\omega)$ is derived from the generating function $G(t)$
as:
\begin{equation}
  \label{eq:opt_spectral_func_gen}
  A(\hbar\omega) = \frac{1}{2 \pi} \int_{-\infty}^{\infty}{G(t) e^{-\gamma |t|} e^{- i (E_{\mathrm{ZPL}} / \hbar - \omega) t} \mathrm{d}t},
\end{equation}
where the $e^{-\gamma |t|}$ term is introduced to phenomenologically account for
homogeneous Lorentzian broadening of the ZPL not captured by the present
theoretical framework, while also effectively incorporating inhomogeneous
broadening effects. The broadening parameter $\gamma$ is chosen to reproduce the
experimentally observed ZPL linewidth and is set to $\gamma = 0.30$~meV in all
calculations. Within the equal-mode approximation, the generating function
$G(t)$ takes the form
\begin{equation}
  \label{eq:gen_func_with_sum}
  G(t) = \exp \left[ -S_{\mathrm{tot}} + \sum_{k} S_{k} e^{\pm i \omega_{k} t} \right],
\end{equation}
where the plus (minus) sign corresponds to luminescence (absorption), $S_k$
denotes the partial Huang--Rhys (HR) factor, and the summation runs over all
vibrational modes in the system. The total electron--phonon coupling strength is
defined as $S_{\mathrm{tot}} = \sum_{k} S_{k}$. The partial HR factor $S_k$
quantifies the average number of phonons excited during an optical
transition~\cite{Huang_1950} and is given by
\begin{equation}
  \label{eq:part_hr_factor}
  S_k = \frac{\omega_k \Delta Q_{k}^{2}}{2 \hbar}.
\end{equation}
Here, $\Delta Q_{k}$ represents the ionic displacement along the $k$-th normal
mode induced by the optical transition. More explicitly, $\Delta Q_k$ is defined
as the projection of the mass-weighted displacement between the ground- and
excited-state equilibrium configurations onto the normalized phonon eigenvector
$\bm{\eta}_{k}$:
\begin{equation}
  \label{eq:delta_q}
  \Delta Q_{k} = \sum_{m\alpha}{\sqrt{M_{m}} \Delta R_{m\alpha} \eta_{k; m\alpha}}.
\end{equation}
In this expression, $\Delta R_{m\alpha}$ denotes the displacement of atom $m$
along the Cartesian direction $\alpha$, and $M_{m}$ is the corresponding atomic
mass.

To describe the generating function in Eq.~\eqref{eq:gen_func_with_sum} for
extended systems, there is a need to account for a continuum of vibrational
frequencies. We introduce the spectral density of the electron--phonon coupling
(also referred to as the spectral function of electron--phonon coupling):
\begin{equation}
  \label{eq:spectral_function_of_ep_coupling}
  S(\hbar\omega) \equiv \sum_{k} S_{k} \delta(\hbar\omega_{k} - \hbar\omega).
\end{equation}
With this definition the generating function can be recast in the integral form
\begin{equation}
  \label{eq:gen_func_with_integral}
  G(t) = \exp \left[-S_{\mathrm{tot}} + \int S(\hbar\omega) e^{\pm i \omega t} \,\mathrm{d}\omega \right].
\end{equation}
The Dirac $\delta$-functions in Eq.~\eqref{eq:spectral_function_of_ep_coupling}
are approximated using Gaussian functions, yielding a smooth representation of
the spectral function of electron--phonon coupling. The Gaussian widths $\sigma$
are chosen to decrease gradually from $1.5$~meV at zero frequency to $0.5$~meV
at the maximum phonon frequency.

Within the supercell approach, a particular challenge arises when trying to determine the
relaxation profile after an optical transition. The finite size of the supercell
in direct DFT calculations limits the long-wavelength components of
$\Delta R_{m\alpha}$ in Eq.~\eqref{eq:delta_q}. To capture the relaxation
profile in the dilute limit, we compute $\Delta Q_{k}$ for each vibrational mode
of a large supercell constructed using the force-constant embedding
methodology (described in the following subsection).  This approach
exploits the linear relation between forces and atomic displacements within
the harmonic approximation. The relaxation component $\Delta Q_k$ is then
obtained as follows:
\begin{equation}
  \label{eq:delta_Q_forces}
  \Delta Q_{k}
  = \frac{1}{\omega_{k}^{2}} \sum_{m\alpha}\frac{F_{m\alpha}}{\sqrt{M_{m}}} \eta_{k; m\alpha},
\end{equation}
where $F_{m\alpha}$ is the force acting on atom $m$ along direction $\alpha$
when the system is in the final electronic state, but maintains the equilibrium
geometry of the initial state, as calculated in the directly accessible
supercell. $\eta_{k; m\alpha}$ denotes the vibrational mode shape obtained from
the embedded supercell. By using forces that already converged in the
computationally tractable supercell, this approach, combined with the embedding
methodology, captures electron--phonon coupling for low-frequency modes (and
subsequently vibrational resonances) and provides an accurate description of
optical lineshapes.

In our calculations, the vibrational mode frequencies are in very good agreement
with experimental data. However, when employing the r$^2$SCAN functional, we
observe a systematic underestimation of atomic relaxations. We assume that the
overall shape of the calculated spectral densities $S(\hbar\omega)$ closely
reflects the peak positions and the general description of the vibrational
structure. To compensate for the underestimated atomic relaxations, we therefore
apply a linear scaling factor $\zeta=1.25$, such that
$S'(\hbar\omega) = \zeta S(\hbar\omega)$ is used in all lineshape calculations.

\subsection{Force-constant embedding methodology}
\label{SM_emb}

The force-constant embedding methodology described in
Refs.~\onlinecite{Alkauskas_2014, Razinkovas_2021} was employed to model the
vibrational properties of large defect-containing supercells. This approach
exploits the short-range nature of interatomic forces in semiconductors to
construct an effective Hessian matrix for systems comprising thousands of atoms.
In this framework, Hessian matrix elements are assigned based on the relative
positions of the atoms: for pairs lying within a specified bulk cutoff radius
$r_{b}$, the corresponding elements are taken from the bulk supercell Hessian;
for atom pairs located within a cutoff radius $r_{d}$ of the defect, the
elements are taken from the defect-containing supercell; in all remaining cases,
the matrix elements are set to zero.

The bulk 4H-SiC supercell had dimensions of $8\times 8\times 3$, comprising 1536
atomic sites, whereas the defect-containing supercell had dimensions of
$6\times 6\times 2$ and contained 576 atomic sites. The cutoff radius for the
bulk system was $r_{b} = 12.200~\text{\r{A}}$, while the defect cutoff radius was
$r_{d} = 9.228~\text{\r{A}}$. The Hessian matrix was constructed using a
$25\times25\times8$ supercell comprising approximately 40\,000 atomic sites. By
employing the embedding methodology to mitigate finite-size effects, the
vibrational basis is expanded from 1\,722 modes in the $6\times6\times2$ defect
supercell to a total of 119\,994 vibrational modes, corresponding to an increase
by a factor of approximately $\sim 70$. This substantial enlargement of the
vibrational space enables the use of smaller Gaussian smoothing widths~$\sigma$,
thereby significantly improving the spectral resolution. Such enhanced
resolution is essential for resolving fine spectral features, such as the
splitting of the first phonon sideband peak in the emission lineshape of
$\mathrm{V_{Si}^{-}}(k)$. Moreover, this methodology grants access to an
extensive set of vibrational modes, including low-energy acoustic phonons, and
yields a relaxation profile representative of the dilute-defect limit (see
Eq.~\eqref{eq:delta_Q_forces}).

\subsection{Simple model for non-adiabatic broadening}
\label{SM_nonadiabatic}

In this Supplementary Note, we develop a simplified theoretical model describing
non-adiabatic broadening of a ZPL arising from electron--phonon coupling between two
electronic states. Electronic states are defined as solutions of the electronic
Schrödinger equation at fixed ionic coordinates. Within the adiabatic
approximation, these states generate potential energy surfaces (PES) governing
nuclear motion. When the electronic energy separation becomes comparable to
phonon energies, vibrational motion induces non-adiabatic mixing between the
states, leading to vibronic broadening of optical transitions. Hartree atomic
units are used throughout.

\paragraph*{Adiabatic model.}

The complete electron–ion Hamiltonian is expressed as the sum of the electronic
kinetic energy, the ionic kinetic energy, and the total potential energy:
\begin{align}
  H & = T_e + T_n + U(\mathbf{r},\mathbf{Q}) \\
    & = T_e + T_n + U(\mathbf{r},\mathbf{Q}_0)
        + \Delta U(\mathbf{r},\mathbf{Q}),
\end{align}
where $\mathbf{Q}$ denotes the ionic normal coordinates and $\mathbf{Q}_0$
their equilibrium configuration. The term $U(\mathbf{r},\mathbf{Q}_0)$
represents the electronic potential energy with the ions frozen at equilibrium,
while $\Delta U(\mathbf{r},\mathbf{Q})$ accounts for all modifications induced by
ionic displacements. The electronic eigenstates $\ket{\psi_i}$ of the frozen-ion
Hamiltonian $H_{\mathrm{el}} = T_e + U(\mathbf{r},\mathbf{Q}_0)$ form a convenient
basis for the expansion of the total vibronic wavefunction,
\[
\Psi(\mathbf{r},\mathbf{Q}) = \sum_i \chi_i(\mathbf{Q}) \ket{\psi_i}.
\]

Restricting the electronic basis to two electronic states, $\ket{\psi_1}$ and
$\ket{\psi_2}$, with energies $\varepsilon_1 = 0$ and $\varepsilon_2 = \Delta$,
the adiabatic Hamiltonian is diagonal in the electronic subspace. Under the
harmonic approximation, the PES are taken to be identical,
$V_{11}(\mathbf{Q}) = V_{22}(\mathbf{Q}) = \sum_k \tfrac{1}{2}\omega_k^2 Q_k^2$,
where $V_{ii} = \langle \psi_i| \Delta U(\mathbf{r},\mathbf{Q})| \psi_i\rangle$. The
vibrational wavefunctions are phonon Fock states $\ket{n_1,n_2,\ldots}$.

\paragraph*{Non-adiabatic coupling.}

Non-adiabatic effects arise from off-diagonal matrix elements of $\Delta U$,
which are assumed to be real and symmetric. Expanding the coupling linearly in normal
coordinates yields
\begin{equation}
  V_{12}(\mathbf{Q}) = \sum_k C_k Q_k
  = \sum_k K_k \omega_k (a_k^\dagger + a_k),
\end{equation}
where $K_k$ are dimensionless vibronic coupling constants. The resulting
vibronic Hamiltonian in second-quantized form reads
\begin{equation}
H =
\begin{pmatrix}
0 & 0 \\
0 & \Delta
\end{pmatrix}
+
\sum_k \frac{1}{2}\omega_k
\begin{pmatrix}
a_k^\dagger a_k + \tfrac{1}{2} & 0\\
0 & a_k^\dagger a_k + \tfrac{1}{2}
\end{pmatrix}
+
\sum_k K_k\omega_k
\begin{pmatrix}
0 & a_k^\dagger + a_k \\
a_k^\dagger + a_k & 0
\end{pmatrix}.
\label{eq:nonadiab_H}
\end{equation}

The eigenstates are expanded in a truncated phonon basis with a cutoff on the
total number of excited phonons, $\sum_k n_k \le N_{\text{max}}$.
Diagonalization of Eq.~\eqref{eq:nonadiab_H} yields vibronic eigenstates of the
form
\begin{equation}
\ket{\Psi_l}
=
\sum_{i=1}^{2}
\sum_{\substack{n_1,\ldots,n_N \\ \sum_k n_k \le N_{\text{max}}}}
c^{\,l}_{i;n_1\ldots n_N}
\ket{n_1,\ldots,n_N}\ket{\psi_i}.
\label{eq:vibronic}
\end{equation}

\paragraph{Optical lineshape.}

To calculate the optical lineshape, we consider transitions from an initial
adiabatic electronic state in the vibrational ground configuration
$\ket{\psi_{0}}\ket{0\cdots0}$ to the vibronic eigenstates defined in
Eq.~\eqref{eq:vibronic}. In addition to the non-adiabatic mixing contained in the
vibronic eigenstates, further electron--phonon coupling is incorporated through
vibrational overlap factors. This allows phonon sidebands to be generated on top
of the broadened zero-phonon line and enables direct visualization of changes in
the vibronic structure.

The vibrational overlaps are evaluated within the Huang--Rhys formalism for
shifted harmonic oscillators, characterized by Huang--Rhys factors
$S_i=\lambda_i^2$,
\[
\braket{\bm{n}|\bm{0}}
\equiv
\braket{n_{1}\cdots n_{N}|0\cdots0}
=
\prod_{i=1}^{N}
e^{-\lambda_i^{2}/2}
\frac{(-\lambda_i)^{n_i}}{\sqrt{n_i!}} .
\]

The absorption lineshape is then given by
\begin{align}
A(\hbar\omega)
&=
\sum_{i}
\left|
\sum_{\bm{n}}
\left(
c_{1;\bm{n}}^{i}\,\braket{\bm{n}|\bm{0}}
\braket{\psi_{1}|\bm{\mu}|\psi_{0}}
+
c_{2;\bm{n}}^{i}\,\braket{\bm{n}|\bm{0}}
\braket{\psi_{2}|\bm{\mu}|\psi_{0}}
\right)
\right|^{2}
\nonumber\\[4pt]
&\quad\times
\delta(E_{\mathrm{ZPL}}+\varepsilon_{\bm{0}}+\varepsilon_{i}
  -\hbar\omega).
  \label{eq:nonad_lineshape}
\end{align}
Because the phonon spectrum is discrete, the energy-conserving delta function is
replaced in practice by a Gaussian of finite width $\sigma$.

When the optical transitions $\psi_0 \rightarrow \psi_1$ and
$\psi_0 \rightarrow \psi_2$ couple to orthogonal optical polarization
directions, the polarization-resolved absorption lineshapes are obtained by
projecting the dipole operator onto the selected polarization. In
Eq.~\eqref{eq:nonad_lineshape}, this is implemented by setting the electronic
transition matrix element associated with the non-selected polarization to zero.
For example, when evaluating the lineshape for the polarization corresponding to
the $\psi_0 \rightarrow \psi_1$ transition, the matrix element
$\braket{\psi_2|\boldsymbol{\mu}|\psi_0}$ is set to zero, and vice versa. This
procedure is applied in the numerical calculations presented below.

\paragraph{Numerical simulations.}

\begin{figure}
  \centering
  \includegraphics[width=0.7\textwidth]{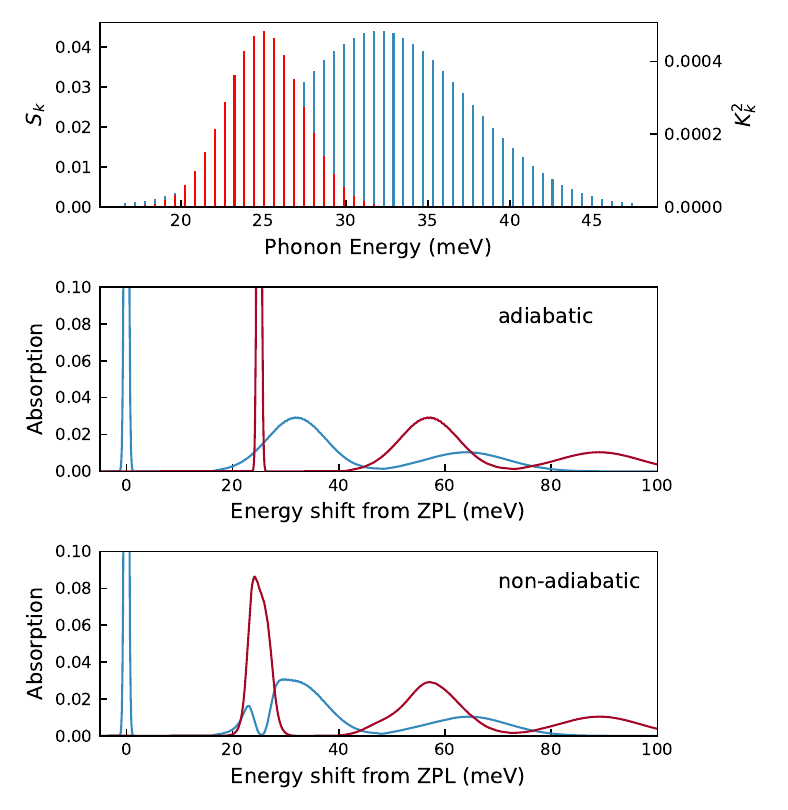}
  \caption{\textbf{Minimal model for non-adiabatic ZPL broadening.}
(a) Distributions of Huang--Rhys factors $S_k$ (blue) and non-adiabatic coupling
strengths $K_k^2$ (red) over 52 effective vibrational modes. The parameters are
chosen such that $S_{\mathrm{tot}}=\sum_k S_k=1$ and
$K^2_{\mathrm{tot}}=\sum_k K_k^2=0.005$, with non-adiabatic coupling restricted
to modes in the vicinity of the second ZPL at 25~meV.
(b) Calculated absorption lineshapes for the two optical transitions in the
absence of non-adiabatic coupling ($K_{\mathrm{tot}}=0$), assuming orthogonal
polarization directions for transitions involving $\psi_1$ and $\psi_2$.
(c) Absorption lineshapes obtained from the full vibronic Hamiltonian including
non-adiabatic coupling, illustrating the resulting broadening of the
zero-phonon line. The model is intended to capture the mechanism of
non-adiabatic ZPL broadening rather than to reproduce the full phonon sideband
structure quantitatively.
\label{fig:nonad}}

\end{figure}

The numerical implementation of the model is illustrated in
Fig.~\ref{fig:nonad}. Figure~\ref{fig:nonad}a shows the distributions of the
non-adiabatic coupling strengths $K_k^2$ (red) and Huang--Rhys factors $S_k$ (blue)
over 52 vibrational modes. The accumulated Huang--Rhys factor is
$S_{\mathrm{tot}} = \sum_k S_k = 1$, while the total non-adiabatic coupling
strength is comparatively small, $K^2_{\mathrm{tot}} = \sum_k K_k^2 = 0.005$.
Non-adiabatic coupling is included only for modes in the vicinity of the second
zero-phonon line at 25~meV, modeled by a Gaussian distribution centered at
25~meV with a width of $\sigma = 3.3$~meV. In contrast, the Huang--Rhys factors
follow a broader Gaussian distribution centered at 32~meV with
$\sigma = 5.5$~meV, chosen to approximately reproduce the first experimentally
observed phonon sideband peak discussed in the main text.

Figure~\ref{fig:nonad}b shows the calculated absorption lineshapes for two
orthogonal polarization directions associated with the two optical transitions
in the absence of non-adiabatic coupling ($K_{\mathrm{tot}}=0$), assuming that
transitions involving $\psi_1$ and $\psi_2$ couple to orthogonal polarizations.
Finally, Fig.~\ref{fig:nonad}c presents the absorption lineshapes obtained from
the full vibronic Hamiltonian. The phonon basis is truncated by imposing a
maximum total number of excited phonons $N_{\mathrm{max}} = 2$.

The dip in the absorption sideband visible in the spectral overlap region in
Fig.~\ref{fig:nonad}c should not be interpreted as a physical feature expected
in experiment. It originates from the restricted set of Jahn--Teller--active
modes included in the present model and reflects a redistribution of spectral
weight rather than a true suppression of absorption. In an experimental
spectrum, this region would be filled by contributions from additional
vibrational modes, including those coupled non-adiabatically, leading to a
smooth overlap of spectral features. The purpose of the present model is
therefore not to reproduce the full phonon sideband in detail, but to
demonstrate the mechanism by which non-adiabatic coupling leads to a broadening
of the ZPL.

\subsection{Quantum embedding for electronic structure}
\label{subsection:QuantumEmbedding}

Quantum embedding calculations were performed based on constrained
random phase approximation (cRPA). In this method an effective Hamiltonian for a subset of states
\begin{equation}
    h = -\sum_{ij}t_{ij}c_i^{\dagger}c_j + \frac{1}{2}\sum_{ijkl}u_{ijkl}c_i^{\dagger}c_j^{\dagger}c_lc_k,
  \end{equation}
is parameterized using a combination of density functional theory and cRPA \cite{Muechler-Dreyer-embedding-2022}.
The cRPA calculations were performed using the projector augmented wave method as implemented in \texttt{VASP}.\cite{Kresse1996a} The parameterized Hamiltonian was then diagonalized using an in house implementation of full configuration
interaction. 
The Hamiltonian was constructed according
to the methodology outlined in Ref.~\citenum{Muechler-Dreyer-embedding-2022},
including double counting corrections for the mean field Coulomb energy. To
recap it briefly, spin polarized DFT is used to obtain the geometry upon which a
spin paired calculation is performed. Here, based on the localization of the
Bloch states, the activate space is chosen. In the case of the silicon vacancy,
the activate space consists of 4 spatial orbitals, corresponding to the dangling
bonds. The one-body matrix elements ($t$) are obtained as the Kohn--Sham eigenenergies
of the activate space states. Next, a cRPA calculation is performed to obtain
the two-body matrix elements ($u$) in the active space. The double counting
corrections is then computed from the two-body matrix elements
\cite{Muechler-Dreyer-embedding-2022}, which are then applied to the one-body
matrix elements. 

\subsection{Electronic states and crystal-field splitting: group theory analysis}

\label{SM_group_theory}

\def\ket#1{\left|#1\right\rangle}
\def\t#1#2{t_{#1#2}}
\def\bt#1#2{\bar{t}_{#1#2}}
\def\a1{a_1}
\def\ba1{\bar{a}_1}

As discussed in the main text, we describe the electronic structure of the
silicon vacancy within the ideal tetrahedral ($T_d$) symmetry. The relevant
many-body configurations are formulated in terms of three holes occupying the
defect-centered $t_2$ and $a_1$ orbitals. We use the conventional representation
found in multiplet-theory literature, in which the three components of the $t_2$
manifold transform as Cartesian vectors under $T_d$ symmetry operations, i.e.,
the orbitals $t_{2x}$, $t_{2y}$, and $t_{2z}$ transform as the $x$, $y$, and $z$
components of a vector in a standardized coordinate system where the threefold
(trigonal) rotational axes are oriented along directions equivalent to $x+y+z$.
The choice of coordinate system is not aligned with the crystallographic axes of
hexagonal SiC; specifically, the $x + y + z$ direction coincides with the
crystallographic $c$-axis and is perpendicular to the basal ($c$) plane. In this
representation, the symmetry-adapted many-body wavefunctions for a $t_2^3$ and
$a_1t_2^2$ configuration are well established from classical multiplet theory,
as summarized, for example, in the monograph by
Sugano--Tanabe--Kamimura.\cite{sugano1970} For completeness and to establish
notation, we explicitly list below the symmetry-adapted wavefunctions relevant
for the quartet and doublet manifolds.

\subsubsection{Symmetry-adapted many-body wavefunctions.}
\label{SM_sym}

\paragraph*{Quartet states.}

For the $S=3/2$ quartet manifold, the optical ground state is $|^4A_2\rangle$ where three electrons (or equivalently three holes) occupy the
$t_{2}$ orbital and transforms according to the $A_2$ irreducible representation.
The wavefunctions of all four spin projections $m_s=\{\pm\frac{3}{2},\pm\frac{1}{2}\}$ can be written as:
\begin{equation}
|^{4}A_{2}\rangle=\left\{ \begin{array}{cc}
m_{s}=+3/2\text{,} & |\t 2xt_{2y}t_{2z}\rangle\\
m_{s}=+1/2\text{,} & (|t_{2x}t_{2y}\bar{t}_{2z}\rangle+|t_{2x}t_{2y}\bar{t}_{2z}\rangle+|t_{2x}t_{2y}\bar{t}_{2z}\rangle)/\sqrt{3}\\
m_{s}=-1/2\text{,} & (|\bar{t}_{2x}\bar{t}_{2y}t_{2z}\rangle+|\bar{t}_{2x}t_{2y}\bar{t}_{2z}\rangle+|t_{2x}\bar{t}_{2y}\bar{t}_{2z}\rangle)/\sqrt{3}\\
m_{s}=-3/2\text{,} & |\bar{t}_{2x}\bar{t}_{2y}\bar{t}_{2z}\rangle
\end{array}\right.\begin{aligned}\underset{{\displaystyle =...\,\text{,}}}{\underset{}{\underset{{\displaystyle =\quad\,\;}}{}}}\frac{1}{\sqrt{6}}\begin{vmatrix}t_{2x}^{\uparrow}(\vec{x}_{1}) & t_{2y}^{\uparrow}(\vec{x}_{1}) & t_{2z}^{\uparrow}(\vec{x}_{1})\\
t_{2x}^{\uparrow}(\vec{x}_{2}) & t_{2y}^{\uparrow}(\vec{x}_{2}) & t_{2z}^{\uparrow}(\vec{x}_{2})\\
t_{2x}^{\uparrow}(\vec{x}_{3}) & t_{2y}^{\uparrow}(\vec{x}_{3}) & t_{2z}^{\uparrow}(\vec{x}_{3})
\end{vmatrix}\text{,}\\
\overset{{\displaystyle =...\,\text{,}}}{\overset{}{\overset{}{\overset{}{\overset{{\displaystyle =\quad\,\;}}{}}}}}\frac{1}{\sqrt{6}}\begin{vmatrix}t_{2x}^{\downarrow}(\vec{x}_{1}) & t_{2y}^{\downarrow}(\vec{x}_{1}) & t_{2z}^{\downarrow}(\vec{x}_{1})\\
t_{2x}^{\downarrow}(\vec{x}_{2}) & t_{2y}^{\downarrow}(\vec{x}_{2}) & t_{2z}^{\downarrow}(\vec{x}_{2})\\
t_{2x}^{\downarrow}(\vec{x}_{3}) & t_{2y}^{\downarrow}(\vec{x}_{3}) & t_{2z}^{\downarrow}(\vec{x}_{3})
\end{vmatrix}\text{,}
\end{aligned}
\label{eq:4A2_wave}
\end{equation}
where $\ket{\cdots}$ denotes a Slater determinant where $\vec{x}_{1}$, $\vec{x}_{2}$, $\vec{x}_{3}$ variables are the coordinates of three particles and we introduce the usual $t_{2\alpha}^{\uparrow}=t_{2\alpha}$, $ t_{2\alpha}^{\downarrow}=\bar{t}_{2\alpha}$ notation for spin degrees of freedom. 
From now on for simplicity, we consider only the states exhibiting maximal spin projection $m_s = +\frac{3}{2}$ $( +\frac{1}{2})$ for spin quartets (doublets) unless noted explicitly.

Exciting one electron from $a_1$ orbital into one of the three $t_{2x}$, $t_{2y}$, $t_{2z}$ triply degenerate levels gives rise the $|^{4}T_{1}\rangle$ multiplet that transforms as $T_1$. 
The corresponding symmetry-adapted components can be written in electron or hole pictures as follows
\begin{equation}
\underset{\text{electron picture}}{\underbrace{|^{4}T_{1i}\rangle=\begin{cases}
|\bar{a}_{1}\bar{t}_{2x}\bar{t}_{2y}\bar{t}_{2z}t_{2x}\rangle, & i=x,\\
|\bar{a}_{1}\bar{t}_{2x}\bar{t}_{2y}\bar{t}_{2z}t_{2y}\rangle, & i=y,\\
|\bar{a}_{1}\bar{t}_{2x}\bar{t}_{2y}\bar{t}_{2z}t_{2z}\rangle, & i=z.
\end{cases}}}\qquad\Longleftrightarrow\qquad\underset{\text{hole picture}}{\underbrace{|^{4}T_{1i}\rangle=\begin{cases}
|a_{1}t_{2y}t_{2z}\rangle, & i=x,\\
|a_{1}t_{2z}t_{2x}\rangle, & i=y,\\
|a_{1}t_{2x}t_{2y}\rangle, & i=z.
\end{cases}}}
\label{eq:4T1_wave}
\end{equation}
where we now also need to depict the occupation of the $a_{1}$ orbitals.
One may notice that the electron picture is now a five-particle wavefunction, in fact even Eq.~\eqref{eq:4A2_wave} should have an additional $|a_{1}\bar{a}_{1}\rangle$ closed shell that we neglected that time.
On the other hand, the hole picture is much more simple.
One can describe $|^{4}T_{1}\rangle$ by exciting one hole into the $a_1$ orbital from one of the three $t_{2x}$, $t_{2y}$, $t_{2z}$ orbitals by a three-particle wavefunction.
The two pictures are equivalent; however, for simplicity, we opt for the three particle language due to its much simpler and straightforward form.
Here, we remind the reader again that Eq.~\eqref{eq:4T1_wave} only spans the $m_s=+\frac{3}{2}$ maximal spin projection.
However, we will depict the $m_s=+\frac{1}{2}$ spin projection of $|^{4}T_{1}\rangle$ in Eq.~\eqref{eq:4T1_ms12} later.
It is worth to mention that, the $x$, $y$, and $z$ components of the $T_1$ representation transform as axial vectors oriented along [100], [010] and [001] crystallographic directions in the
same manner as the angular-momentum operators $\hat{L}_x$, $\hat{L}_y$, and $\hat{L}_z$.

\paragraph*{Doublet states.}
We next consider the doublet manifold in spin projection $m_s = 1/2$. 
The $|^2 T_1\rangle$ multiplet is spanned by the following
combinations,
\begin{equation}
|^{2}T_{1i}\rangle=\frac{1}{\sqrt{2}}\begin{cases}
\bigl|t_{2x}t_{2y}\bar{t}_{2y}\bigr\rangle-\bigl|t_{2x}t_{2z}\bar{t}_{2z}\bigr\rangle\text{,} & i=x,\\
\bigl|t_{2y}t_{2z}\bar{t}_{2z}\bigr\rangle-\bigl|t_{2y}t_{2x}\bar{t}_{2x}\bigr\rangle\text{,} & i=y,\\
\bigl|t_{2z}t_{2x}\bar{t}_{2x}\bigr\rangle-\bigl|t_{2z}t_{2y}\bar{t}_{2y}\bigr\rangle\text{,} & i=z.
\end{cases}
\end{equation}
The doubly degenerate $|^2 E\rangle$ manifold is given by
\begin{equation}
|^{2}E_{i}\rangle=\left\{ \begin{array}{ll}
\frac{1}{\sqrt{2}}\bigl(\bigl|t_{2x}\bar{t}_{2y}t_{2z}\bigr\rangle-\bigl|\bar{t}_{2x}t_{2y}t_{2z}\bigr\rangle\bigr)\text{,} & i=x^{2}-y^{2}\text{,}\\
\frac{1}{\sqrt{6}}\bigl(2\bigl|t_{2x}t_{2y}\bar{t}_{2z}\bigr\rangle-\bigl|t_{2x}\bar{t}_{2y}t_{2z}\bigr\rangle-\bigl|\bar{t}_{2x}t_{2y}t_{2z}\bigr\rangle\bigr)\text{,} & i=2z^{2}-x^{2}-y^{2}\text{.}
\end{array}\right.
\label{eq:2E_wave}
\end{equation}
Finally, the $|^2 T_2\rangle$ multiplet is spanned by
\begin{equation}
|^{2}T_{2i}\rangle=\frac{1}{\sqrt{2}}\begin{cases}
\bigl|t_{2x}t_{2y}\bar{t}_{2y}\bigr\rangle+\bigl|t_{2x}t_{2z}\bar{t}_{2z}\bigr\rangle\text{,} & i=x,\\
\bigl|t_{2y}t_{2z}\bar{t}_{2z}\bigr\rangle+\bigl|t_{2y}t_{2x}\bar{t}_{2x}\bigr\rangle\text{,} & i=y,\\
\bigl|t_{2z}t_{2x}\bar{t}_{2x}\bigr\rangle+\bigl|t_{2z}t_{2y}\bar{t}_{2y}\bigr\rangle\text{,} & i=z.
\end{cases}
\end{equation}
Here we note that, in contrast to $T_1$ previously, $x$, $y$, $z$ components of the $T_2$ representation transform as polar (real) vectors oriented along [100], [010], [001] crystallographic directions.

The energies of these multiplets can be parametrized entirely in terms of
two-body Coulomb and exchange integrals within the
Sugano--Tanabe--Kamimura formalism.\cite{sugano1970} In this parametrization, the
multiplet energies are given by:
\begin{equation}
\begin{aligned}
&
E\big[|^4A_2\rangle \big] = 3J(xy) - 3K(xy)\text{,} \qquad
&&
E\big[|^2T_1\rangle \big] = 2J(xy) + J(zz) - 2K(xy)\text{,}
\\
&
E\big[|^2E\rangle \big] = 3J(xy)\text{,} \qquad
&&
E\big[|^2T_2\rangle \big] = 2J(xy) + J(zz)\text{,}
\end{aligned}
\end{equation}
where $J$ and $K$ denote the relevant Coulomb and exchange integrals between
pairs of $t_{2x}$, $t_{2y}$ or $t_{2z}$ orbitals depicted by $x$, $y$, $z$ labels. Taking into account that $K(xy) > 0$ and
$J(zz) > J(xy)$, the lowest-lying state is predicted to be $|^4A_2\rangle$, while the
highest is $^2T_2$. However, the relative ordering of the $|^2E\rangle$ and $|^2T_2\rangle$
states cannot be determined from these considerations alone.

To place this ambiguity in context, we note that in the well-studied case of the
negatively charged vacancy in diamond,\cite{PhysRevB.49.11010} the experimentally
and theoretically accepted level ordering is $|^2T_1\rangle$, $|^2E\rangle$,
and $|^2T_2\rangle$ from lowest to highest energy (see Fig.~2 therein). While
this analogy does not by itself fix the ordering in the present system, it
provides a physically motivated reference for the expected hierarchy of the
doublet states.

\subsubsection{Crystal-field Hamiltonian and symmetry lowering for the quartet excited state.\label{SM_LCFH}}

We describe the effect of symmetry lowering from the ideal $T_{d}$ environment
by introducing an effective crystal-field perturbation $H_{\mathrm{cf}}$ that
lowers the symmetry to the trigonal $C_{3v}$ symmetry. The relevant distortion
corresponds to an uniaxial stress projected from the $[111]$ direction. In the
single-particle basis $\{t_{2x},t_{2y},t_{2z}\}$ of electronic orbitals in
$T_{d}$, the corresponding crystal-field operator can be written
as\cite{Udvarhelyi_2020}
\begin{equation}
H_{\mathrm{cf}}^{(1)}=-\frac{\delta}{3}\Bigl(|t_{2x}\rangle\langle t_{2y}|+|t_{2y}\rangle\langle t_{2z}|+|t_{2z}\rangle\langle t_{2x}|+\mathrm{h.c.}\Bigr)=-\frac{\delta}{3}\begin{pmatrix}0 & 1 & 1\\
1 & 0 & 1\\
1 & 1 & 0
\end{pmatrix}\text{,}
\label{eq:crystalpertop}
\end{equation}
where $\delta$ parameterizes the strength of the symmetry-lowering
distortion. When acting on the $|a_{1}\rangle$ orbital, $\hat{H}_{\mathrm{cf}}$
merely produces a constant energy shift, which is not relevant for
the level splittings discussed here and is therefore omitted.

Using the Slater--Condon rules, one can straightforwardly show that the
effective crystal-field Hamiltonian within the many-body $|^4 T_1\rangle$ manifold has
the same matrix structure as in the single-particle $|t_{2i}\rangle$ basis:
\begin{equation}
\langle{}^{4}T_{1x}|H_{\mathrm{cf}}^{(1)}|{}^{4}T_{1y}\rangle=\langle\underset{=\langle^{5}A_{2}|}{\underbrace{\bar{a}_{1}\bar{t}_{2x}\bar{t}_{2y}\bar{t}_{2z}}}t_{2x}|\hat{H}_{\mathrm{cf}}^{(1)}|\underset{=|^{5}A_{2}\rangle}{\underbrace{\bar{a}_{1}\bar{t}_{2x}\bar{t}_{2y}\bar{t}_{2z}}}t_{2y}\rangle=\langle^{5}A_{2}|^{5}A_{2}\rangle\langle t_{2x}|\hat{H}_{\mathrm{cf}}^{(1)}|t_{2y}\rangle=-\frac{\delta}{3}
\text{.}
\end{equation}
However, there is opposite sign in the hole picture:
\begin{equation}
\langle{}^{4}T_{1x}|H_{\mathrm{cf}}^{(1)}|{}^{4}T_{1y}\rangle=\langle a_{1}t_{2y}t_{2z}|\hat{H}_{\mathrm{cf}}^{(1)}|a_{1}t_{2z}t_{2x}\rangle=\langle t_{2y}t_{2z}|\hat{H}_{\mathrm{cf}}^{(1)}|t_{2z}t_{2x}\rangle=-\langle t_{2z}t_{2y}|\hat{H}_{\mathrm{cf}}^{(1)}|t_{2z}t_{2x}\rangle=-\langle t_{2y}|\hat{H}_{\mathrm{cf}}^{(1)}|t_{2x}\rangle=+\frac{\delta}{3}
\text{.}
\label{eq:holecrystal}
\end{equation}

Therefore, one would need to take into account that the crystal field parameter is inverted for holes because both pictures describe the exact same system. 
On the other hand, it is convenient to transform the coordinate system towards the $[111]$  $C_{3v}$ symmetry axis by $|^{4}A_{2}^{\prime}\rangle=\frac{1}{\sqrt{3}}\bigl(|^{4}T_{1x}\rangle+|^{4}T_{1y}\rangle+|^{4}T_{1z}\rangle\bigr)\text{,}|^{4}E_{y}\rangle=\frac{1}{\sqrt{2}}\bigl(|^{4}T_{1x}\rangle-|^{4}T_{1y}\rangle\bigr)\text{,}|^{4}E_{x}\rangle=\frac{1}{\sqrt{6}}\bigl(|^{4}T_{1x}\rangle+|^{4}T_{1y}\rangle-2|^{4}T_{1z}\rangle\bigr)$
that transforms the crystal field Hamiltonian as follows:

\begin{equation}
\hat{H}_{\mathrm{cf},[111]}^{(^{4}T_{1})}=\frac{\delta}{3}\Bigl(|^{4}E_{x}\rangle\langle^{4}E_{x}|+|^{4}E_{y}^{\prime}\rangle\langle^{4}E_{y}|-|^{4}A_{2}^{\prime}\rangle\langle^{4}A_{2}^{\prime}|\Bigr)=U\hat{H}_{\mathrm{cf}}^{(1)}U^{\dagger}=\frac{\delta}{3}\begin{pmatrix}+1 & 0 & 0\\
0 & +1 & 0\\
0 & 0 & -2
\end{pmatrix},
\end{equation}
where the $U$ unitary transformation depicts the change of natural basis that of $T_d$ and $C_{3v}$:

\begin{equation}
U=\begin{pmatrix}\langle{}^{4}E_{x}|\\
\langle{}^{4}E_{y}|\\
\langle{}^{4}A_{2}^{\prime}|
\end{pmatrix}\begin{pmatrix}|^{4}T_{1x}\rangle & |^{4}T_{1y}\rangle & |^{4}T_{1z}\rangle\end{pmatrix}=\begin{pmatrix}\frac{1}{\sqrt{6}} & \frac{1}{\sqrt{6}} & \frac{-2}{\sqrt{6}}\\
\frac{1}{\sqrt{2}} & \frac{-1}{\sqrt{2}} & 0\\
\frac{1}{\sqrt{3}} & \frac{1}{\sqrt{3}} & \frac{1}{\sqrt{3}}
\end{pmatrix}\text{.}
\end{equation}
Therefore, the parameter $\delta$ corresponds directly to the energy splitting
between $V_1$ and $V_1^{\prime}$ (or, analogously, $V_2$ and $V_2^{\prime}$)
optical transitions.

\subsubsection{Quenching of crystal-field splitting in the doublet manifolds}
\label{sec:crystalquenching}
Having established the value of $\delta$, we now estimate the crystal-field
splitting of the doublet $T_1$ and $T_2$ electronic states. We treat
$H_{\mathrm{cf}}$ as a spin-independent one-body operator acting within the
$\{t_{2x},t_{2y},t_{2z}\}$ orbital subspace, and evaluate its matrix elements in
the symmetry-adapted many-body basis using the Slater--Condon rules.

In the basis $\{|{}^{2}T_{1x}\rangle,|^{2}T_{1y}\rangle,|^{2}T_{1z}\rangle\}$
(as defined above), the diagonal matrix elements are all zeros because
$\hat{H}_{\mathrm{cf}}^{(1)}$ can only excite a single electron:
\begin{equation}
\begin{array}{l}
\langle^{2}T_{1x}|\hat{H}_{\mathrm{cf}}^{(1)}|^{2}T_{1x}\rangle=\frac{1}{2}\bigl(\langle t_{2x}t_{2y}\bar{t}_{2y}|-\langle t_{2x}t_{2z}\bar{t}_{2z}|\bigr)\hat{H}_{\mathrm{cf}}^{(1)}\bigl(\bigl|t_{2x}t_{2y}\bar{t}_{2y}\bigr\rangle-\bigl|t_{2x}t_{2z}\bar{t}_{2z}\bigr\rangle\bigr)\\
={\displaystyle \frac{1}{2}\Bigl(\underset{=0\quad(i)}{\underbrace{\langle t_{2x}t_{2y}\bar{t}_{2y}|\hat{H}_{\mathrm{cf}}^{(1)}\bigl|t_{2x}t_{2y}\bar{t}_{2y}\bigr\rangle}}-\underset{=0\quad(ii)}{\underbrace{\langle t_{2x}t_{2y}\bar{t}_{2y}|\hat{H}_{\mathrm{cf}}^{(1)}\bigl|t_{2x}t_{2z}\bar{t}_{2z}\bigr\rangle}}-\underset{=0\quad(iii)}{\underbrace{\langle t_{2x}t_{2z}\bar{t}_{2z}|\hat{H}_{\mathrm{cf}}^{(1)}\bigl|t_{2x}t_{2y}\bar{t}_{2y}\bigr\rangle}}+\underset{=0\quad(iv)}{\underbrace{\langle t_{2x}t_{2z}\bar{t}_{2z}|\hat{H}_{\mathrm{cf}}^{(1)}\bigl|t_{2x}t_{2z}\bar{t}_{2z}\bigr\rangle}}\Bigr)=0}
\end{array}
\label{eq:vanish1}
\end{equation}
where $(i)$ and $(iv)$ terms try to connect the same configurations, while
$(ii)$ and $(iii)$ connecting configurations differ by two particles, thus all zeroes. Similarly, we
determine the off-diagonal matrix elements as
\begin{equation}
\begin{array}{l}
\langle^{2}T_{1x}|\hat{H}_{\mathrm{cf}}^{(1)}|^{2}T_{1y}\rangle=\frac{1}{2}\bigl(\langle t_{2x}t_{2y}\bar{t}_{2y}|-\langle t_{2x}t_{2z}\bar{t}_{2z}|\bigr)\hat{H}_{\mathrm{cf}}^{(1)}\bigl(\bigl|t_{2y}t_{2z}\bar{t}_{2z}\bigr\rangle-\bigl|t_{2y}t_{2x}\bar{t}_{2x}\bigr\rangle\bigr)\\
\begin{array}{ccccccc}
={\displaystyle \frac{1}{2}\Bigl(} & {\displaystyle \underset{=0}{\underbrace{\langle t_{2x}t_{2y}\bar{t}_{2y}|\hat{H}_{\mathrm{cf}}^{(1)}\bigl|t_{2y}t_{2z}\bar{t}_{2z}\bigr\rangle}}} & -\langle t_{2x}t_{2y}\bar{t}_{2y}|\hat{H}_{\mathrm{cf}}^{(1)}\bigl|t_{2y}t_{2x}\bar{t}_{2x}\bigr\rangle & -\langle t_{2x}t_{2z}\bar{t}_{2z}|\hat{H}_{\mathrm{cf}}^{(1)}\bigl|t_{2y}t_{2z}\bar{t}_{2z}\bigr\rangle & +\underset{=0}{\underbrace{\langle t_{2x}t_{2z}\bar{t}_{2z}|\hat{H}_{\mathrm{cf}}^{(1)}\bigl|t_{2y}t_{2x}\bar{t}_{2x}\bigr\rangle}} & \Bigr)\\
={\displaystyle \frac{1}{2}\Bigl(} & 0 & +\underset{=\delta/3}{\underbrace{\langle\bar{t}_{2y}|\hat{H}_{\mathrm{cf}}^{(1)}\bigl|\bar{t}_{2x}\bigr\rangle}} & \underset{=-\delta/3}{\underbrace{-\langle t_{2x}|\hat{H}_{\mathrm{cf}}^{(1)}\bigl|t_{2y}\bigr\rangle}} & +0 & \Bigr) & =0
\end{array}
\end{array}
\label{eq:vanish2}
\end{equation}
where even the innermost two terms terms are annihilating each other due to
Slater--Condon rules: $-\langle t_{2x}t_{2y}\bar{t}_{2y}|\hat{H}_{\mathrm{cf}}^{(1)}\bigl|t_{2y}t_{2x}\bar{t}_{2x}\bigr\rangle=+\langle t_{2y}t_{2x}\bar{t}_{2y}|\hat{H}_{\mathrm{cf}}^{(1)}\bigl|t_{2y}t_{2x}\bar{t}_{2x}\bigr\rangle=+\langle\bcancel{t_{2y}t_{2x}}\bar{t}_{2y}|\hat{H}_{\mathrm{cf}}^{(1)}\bigl|\bcancel{t_{2y}t_{2x}}\bar{t}_{2x}\bigr\rangle=\langle\bar{t}_{2y}|\hat{H}_{\mathrm{cf}}^{(1)}\bigl|\bar{t}_{2x}\bigr\rangle=\delta/3$.
Therefore, the crystal field is quenched for the doublets at the zeroth
order of perturbation.
\subsection{Crystal-field splitting by means of multi-configurational corrections from quantum embedding}
\label{section:s2f}

In this section, we shall try to determine the crystal field splitting for the correlated $|^2T_1\rangle$, $|^2T_2\rangle$ multiplets from the following ingredients:
\begin{itemize}
\item Crystal field splitting parameter ($\delta$) observed in the spin quartet transition: $|^{4}A_{2}\rangle\overset{\sim1.4\:\mathrm{eV}}{\longleftrightarrow}|^{4}T_{1}\rangle$.
\item Multiconfigurational expansion of the doublet states up to first and second excited Slater determinants as obtained from quantum embedding.
\end{itemize}
Then, we will evaluate the $H_{\mathrm{cf}}^{(1)}=-\frac{\delta}{3}\bigl(|t_{2x}\rangle\langle t_{2y}|+... \bigl)$ single particle crystal field operator on the multiconfigurational expansion of $|^{2}T_{1}\rangle$ and $|^{2}T_{2}\rangle$ doublets:
\begin{equation}
|^{2}T_{1}^{\text{corr}}\rangle=a\underset{a_{1}^{2}t_{2}^{3}}{\underbrace{|^{2}T_{1}\rangle}}+b\underset{a_{1}^{1}t_{2}^{4}}{\underbrace{|^{2}T_{1}^{\prime}\rangle}},\qquad|^{2}T_{2}^{\text{corr}}\rangle=c\underset{a_{1}^{2}t_{2}^{3}}{\underbrace{|^{2}T_{2}\rangle}}+d\underset{a_{1}^{1}t_{2}^{4}}{\underbrace{|^{2}T_{2}^{\prime}\rangle}}+e\underset{a_{1}^{0}t_{2}^{5}}{\underbrace{|^{2}T_{2}^{\prime\prime}\rangle}},\qquad|^{2}E^{\text{corr}}\rangle=f\underset{a_{1}^{2}t_{2}^{3}}{\underbrace{|^{2}E\rangle}}+g\underset{a_{1}^{1}t_{2}^{4}}{\underbrace{|^{2}E^{\prime}\rangle}}\text{.}
    \label{eq:correlation_expansion}
\end{equation}
Frist, we need to obtain the symmetry adapted form of the first excited $|^{2}T_{1}^{\prime}\rangle$, $|^{2}T_{2}^{\prime}\rangle$ and second excited $|^{2}T_{2}^{\prime\prime}\rangle$ configurations.

We estimate each configuration's expansion directly from our quantum embedding calculations for the $V_k$ site thus we obtain the following $a^2=93(2)\%$, $b^2=6(1)\%$, $c^2=85(5)\%$, $d^2=11(6)\%$, $e^2=4(1)\%$, $f^2=78(1)\%$, $g^2=20(2)\%$ factors where the error bars are originating from the inherent crystal field of a 128-atom model.
The small supercell size gives rise to severe numerical inaccuracies in the quantum embedding calculations as well, a system size is so small cannot capture the finite size effects of $\delta$ correctly.
Thus, the value of parameters $a$-$e$  as well as the energy position of $|^4T_1\rangle$, $|^2T_2^{\text{corr}}\rangle$, $|^2E^{\text{corr}}\rangle$, $|^2T_1^{\text{corr}}\rangle$ are not exactly are the same see Fig.~\ref{fig:mbspectra_pbe_k} or Fig.~4(a) of the main text where degenerate levels are not exactly degenerate.
Therefore, we rely only the average of these parameters and take the $\delta$ crystal field parameter from the $|^4T_2\rangle$ excited quartet.

\subsubsection{Excited symmetry-adapted three-hole wavefunctions}

The first excited determinants originate from the $(a_{1}^{1}t_{2}^{4})$ single particle occupation spanning total 5 electrons in the configurational space.
In contrast, it is far more convenient to describe the system in the equivalent hole picture that can be represented by three holes: $(a_{1}t_{2}^{2})$.

According to Fig. 2.2 in the book of Sugano\cite{sugano1970}, we can expand $t_2^2$ for the first excited configurations $(a_{1}^{1}t_{2}^{2})={}^{2}A_{1}\otimes({}^{3}T_{1}\oplus{}^{1}E\oplus{}^{1}T_{2}\oplus{}^{1}A_{1})={}^{4}T_{1}\oplus{}^{2}T_{1}\oplus{}^{2}E\oplus{}^{2}T_{2}\oplus{}^{2}A_{1}$ thus we can identify both $|^{2}T_{1}^{\prime}\rangle$ and $|^{2}T_{2}^{\prime}\rangle$ configurations in Eq.~\eqref{eq:correlation_expansion} . 
Specifically, we can identify that both $|^{4}T_{2}\rangle$ from Eq.~\eqref{eq:4T1_wave} and $|^{2}T^\prime_{2}\rangle$ multiplets originates by adding an additional $a_1$ particle to the two particle construction: $|^{3}T_{2i}\rangle=\{|t_{2y}t_{2z}\rangle,|t_{2z}t_{2x}\rangle,|t_{2x}t_{2y}\rangle\}$ as defined in Table 2.2 of the book\cite{sugano1970}.
However, determining the wavefunction that of $|^{2}T^\prime_{2}\rangle$ is not a trivial task, thus we shall proceed on an alternative tour.
What we know for sure is that the spin projection $m_s=+\frac{1}{2}$ of $|^{4}T_{2}\rangle$ can be written as
\begin{equation}
|^{4}T_{1i}^{m_s=+\frac{1}{2}}\rangle=\frac{1}{\sqrt{3}}\begin{cases}
|\bar{a}_{1}t_{2y}t_{2z}\rangle+|a_{1}\bar{t}_{2y}t_{2z}\rangle+|a_{1}t_{2y}\bar{t}_{2z}\rangle, & i=x,\\
|\bar{a}_{1}t_{2z}t_{2x}\rangle+|a_{1}\bar{t}_{2z}t_{2x}\rangle+|a_{1}t_{2z}\bar{t}_{2x}\rangle, & i=y,\\
|\bar{a}_{1}t_{2x}t_{2y}\rangle+|a_{1}\bar{t}_{2x}t_{2y}\rangle+|a_{1}t_{2x}\bar{t}_{2y}\rangle, & i=z,
\end{cases}
    \label{eq:4T1_ms12}
\end{equation}
that expression can be obtained by applying the $\hat{S}_{-}$  pin lowering operator on Eq.~\eqref{eq:4T1_wave} 's $m_s=+\frac{3}{2}$ case.
At this point, one may notice that both $|^{2}T^\prime_{2}\rangle$ and $|^{4}T_{1i}^{m_s=+\frac{1}{2}}\rangle$ have to be composed from the same $(a_{1}^{1}t_{2\alpha}^{1}t_{2\beta}^{1})$ orbitals where $\alpha\neq\beta$.
Therefore, we can construct the following additional six symmetry adapted configurations made from $|a_{1}t_{2\alpha}t_{2\beta}\rangle$'s that are orthogonal to the wavefunctions from Eq.~\eqref{eq:4T1_ms12}:
\begin{equation}
|^{2}T_{1i}^{\prime}\rangle=\frac{1}{\sqrt{6}}\begin{cases}
2|\bar{a}_{1}t_{2y}t_{2z}\rangle-|a_{1}\bar{t}_{2y}t_{2z}\rangle-|a_{1}t_{2y}\bar{t}_{2z}\rangle, & i=x,\\
2|\bar{a}_{1}t_{2z}t_{2x}\rangle-|a_{1}\bar{t}_{2z}t_{2x}\rangle-|a_{1}t_{2z}\bar{t}_{2x}\rangle, & i=y,\\
2|\bar{a}_{1}t_{2x}t_{2y}\rangle-|a_{1}\bar{t}_{2x}t_{2y}\rangle-|a_{1}t_{2x}\bar{t}_{2y}\rangle, & i=z
\end{cases}\quad\text{and}\quad|^{2}T_{2i}^{\prime}\rangle=\frac{1}{\sqrt{2}}\begin{cases}
|a_{1}\bar{t}_{2y}t_{2z}\rangle-|a_{1}t_{2y}\bar{t}_{2z}\rangle, & i=x,\\
|a_{1}\bar{t}_{2z}t_{2x}\rangle-|a_{1}t_{2z}\bar{t}_{2x}\rangle, & i=y,\\
|a_{1}\bar{t}_{2x}t_{2y}\rangle-|a_{1}t_{2x}\bar{t}_{2y}\rangle, & i=z.
\end{cases}
    \label{eq:2T1p_2T2p}
\end{equation}
One can identify the first orbital triplet as a $|^{2}T^\prime_{1}\rangle$ because it can be proven that it is transforming as the $T_1$ irreducible representation.
Surprisingly the other $|^{2}T^\prime_{2}\rangle$ configuration is also a configuration we are looking for.
It is trivial that, the $C_3$ threefold rotations will just permute the $x$, $y$, $z$ coordinates for both $|^{2}T_{1i}^{\prime}\rangle$ and $|^{2}T_{2i}^{\prime}\rangle$.
However, we still need to prove that ${\sigma}_{v}$ mirror planes do not mix $|^{2}T_{1i}^{\prime}\rangle$ and $|^{2}T_{2i}^{\prime}\rangle$ thus they will form individual $T_1$ and $T_2$ representations.
The first mirror plane ($\sigma_v^{(1)}$) will interchange the first two coordinates: $x\leftrightarrow y$ while it will leave $z$ intact.
Therefore its effect will be $\sigma_{v}^{(1)}|^{2}T_{1x}^{\prime}\rangle=\hat{\sigma}_{v}^{(1)}(2|\bar{a}_{1}t_{2y}t_{2z}\rangle-|a_{1}\bar{t}_{2y}t_{2z}\rangle-|a_{1}t_{2y}\bar{t}_{2z}\rangle)=(2|\bar{a}_{1}t_{2x}t_{2z}\rangle-|a_{1}\bar{t}_{2x}t_{2z}\rangle-|a_{1}t_{2x}\bar{t}_{2z}\rangle)=-|^{2}T_{1y}^{\prime}\rangle$ on the first wavefunction. We leave the case of $|^{2}T_{1y}^{\prime}\rangle$ and $|^{2}T_{1z}^{\prime}\rangle$ to the reader and the case of the other 5 remaining ($\sigma_{v}^{(2-6)}$) mirror planes.
The final transformation law will be $\sigma_{v}^{(1)}|^{2}T_{1i}^{\prime}\rangle=\Bigl(\begin{smallmatrix}0 & -1 & 0\\
-1 & 0 & 0\\
0 & 0 & -1
\end{smallmatrix}\Bigr)|^{2}T_{1i}^{\prime}\rangle$ for $\sigma_{v}^{(1)}$, thus the trace of the operation will be: $\text{Tr}({\sigma}_{v}^{(1)})=-1$. This is the same value that appears in the character table of $T_d$ at the column for the six $6\sigma_v$ mirror planes at the $T_1$ irreducible representation's row.
Similarly, in the case of $|^{2}T_{2i}^{\prime}\rangle$ the trace of the operation would be $\text{Tr}({\sigma}_{v}^{(1)})=+1$ therefore, we correctly assumed in Eq.~\eqref{eq:2T1p_2T2p} that it is transforming as $T_2$.

The excited $|^2E^\prime\rangle$ can be generated by multiplying the $|^1E\rangle$ ($t_2^2$) two-hole wavefunction (see Table 2.2 in the book\cite{sugano1970} of Sugano) by an $(a_1^1)=|^2A_1\rangle $ hole:
\begin{equation}
|^{2}E^{\prime}\rangle=|a_{1}\rangle\otimes|^{1}E\rangle=\begin{cases}
\frac{1}{\sqrt{6}}\left(2|a_{1}\bar{t}_{2z}t_{2z}\rangle-|a_{1}\bar{t}_{2x}t_{2x}\rangle-|a_{1}\bar{t}_{2y}t_{2y}\rangle\right),\\
\frac{1}{\sqrt{2}}\left(|a_{1}\bar{t}_{2x}t_{2x}\rangle-|a_{1}\bar{t}_{2y}t_{2y}\rangle\right),
\end{cases}
    \label{eq:2Ep}
\end{equation}

Finally, only one configuration is possible by occupation $(a_1^0t_2^5)$:
\begin{equation}
|^{2}T_{2i}^{\prime\prime}\rangle=\begin{cases}
|\bar{a}_{1}a_{1}t_{2x}\rangle, & i=x\text{,}\\
|\bar{a}_{1}a_{1}t_{2y}\rangle, & i=y\text{,}\\
|\bar{a}_{1}a_{1}t_{2z}\rangle, & i=z\text{,}
\end{cases}
    \label{eq:4T2pp}
\end{equation}
thus $|^{2}T_{1}^{\prime\prime}\rangle$, $|^{2}E^{\prime\prime}\rangle$ terms are not possible in the configuration expansion that of Eq.~\eqref{eq:correlation_expansion}.
Now, with all ingredients at hand, our next step will be to determine the crystal field matrix elements for $|^{2}T_{1}^{\text{corr}}\rangle$ and $|^{2}T_{2}^{\text{corr}}\rangle$ in the next subsection.

\subsubsection{First order multi-configurational crystal field corrections for $|^2T_1\rangle$ and $|^2T_2\rangle$}
In this subsection we will determine the leading terms due to $\hat{H}_{\mathrm{cf}}^{(1)}$ from Eq. Eq.~\eqref{eq:crystalpertop} by means of first order perturbation theory.
Firstly, we determine the expansion for the lower  $|^{2}T_{1}^{\text{corr}}\rangle$ doublet.
One may notice that $H_{\mathrm{cf}}^{(1)}$ does not contain any excitation $|a_{1}\rangle\langle t_{2i}|$ term, thus $\langle^{2}T_{1i}|H_{\mathrm{cf}}^{(1)}|^{2}T_{1j}^{\prime}\rangle$ is zero.

Therefore, we can write simply that only the third term matters in the multi-configurational expansion:
$\langle^{2}T_{1i}^{\text{corr}}|H_{\mathrm{cf}}^{(1)}|^{2}T_{1j}^{\text{corr}}\rangle=\cancel{a^{2}\langle^{2}T_{1i}|H_{\mathrm{cf}}^{(1)}|^{2}T_{1j}\rangle}+\cancel{2ab\langle^{2}T_{1i}|H_{\mathrm{cf}}^{(1)}|^{2}T_{1j}^{\prime}\rangle}+b^{2}\langle^{2}T_{1i}^{\prime}|H_{\mathrm{cf}}^{(1)}|^{2}T_{1j}^{\prime}\rangle$. According to the Slater--Condon rules the matrix element will be:

\begin{equation}
\begin{array}{l}
\langle^{2}T_{1x}^{\prime}|H_{\mathrm{cf}}^{(1)}|^{2}T_{1y}^{\prime}\rangle=\frac{1}{6}\bigl(2\langle\bar{a}_{1}t_{2y}t_{2z}|-\langle a_{1}\bar{t}_{2y}t_{2z}|-\langle a_{1}t_{2y}\bar{t}_{2z}|\bigr)\hat{H}_{\mathrm{cf}}^{(1)}\bigl(2|\bar{a}_{1}t_{2y}t_{2z}\rangle-|a_{1}\bar{t}_{2y}t_{2z}\rangle-|a_{1}t_{2y}\bar{t}_{2z}\rangle\bigr)=\\
\begin{array}{cccccc}
={\displaystyle \frac{1}{6}\Bigl(} & \underset{=+4\delta/3}{\underbrace{+4\langle\bar{a}_{1}t_{2y}t_{2z}|\hat{H}_{\mathrm{cf}}^{(1)}|\bar{a}_{1}t_{2z}t_{2x}\rangle}} & -2\underset{=0}{\underbrace{\langle\bar{a}_{1}t_{2y}t_{2z}|\hat{H}_{\mathrm{cf}}^{(1)}|a_{1}\bar{t}_{2z}t_{2x}\rangle}} & -2\underset{=0}{\underbrace{\langle\bar{a}_{1}t_{2y}t_{2z}|\hat{H}_{\mathrm{cf}}^{(1)}|a_{1}t_{2z}\bar{t}_{2x}\rangle}}\\
 & -2\underset{=0}{\underbrace{\langle a_{1}\bar{t}_{2y}t_{2z}|\hat{H}_{\mathrm{cf}}^{(1)}|\bar{a}_{1}t_{2z}t_{2x}\rangle}} & +\underset{=0}{\underbrace{\langle a_{1}\bar{t}_{2y}t_{2z}|\hat{H}_{\mathrm{cf}}^{(1)}|a_{1}\bar{t}_{2z}t_{2x}\rangle}} & +\underset{=+\delta/3}{\underbrace{\langle a_{1}\bar{t}_{2y}t_{2z}|\hat{H}_{\mathrm{cf}}^{(1)}|a_{1}t_{2z}\bar{t}_{2x}\rangle}}\\
 & -2\underset{=0}{\underbrace{\langle a_{1}t_{2y}\bar{t}_{2z}|\hat{H}_{\mathrm{cf}}^{(1)}|\bar{a}_{1}t_{2z}t_{2x}\rangle}} & +\underset{=\delta/3}{\underbrace{\langle a_{1}t_{2y}\bar{t}_{2z}|\hat{H}_{\mathrm{cf}}^{(1)}|a_{1}\bar{t}_{2z}t_{2x}\rangle}} & +\underset{=0}{\underbrace{\langle a_{1}t_{2y}\bar{t}_{2z}|\hat{H}_{\mathrm{cf}}^{(1)}|a_{1}t_{2z}\bar{t}_{2x}\rangle}} & \Bigr) & ={\displaystyle \frac{\delta}{3}\times\frac{4+1+1}{6}=+\frac{\delta}{3}}
\end{array}
\end{array}
\text{.}
    \label{eq:2T2_2T1_crystalfield}
\end{equation}
that is the exact same that we seen for the spin quartet $|^{4}T_{1}\rangle$. Therefore, $|^{2}T_{1}\rangle$ will split into $|^{2}A_{2}\rangle$ and $|^{2}E\rangle$ where the $|^{2}A_{2}\rangle$ is the lowest followed by $|^{2}E\rangle$ with an energy value of $\kappa_{\text{theory}}$.
However, the $|^{2}T_{1z}^{\prime}\rangle$ configuration is only present with a small probability of $b^2$ thus the crystal field splitting for $|^{2}T^\text{corr}_{1}\rangle$ will be:
\begin{equation}
\boxed{\, \kappa^{(1)}=b^{2}\delta \,}
    \label{eq:kappa_1st}
    \text{ .}
\end{equation}

In the case of $|^{2}T_{2}^{\text{corr}}\rangle$ the configuration decomposition is the following: $\langle^{2}T_{2i}^{\text{corr}}|H_{\mathrm{cf}}^{(1)}|^{2}T_{2}^{\text{corr}}\rangle=d^{2}\langle^{2}T_{2i}|H_{\mathrm{cf}}^{(1)}|^{2}T_{2j}^{\prime}\rangle+e^{2}\langle^{2}T_{2i}^{\prime}|H_{\mathrm{cf}}^{(1)}|^{2}T_{2j}^{\prime\prime}\rangle$ because the $\langle^{2}T_{1i}|H_{\mathrm{cf}}^{(1)}|^{2}T_{1j}^{}\rangle$, $\langle^{2}T_{1i}|H_{\mathrm{cf}}^{(1)}|^{2}T_{1j}^{\prime}\rangle$, $\langle^{2}T_{2i}^{\prime}|H_{\mathrm{cf}}^{(1)}|^{2}T_{2j}^{\prime\prime}\rangle$ matrix elements are all zeroes. The matrix elements for the first excited configuration is:

\begin{equation}
\begin{array}{l}
\langle^{2}T_{2x}^{\prime}|H_{\mathrm{cf}}^{(1)}|^{2}T_{2y}^{\prime}\rangle=\frac{1}{2}\bigl(\langle a_{1}\bar{t}_{2y}t_{2z}|-\langle a_{1}t_{2y}\bar{t}_{2z}|\bigr)\hat{H}_{\mathrm{cf}}^{(1)}\bigl(|a_{1}\bar{t}_{2z}t_{2x}\rangle-|a_{1}t_{2z}\bar{t}_{2x}\rangle\bigr)=\\
\begin{array}{ccccc}
={\displaystyle \frac{1}{2}\Bigl(} & \underset{=0}{\underbrace{+\langle a_{1}\bar{t}_{2y}t_{2z}|\hat{H}_{\mathrm{cf}}^{(1)}|a_{1}\bar{t}_{2z}t_{2x}\rangle}} & \underset{=-\delta/3}{\underbrace{-\langle a_{1}\bar{t}_{2y}t_{2z}|\hat{H}_{\mathrm{cf}}^{(1)}|a_{1}t_{2z}\bar{t}_{2x}\rangle}}\\
 & \underset{=-\delta/3}{\underbrace{-\langle a_{1}t_{2y}\bar{t}_{2z}|\hat{H}_{\mathrm{cf}}^{(1)}|a_{1}\bar{t}_{2z}t_{2x}\rangle}} & \underset{=0}{\underbrace{+\langle a_{1}t_{2y}\bar{t}_{2z}|\hat{H}_{\mathrm{cf}}^{(1)}|a_{1}t_{2z}\bar{t}_{2x}\rangle}} & \Bigr) & ={\displaystyle \frac{\delta}{3}\times\frac{-1-1}{2}=-\frac{\delta}{3}}
\end{array}
\end{array}
\text{,}
    \label{eq:crystalfield2T2p}
\end{equation}
while the second excited configuration will be:
\begin{equation}
\langle^{2}T_{2x}^{\prime\prime}|H_{\mathrm{cf}}^{(1)}|^{2}T_{2y}^{\prime\prime}\rangle=\langle\bar{a}_{1}a_{1}t_{2x}|H_{\mathrm{cf}}^{(1)}|\bar{a}_{1}a_{1}t_{2y}\rangle=\langle t_{2x}|H_{\mathrm{cf}}^{(1)}|t_{2y}\rangle=-\frac{\delta}{3}
    \label{eq:2T1_delta}
    \text{ .}
\end{equation}
Therefore the $|^2E^{\prime\prime}\rangle$ orbital doublet will precede the $|^2A_1\rangle$ orbital singlet in $|^{2}T_{1}^{\text{corr}}\rangle$'s crystal field splitting by the following energy:
\begin{equation}
\boxed{\:\eta^{(1)}=-(d^{2}+e^{2})\delta\:}
    \label{eq:kappa_1st}
    \text{ .}
\end{equation}
\subsubsection{Second order corrections}

In order to determine the second order corrections due to $H_{\mathrm{cf}}^{(1)}$, first, we rewrite the projector to the $|^2E\rangle$ doublet that of Eq.~\eqref{eq:2E_wave} as follows:
\begin{equation}
|^{2}E\rangle\langle^{2}E|=\!\!\!\!\!\!\!\!\!\sum_{\begin{smallmatrix}\alpha=\{x^{2}-y^{2}\\
,2z^{2}-x^{2}-y^{2}\}
\end{smallmatrix}}\!\!\!\!\!\!\!\!\!|^{2}E_{\alpha}\rangle\langle^{2}E_{\alpha}|=\frac{2}{3}\mathrm{Re}\biggr[\underset{{\displaystyle \times\left(\langle t_{2x}t_{2y}\bar{t}_{2z}|+\langle t_{2x}\bar{t}_{2y}t_{2z}|e^{-i2{\pi}/3}+\langle\bar{t}_{2x}t_{2y}t_{2z}|e^{+i2{\pi}/3}\right)\biggr]\text{,}}}{\left(|t_{2x}t_{2y}\bar{t}_{2z}\rangle+e^{i2{\pi}/3}|t_{2x}\bar{t}_{2y}t_{2z}\rangle+e^{-i2{\pi}/3}|\bar{t}_{2x}t_{2y}t_{2z}\rangle\right)}
\label{eq:projector2E}
\end{equation}
where one may notice that the perturbation jumps to an energetically lower state $|^{2}E\rangle$, thus the $\Delta$ energy difference is negative.
We do this to get a more compact form for the second order perturbation through the middle $|^2E\rangle$ doublet:
\begin{equation}
\kappa_{E}^{(2)}=3\frac{\langle^{2}T_{2x}^{\text{corr}}|\hat{H}_{\mathrm{cf}}^{(1)}|^{2}E^{\text{corr}}\rangle\langle{}^{2}E^{\text{corr}}|\hat{H}_{\mathrm{cf}}^{(1)}|{}^{2}T_{2y}^{\text{corr}}\rangle}{\underset{=-\Delta}{\underbrace{E\bigl[|^{2}E\rangle\bigr]-E\bigl[|^{2}T_{2}\rangle\bigr]}}}=-\frac{3c^{2}f^{2}}{\Delta\sqrt{2}\sqrt{2}}\underset{\times{\displaystyle \hat{H}_{\mathrm{cf}}^{(1)}\left(\bigl|t_{2y}t_{2z}\bar{t}_{2z}\bigr\rangle+\bigl|t_{2y}t_{2x}\bar{t}_{2x}\bigr\rangle\right)\text{.}}}{\left(\langle t_{2x}t_{2y}\bar{t}_{2y}|+\langle t_{2x}t_{2z}\bar{t}_{2z}|\right)\hat{H}_{\mathrm{cf}}^{(1)}|^{2}E\rangle\langle^{2}E|}
\label{eq:2ndpert_2E}
\end{equation}
where the terms containing $d$, $g$ coefficients all vanish because it can be derived that the second term in the transition matrix element is zero: $\langle^{2}T_{2}^{\text{corr}}|\hat{H}_{\mathrm{cf}}^{(1)}|^{2}E^{\text{corr}}\rangle=cf\langle^{2}T_{2}|\hat{H}_{\mathrm{cf}}^{(1)}|^{2}E\rangle+dg\cancel{\langle^{2}T_{2}^{\prime}|\hat{H}_{\mathrm{cf}}^{(1)}|^{2}E^{\prime}\rangle}$.
Now, we insert the decomposition of $|^{2}E\rangle\langle^{2}E|$ projector:
\begin{equation}
\begin{split}\eta_{E}^{(2)}=-\frac{3}{\Delta}\Bigl(\frac{\delta}{3}\Bigr)^{2}\frac{2^{2}c^{2}f^{2}}{2}\times\frac{2}{3}\left(\langle t_{2x}t_{2z}\bar{t}_{2y}|+\langle t_{2x}t_{2y}\bar{t}_{2z}|\right)\times\mathrm{Re}\biggr[|t_{2x}t_{2y}\bar{t}_{2z}\rangle-e^{i2\pi/3}|t_{2x}t_{2z}\bar{t}_{2y}\rangle+e^{-i2\pi/3}|t_{2y}t_{2z}\bar{t}_{2x}\rangle\\
\times\left(\langle t_{2x}t_{2y}\bar{t}_{2z}|-\langle t_{2x}t_{2z}\bar{t}_{2y}|e^{-i2\pi/3}+\langle t_{2y}t_{2z}\bar{t}_{2x}|e^{+i2\pi/3}\right)\biggr]\times\left(\bigl|t_{2y}t_{2x}\bar{t}_{2z}\bigr\rangle+\bigl|t_{2y}t_{2z}\bar{t}_{2x}\bigr\rangle\right)
\end{split}
\label{eq:2ndpert_2Eb}
\end{equation}
\begin{equation}
=-\frac{4\delta^{2}}{9\Delta}\mathrm{Re}\left[\left(1-1e^{i2\pi/3}+0e^{-i2\pi/3}\right)\left(1-1e^{-i2\pi/3}+0e^{+i2\pi/3}\right)\right]=\boxed{\kappa_{E}^{(2)}=-c^{2}f^{2}\frac{4\delta^{2}}{3\Delta}}
\label{eq:2ndpert_2Ec}
\end{equation}

We have seen in Eqs.~\eqref{eq:vanish1} and \eqref{eq:vanish2} that the crystal field for both $|^2T_1\rangle$ and $|^2T_2\rangle$ vanishes. However, the offdiagonal matrix elements does not: $\langle{}^{2}T_{1x}|\hat{H}_{\mathrm{cf}}^{(1)}|^{2}T_{2y}\rangle=\frac{\delta}{3}$. 
This allows us calculate the second order perturbation of the crystal field:
\begin{equation}
\eta^{(2)}_T=\frac{\langle{}^{2}T_{1x}^{\mathrm{corr}}|\hat{H}_{\mathrm{cf}}^{(1)}|^{2}T_{2z}^{\mathrm{corr}}\rangle\langle^{2}T_{2z}^{\mathrm{corr}}|\hat{H}_{\mathrm{cf}}^{(1)}|^{2}T_{1y}^{\mathrm{corr}}\rangle}{\underset{=\Lambda}{\underbrace{E\bigl[|^{2}T_{1}\rangle\bigr]-E\bigl[|^{2}T_{2}\rangle\bigr]}}}=\boxed{\eta^{(2)}_T=-\frac{a^{2}c^{2}\delta^{2}}{\Lambda}}\text{,}
\label{eq:2ndpert_2T12}
\end{equation}
where only the leading terms involving ($a$ and $c$) in multiconfigurational expansion remains because $\langle{}^{2}T_{1x}^{\prime}|\hat{H}_{\mathrm{cf}}^{(1)}|^{2}T_{2y}^{\prime}\rangle$ matrix element for ($b$ and $d$) is zero. We note that the same correction also exists for the lowest triplet with the opposite sign because the energy difference is being flipped: $\boxed{\kappa_{T}^{(2)}=-\eta_{T}^{(2)}}$.

Additional terms 
In summary, we collect all correction terms terms as follows:
\begin{equation}
\boxed{\begin{array}{cc}
\kappa_{\text{theory}}=\kappa^{(1)}+\kappa_{T}^{(2)}+\mathcal{O}(\delta^{3})\approx & \underset{\underset{V_{k}:+1.88\:\text{meV}}{{\scriptscriptstyle V_{h}:+0.45\:\text{meV }}}}{\underbrace{b^{2}\delta}}+\underset{{\scriptscriptstyle \underset{V_{k}:+0.63\:\text{meV}}{V_{h}:+0.04\:\text{meV }}}}{\underbrace{a^{2}c^{2}{\displaystyle \frac{\delta^{2}}{\Lambda}}}}\end{array}}
\end{equation}
and
\begin{equation}
\boxed{\begin{array}{cccc}
\eta_{\text{theory}}=\eta^{(1)}+\eta_{E}^{(2)}+\eta_{T}^{(2)}+\mathcal{O}(\delta^{3})\approx & \underset{\underset{V_{k}:-4.33\:\text{meV}}{{\scriptscriptstyle V_{h}:-1.04\:\text{meV }}}}{\underbrace{-(d^{2}+e^{2})\delta}} & \underset{\underset{V_{k}:-2.01\:\text{meV}}{{\scriptscriptstyle V_{h}:-0.12\:\text{meV }}}}{\underbrace{-c^{2}f^{2}{\displaystyle \frac{4\delta^{2}}{3\Delta}}}} & \underset{\underset{V_{k}:-0.63\:\text{meV}}{{\scriptscriptstyle V_{h}:-0.04\:\text{meV }}}}{\underbrace{-a^{2}c^{2}{\displaystyle \frac{\delta^{2}}{\Lambda}}}}\end{array}}
\end{equation}
to evaluate the the theoretical model. We use the $C_{3v}$ crystal field parameters as calculated by P. Udvarhelyi et al, see Table 1. in Ref.~\cite{udvarhelyi2020vibronic}. Therefore, we use $\delta=\text{7 meV}$ for $V_h$ and $\delta=\text{29 meV}$ for $V_h$.
We note that these values are intrinsic, they are not multiplied by the Ham reduction factor of the Jahn--Teller effect. 
However, similarly to $\hat{H}_{\text{crystal}}$ in Eq.~\eqref{eq:2T2_2T1_crystalfield}, the $\hat{H}_{\text{JT}}$ Jahn--Teller Hamiltonian~\cite{udvarhelyi2020vibronic} is also quenched in the uncorrelated doublet wavefunctions eg. $\langle^{2}T_{1}|\hat{H}_{\text{JT}}|^{2}T_{1}\rangle=\langle^{2}T_{2}|\hat{H}_{\text{JT}}|^{2}T_{2}\rangle=0$. 
Therefore, we neglect the Jahn--Teller Ham factors for $|^{2}T_{2}^{\mathrm{corr}}\rangle$, $|^{2}T_{1}^{\mathrm{corr}}\rangle$ doublet levels for simplicity. 
We substitute the experimentally observed values of $\Delta = 369$ and (371) meV, $\Lambda = 1095$ and (1065) meV for $V_h$ and ($V_k$) respectively.
Therefore, our theory predicts the following crystal field parameters:

\begin{equation}
\boxed{\begin{array}{c}
\:V_{h}:\biggl\{\begin{array}{ccccc}
\kappa_{\text{theory}}= & +0.5\text{ meV} &  & |\kappa_{\text{expt.}}\!|< & \!\!\!\!\!1\:\text{or}>10\text{ meV}\\
\eta_{\text{theory}}= & -1.2\text{ meV} &  & \eta_{\text{expt.}}= & \!\!\!\!\!-4\:\text{meV}\qquad\quad\;
\end{array}\\
V_{h}:\biggl\{\begin{array}{ccccc}
\kappa_{\text{theory}}= & +2.5\text{ meV} &  & |\kappa_{\text{expt.}}\!|< & \!\!\!\!1\:\text{or}>10\text{ meV}\\
\eta_{\text{theory}}= & -7.0\text{ meV} &  & \eta_{\text{expt.}}= & \!\!\!\!\!-10\:\text{meV}\qquad\;
\end{array}
\end{array}}
\label{eq:doublet_model_crystalfieldresults}
\end{equation}
Therefore, based on Eq.~\eqref{eq:doublet_model_crystalfieldresults}, we can make the following conclusions:
\begin{itemize}
\item The sign of the $\eta_{\text{theory}}$ crystal field parameter for $|^{2}T_{2}^{\text{corr}}\rangle$ is in agreement with the experiments. It was identified from optical polarization measurements (see $\mathsf{A}$, $\mathsf{B}$, $\mathsf{C}$ optical peaks in Fig.~4(c-e) of the main text) that the crystal field shifts the orbitally 2$\times$ degenerate $|^{2}E^{\prime\prime}\rangle$ level (of $C_{3v}$ symmetry) downwards while lifts $|^{2}A_{2}\rangle$ upwards thus $\eta_{\text{expt.}}$ is negative.
\item The $\kappa_{\text{theory}}$ crystal field splitting that of $|^{2}T_{2}^{\text{corr}}\rangle$ is significantly smaller than the previous $\eta_{\text{theory}}$. Therefore, the present theoretical result suggests that $|\kappa_{\text{expt.}}\!|>10\text{ meV}$ can not be true. This can be interpreted as $\mathsf{E}$, $\mathsf{F}$ transitions in Fig.~4(c-e) exhibit a splitting smaller than the experimental spectral resolution: $|\kappa_{\text{expt.}}\!|<1\text{ meV}$.
\end{itemize}

\subsection{Optical selection rules for spin doublets}
\label{SM_selection}
\def\tx{x}
\def\ty{y}
\def\ta{a}
\def\txb{\bar{x}}
\def\tyb{\bar{y}}
\def\tab{\bar{a}}

\def\axy{\ta\tx\tyb}%
\def\ayx{\ta\ty\txb}%
\def\xya{\tx\ty\tab}%
\def\aya{\ta\ty\tab}%
\def\axa{\ta\tx\tab}%
\def\xyx{\tx\ty\txb}%
\def\axx{\ta\tx\txb}%
\def\ayy{\ta\ty\tyb}%
\def\xyy{\tx\ty\tyb}%

To determine the optical selection rules of the silicon vacancy center, we work
in a representation of the $T_2$ and $T_1$ manifolds that is simultaneously
adapted to the crystallographic coordinate system of 4H-SiC and to the symmetry
operations of the $C_{3v}$ subgroup. Concretely, we express the representation
matrices of the $T_2$ manifold in a basis aligned with the crystallographic axes,
which also renders the action of the $C_{3v}$ operations block diagonal. This
corresponds to a unitary change of basis from
${t_{2x}, t_{2y}, t_{2z}}$ to ${t_{2,a_1}, t_{2,e_x}, t_{2,e_y}}$, where
$t_{2,a_1}$ transforms as a vector aligned with the crystallographic $c$ axis,
while $t_{2,e_x}$ and $t_{2,e_y}$ transform as vectors lying within the basal
plane perpendicular to $c$. In this basis, the restriction of the $T_2$
representation to the $C_{3v}$ subgroup decomposes as
$T_2 \downarrow C_{3v} = A_1 \oplus E$, and the notation reflects that the selected components of $T_2$ also correspond to 
irreducible representations of $C_{3v}$.

An analogous transformation is performed for the $T_1$ representation. In this
case, the transformed basis functions behave as pseudovectors aligned with the
crystallographic coordinate axes. Accordingly, upon restriction to the $C_{3v}$
subgroup, the $T_1$ representation decomposes as
$T_1 \downarrow C_{3v} = A_2 \oplus E$.

In this symmetry-adapted representation, the many-body electronic states are
constructed using group-theoretical projection operators,\cite{elliott1979b}
applied to the direct product of the orbital and spin representations. The
resulting basis spans the antisymmetrized three-hole configurations transforming
as $T_1$ and $T_2$. In the following, we explicitly list the resulting many-body
wavefunctions corresponding to the $T$ manifolds (to shorten the notation, we
define $a \equiv t_{2,a_1}$, $x \equiv t_{2,e_x}$, and $y \equiv t_{2,e_y}$):
\begin{align*}
\ket{T_{1}(A_{2})} & =\frac{1}{\sqrt{6}}\left(\ket{\axy}-\ket{\ayx}-2\ket{\xya}\right),\\
\ket{T_{1}(E_{x})} & =\frac{1}{\sqrt{3}}\left(-\ket{\axy}-\ket{\ayx}+\frac{1}{\sqrt{2}}\ket{\aya}-\frac{1}{\sqrt{2}}\ket{\xyx}\right),\\
\ket{T_{1}(E_{y})} & =\frac{1}{\sqrt{3}}\left(\frac{1}{\sqrt{2}}\ket{\axa}+\ket{\axx}-\ket{\ayy}+\frac{1}{\sqrt{2}}\ket{\xyy}\right);\\
\ket{T_{2}(A_{1})} & =\frac{1}{\sqrt{2}}\left(\ket{\axx}+\ket{\ayy}\right),\\
\ket{T_{2}(E_{x})} & =\frac{1}{\sqrt{2}}\left(\ket{\axa}+\ket{\xyy}\right),\\
\ket{T_{2}(E_{y})} & =\frac{1}{\sqrt{2}}\left(\ket{\aya}+\ket{\xyx}\right).
\end{align*}

The electric dipole operator transforms as the $T_2$ vector representation. In
the chosen basis, the dipole component along the $c$ axis, $\mu_z$, transforms
as the $a$ component of the $T_2$ representation, while the in-plane components
$\mu_x$ and $\mu_y$ transform as the $x$ and $y$ components of $T_2$,
respectively. Consequently, dipole-mediated optical selection rules follow
directly from the $T_1 \otimes T_2 \otimes T_2$ coupling. In this
representation, the allowed couplings are encoded in the corresponding
Clebsch--Gordan matrices (calculated using GTPack
package\cite{gtpack1,gtpack2}):
\begin{align}
  \label{eq:1}
  C_{a}(\mu_z) = \left(
  \begin{array}{ccc}
    0 & 0 & 0 \\
    0 & 0 & \frac{1}{\sqrt{2}} \\
    0 & -\frac{1}{\sqrt{2}} & 0 \\
  \end{array}
  \right),
  \quad
  C_{x}(\mu_x) = \left(
  \begin{array}{ccc}
    0 & 0 & -\frac{1}{\sqrt{2}} \\
  0 & 0 & 0 \\
  \frac{1}{\sqrt{2}} & 0 & 0 \\
  \end{array}
  \right),
  \quad
  C_{y}(\mu_y) = \left(
\begin{array}{ccc}
 0 & \frac{1}{\sqrt{2}} & 0 \\
 -\frac{1}{\sqrt{2}} & 0 & 0 \\
 0 & 0 & 0 \\
\end{array}
\right),
\end{align}
where the matrix elements correspond to the actual Clebsch--Gordan coefficients, defined as
${(C_l)}_{ij} = \langle T_{2;j} T_{2;l}| T_{1i}\rangle^*$, with $i=1$ corresponding to $a$ and $i=2,3$ corresponding to $x,y$.

These matrices imply that $\mu_z$ polarization induces transitions only between
different components of the $E$ manifolds (e.g.,
$T_1(E_x) \leftrightarrow T_2(E_y)$), whereas in-plane polarizations $\mu_x$ and
$\mu_y$ induce transitions between $T_2(A_1)$ and $T_1(E_{x,y})$, or between
$T_2(E_{x,y})$ and $T_1(A_2)$. These selection rules apply both at the many-body level, where only the total
many-body symmetries are relevant, and at the orbital level, where they follow
from the Slater--Condon rules applied to the wavefunctions listed above.

\clearpage
\section*{Supplemental results and discussion}
\addcontentsline{toc}{section}{Supplemental results and discussion}

\subsection{Transient Absorption Spectroscopy Results}
\label{SM_TA}

As illustrated in Fig.\,\ref{fig:chap6_2D_SiC}-a, a two-dimensional temporal and spectral scan of the sample was conducted, with the sample being pumped at 800\,nm (1.55\,eV), exciting both \vsih and \vsik via their phonon-sideband (PSB), with unpolarized probe light. The results of this scan revealed various distinct features that can be interpreted as follows:

\begin{figure}[h!]
    \centering
\includegraphics[width=0.7\columnwidth]{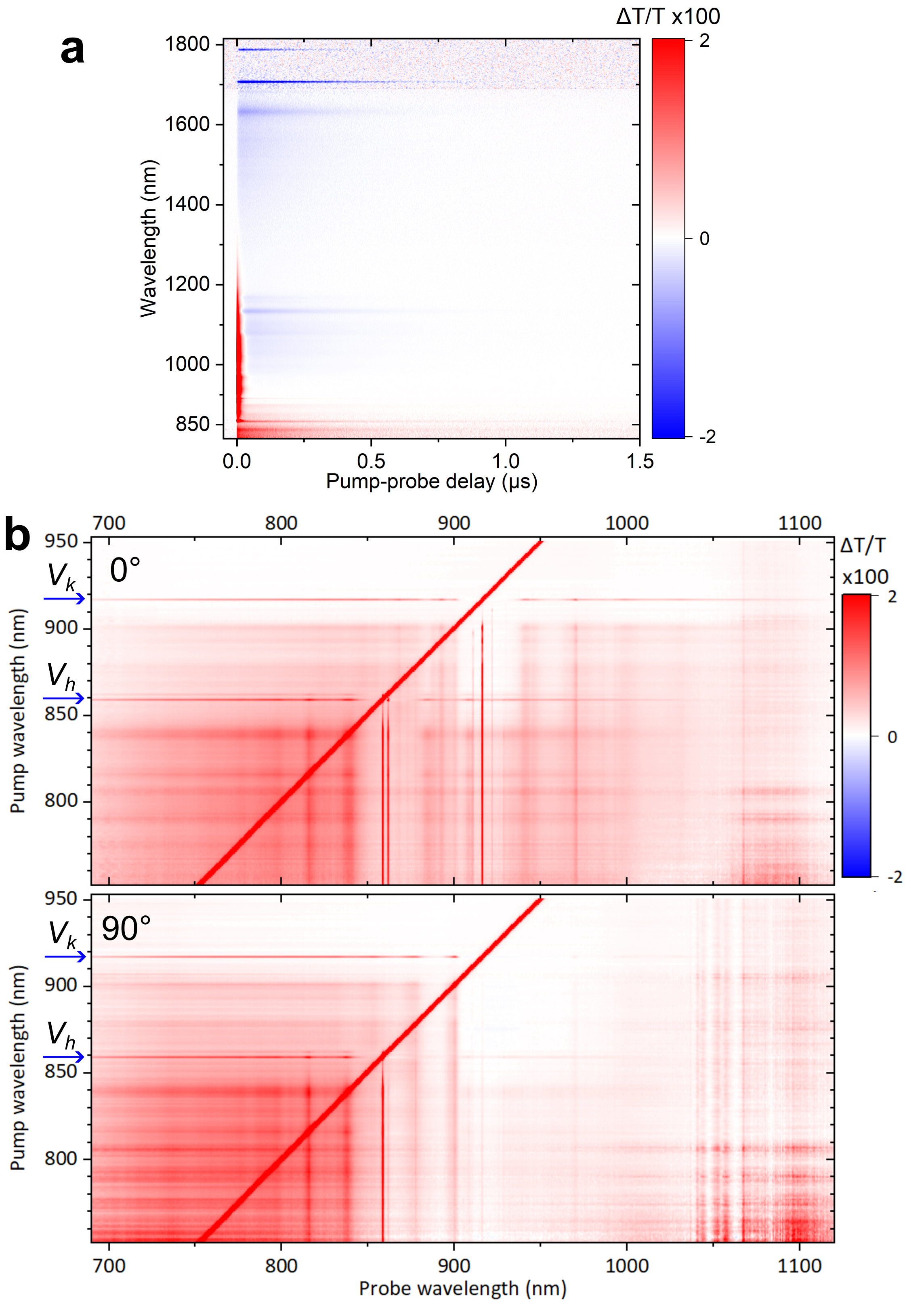}
    \caption{a) Temporal- and spectrally-resolved \dTT\ of \vsi\ in SiC pumped at 800\,nm. b) shows \dTT\ spectra of \vsi\ while scanning the pump wavelength for 0$^\circ$ and 90$^\circ$ polarized probe, respectively. It is measured at a pump-probe delay of approximately 4\,ns.}
    \label{fig:chap6_2D_SiC}
\end{figure}

At energies higher than the ZPL of \vsih (E > 1.44\,eV, $\lambda$ < 860\,nm), there is increased transmission (\dTT\ > 0), due to ground state bleaching (GSB), which resembles steady-state absorption.
GSB arises from population transfer from the ground state to an excited state caused by pump excitation, leading to bleaching of the absorption.
At energies lower than the ZPL (E\,<\,1.44\,eV, $\lambda$\,>\,860\,nm), a positive signal is observed (\dTT\ > 0), which is attributed to  stimulated emission (SE), spectrally resembling photoluminescence (PL). 
Additionally, the contribution from the \vsik excitation can be discerned. At energies higher than the ZPL (E\,>\,1.35\,eV, $\lambda$\,<\,916\,nm) GSB is observed, while at energies lower than the ZPL, SE is present.
Negative signals in the TA spectrum (\dTT\ < 0) generally originate from excited state absorption (ESA). Typically, this signal indicates the absorption from the first excited state (immediately from the pump) to another higher excited state. However, it is also indicative of the population dynamics between the spin-quartet and the spin-doublet channels.
In this particular instance, the ESA signal observed in this spectral region originates from the population in the spin-doublet states, which are populated through ISC and possess a relatively long lifetime (approximately 200\,ns). This is characterized by the delayed appearance of the ESA signal, whose rise time corresponds to the population transfer from the spin-quartet excited states to the spin-doublet states (or "upper ISC" lifetime). The decay of this ESA signal is thus the metastable lifetime of the spin-doublet channel before returning to the spin-quartet ground state ("lower ISC" lifetime). Therefore, the ESA signal time trace allows direct observation of the so-called spin polarization cycle of \vsim\ in 4H-SiC. 

It is noteworthy that the data presented in Fig.\,\ref{fig:chap6_2D_SiC}-a contains the contributions from both \vsih and \vsik due to the excitation into their PSB. 
To differentiate between these features, the wavelength from the ns-OPO is scanned through the absorption range of \vsih and \vsik to excite them selectively. The \dTT\ spectrum while sweeping the pump wavelength at a pump-probe delay of approximately 4\,ns is shown as a 2D contour in Fig.\,\ref{fig:chap6_2D_SiC}-b. Also, a probe polarization dependence was observed in these measurements, which is expected due to the optical selection rules discussed in the manuscript and the SI. 
The 2D spectrum (pump vs. probe wavelength) for the probe polarization rotated by 90$^\circ$ with respect to the previous case is plotted in the lower panel of Fig.\,\ref{fig:chap6_2D_SiC}-b.

\def\vish{$V_{h}$}

\subsection{Photoexcitation dynamics of h-site vacancy $V_h$}
\label{SM_vsih}

\begin{figure}[h!]
    \centering
\includegraphics[width= 0.6 \columnwidth]{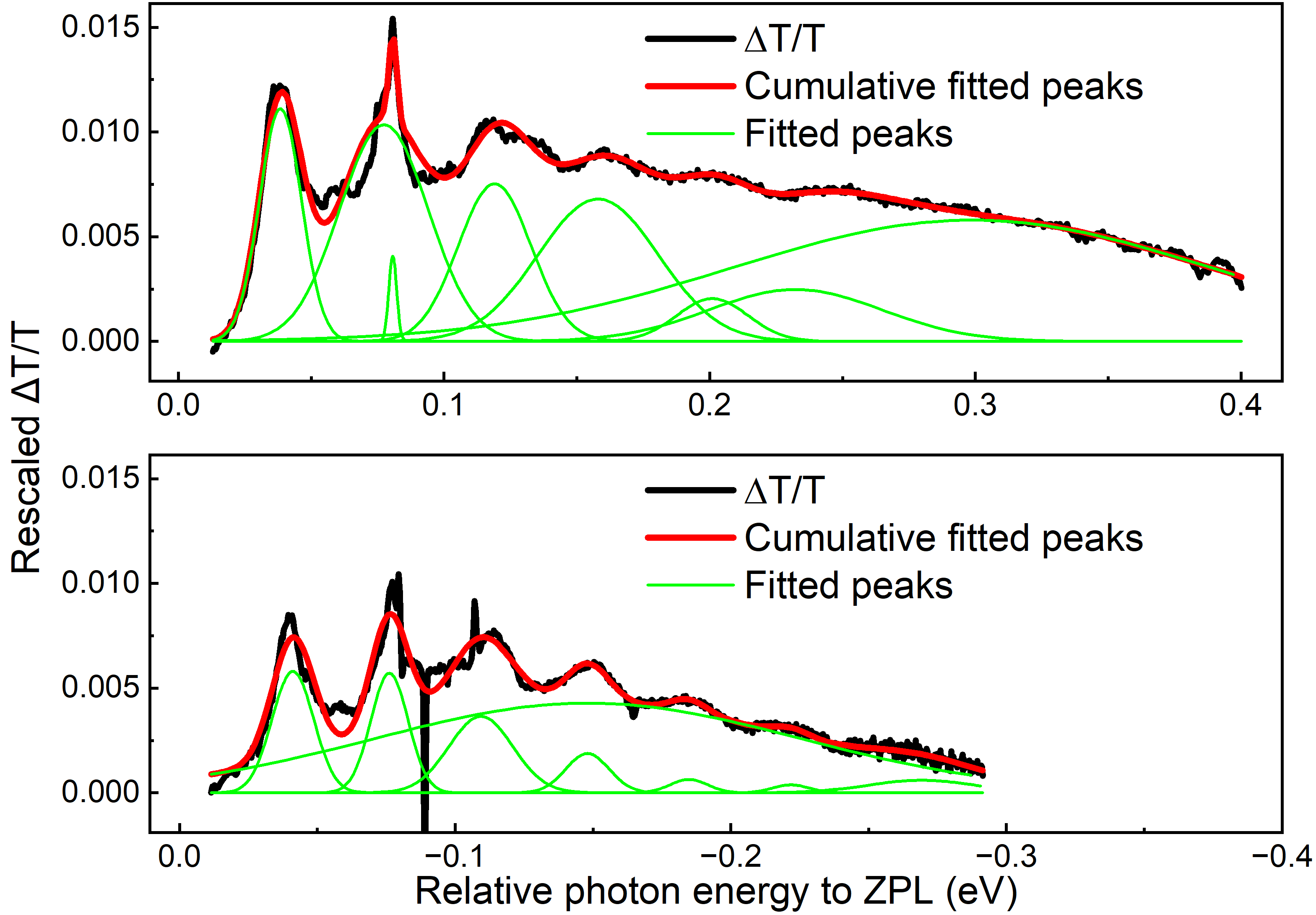}
    \caption{Fitting the GSB and SE of \vsih in the top and bottom plot, respectively.}
    \label{fig:chap6_v1_gsb_se_fit}
\end{figure}

The following spectra are obtained from the top panel of Fig.~\ref{fig:chap6_2D_SiC}b (0°-polarized probe) and indicated with "\vsih". The GSB and SE spectrum of \vsih, shown in Fig.\,\ref{fig2}-b (top panel), are fitted with Gaussian functions to determine the characteristics of the PSB. 
The fitted graphs are plotted in Fig.\,\ref{fig:chap6_v1_gsb_se_fit} and the fitting parameters are listed in Table\,\ref{tab:V1_GSB_SE_param}.  
The separation  between the two phonon lines is approximately 37\,($\pm 3$)\,meV. The ZPL of V1 (\vsih) at 1.438\,eV is also fitted separately with a Lorentzian function (see Section \ref{SM_ZPL}), with the fitting parameters listed in the table. 
Assuming an equal contribution of GSB and SE on the ZPL, and an equal contribution of both V1 and V1' (\vsih), the Debye-Waller factor (DWF) is estimated to be 2.9 and 6.5\,\% for GSB and SE, respectively. 
The value for SE that corresponds to the DWF of PL value is close to previously reported values\,\cite{udvarhelyi2020vibronic,hashemi2021photoluminescence}. 

\begin{table}[htb]
\centering
\renewcommand{\arraystretch}{1.3}
\setlength{\tabcolsep}{6pt}
\small
\begin{tabular}{|c|c|c|c|c|c|c|}
\hline
 & \textbf{Peak Index} & \textbf{Intg. area $\times 10000$} & \textbf{FWHM (meV)} & \textbf{Max height $\times 1000$} & \textbf{Center (meV)} & \textbf{Area ratio (\%)} \\
\hline
\textbf{ZPL} & 0 & 0.9 & 0.7 & 87 & 0 & 2.9 (GSB), 6.49 (SE) \\
\hline
\multirow{8}{*}{\textbf{GSB}} 
& 1 & 2.21 & 19 & 11.1 & 38 & 8.0 \\ \cline{2-7}
& 2 & 4.34 & 39 & 10.3 & 77 & 15.7 \\ \cline{2-7}
& 3 & 0.15 & 3.5 & 4.1 & 81 & 0.6 \\ \cline{2-7}
& 4 & 2.59 & 32 & 7.5 & 119 & 9.4 \\ \cline{2-7}
& 5 & 3.87 & 54 & 6.8 & 158 & 14.0 \\ \cline{2-7}
& 6 & 0.69 & 32 & 2.0 & 201 & 2.5 \\ \cline{2-7}
& 7 & 2.03 & 77 & 205 & 232 & 7.3 \\ \cline{2-7}
& 8 & 11.0 & 213 & 5.8 & 299 & 39.7 \\ 
\hline
\multirow{8}{*}{\textbf{SE}} 
& 1 & 1.10 & 18 & 5.8 & -41 & 8.8 \\ \cline{2-7}
& 2 & 0.97 & 16 & 5.7 & -76 & 7.8 \\ \cline{2-7}
& 3 & 1.06 & 27 & 3.6 & -109 & 8.5 \\ \cline{2-7}
& 4 & 0.37 & 19 & 1.19 & -148 & 3.0 \\ \cline{2-7}
& 5 & 7.70 & 181 & 4.3 & -148 & 61.9 \\ \cline{2-7}
& 6 & 0.11 & 16& 0.64 & -185 & 0.9 \\ \cline{2-7}
& 7 & 0.06 & 16 & 0.38 & -222 & 0.5 \\ \cline{2-7}
& 8 & 0.26 & 46 & 0.60 & -269 & 2.1 \\ 
\hline
\end{tabular}
\caption{Fitting parameters of ZPL, GSB and SE features of \vsih. Intg. area is the area under each fitted function. The area ratio is the ratio of the area under each peak to the entire sideband including ZPL.}
\label{tab:V1_GSB_SE_param}
\end{table}

\begin{figure}[htb]
    \centering
\includegraphics[width= 0.8 \columnwidth]{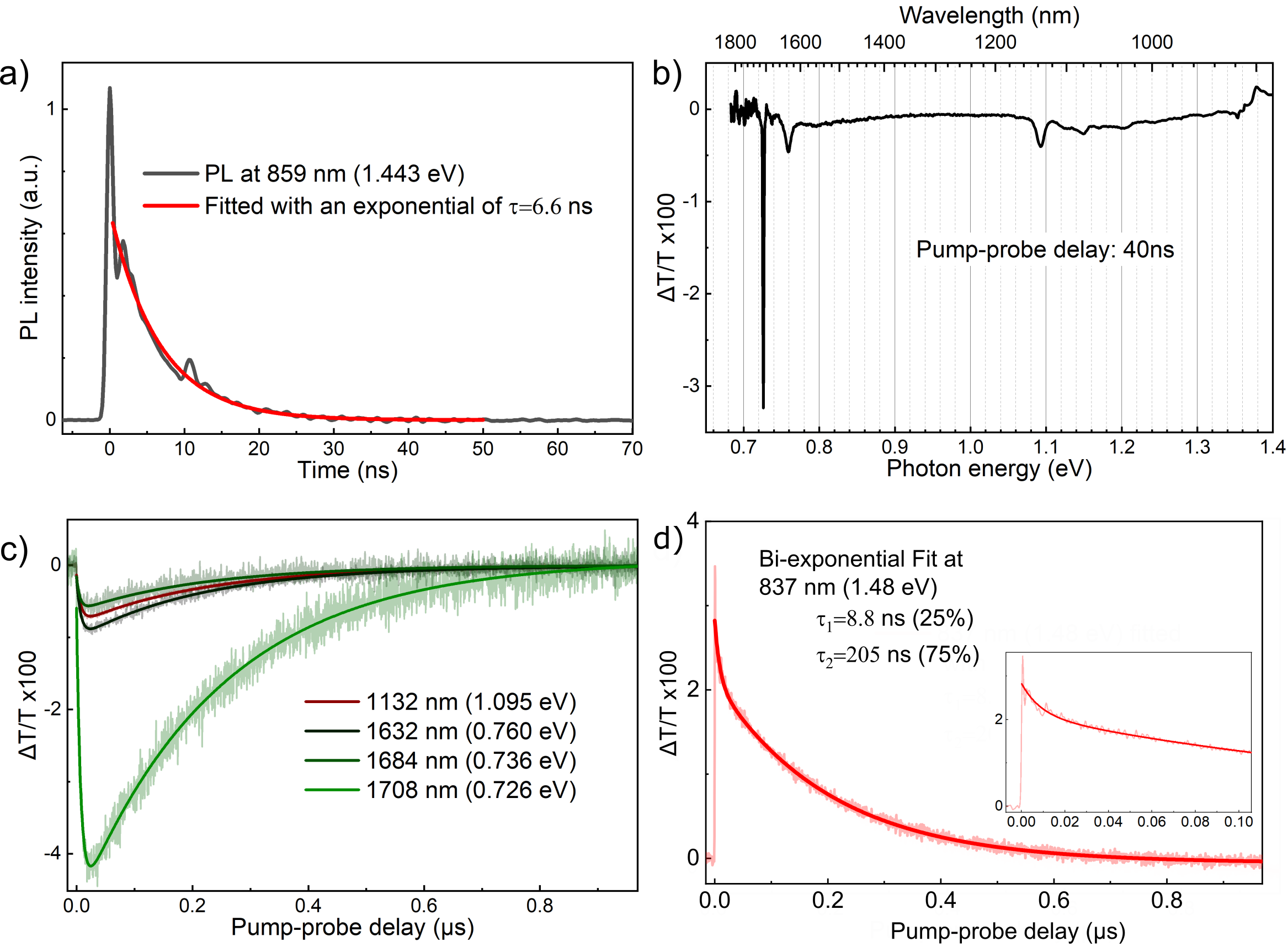}
    \caption{a) The PL decay of \vsih fitted with an exponential function with a lifetime of 6.6\,ns. b) The absorptive features from the lower spin-doublet states to the higher ones of \vsih measured at 40\,ns pump-probe delay. c) Dynamics of the absorption from the spin-doublet states indicating the ISC processes fitted with bi-exponential functions d) the GSB dynamics of \vsih fitted with a bi-exponential function comprising lifetimes of 8.8 and 205\,ns. The percentages indicate the relative amplitude of each exponential component. Inset highlights the electronic signal artifact within the first $\sim$10\,ns.}
    \label{fig:chap6_v1_all}
\end{figure}

Following the excitation from the ground state \gsq, the electrons within the excited state \esqA, relax back to the ground state \gsq\ via two distinct mechanisms. The first pathway involves a radiative decay process, resulting in PL. The second pathway involves ISC through the intermediate doublet-spin states, which is non-radiative. The radiative relaxation lifetime is 6.6\,ns, as demonstrated in Fig.\,\ref{fig:chap6_v1_all}-a. 
Conversely, the non-radiative decay process exhibits a longer duration before reaching the ground state. TA measurements provide direct access to the dynamics associated with this decay process, which is evident as absorptive transitions in the \dTT\ plot, the spectrum of which is shown in Fig.\,\ref{fig:chap6_v1_all}-b. 
The absorption signal in the plot indicates the occupation of the lowest spin-doublet state from which an excitation to higher spin-doublet states can occur. The plot exhibits two regions of absorption: one at 1708\,nm (0.726\,eV) with a narrow linewidth and several peaks with broader linewidths at higher energies.  
This observation suggests that the narrow peak at 1708\,nm (0.726\,eV) may correspond to a ZPL with its PSB located on the blue-shifted side. The other absorption feature contains two peaks around 1100\,nm (1.13\,eV).

As shown in Fig.\,\ref{fig:chap6_v1_all}c, the dynamics of these absorptive peaks and their fitting to a bi-exponential function are analogous, with a similar build-up within approximately 7\,ns (upper ISC time) and a decay with a time constant of 220\,ns (lower ISC time). 
The fitting parameters of these dynamics are listed in Table\,\ref{tab:V1_ISC_fit}. The lower ISC lifetime measured here, around 220\,ns, is in close agreement with previously reported values\,\cite{liu2024silicon}. The GSB dynamics displayed in Fig.\,\ref{fig:chap6_v1_all}d can be fitted to a bi-exponential function, with the time constants shown in the legend. Like in the case of the doublet absorption transients, both time constants correspond to the upper and lower ISC time, respectively. 
The discrepancy between the PL lifetime and the GSB fast time constant of 8.8 ns can be attributed to the electronic artifacts near time-zero (see inset), which is the main source of error for either fits. 
The lifetimes of these states imply that the population of the excited spin-quartet state is either directly transferred to the lower spin-doublet state or indirectly via a rapid decay from a higher spin-doublet state. A similar phenomenon to the latter case has been observed in nitrogen-vacancy centers in diamond\,\cite{luu2024nitrogen}. However, due to the temporal resolution limitations of the ns-$\mu$s TA setup and the possibility of fast decay times (below a nanosecond), the distinction between these two processes remains unresolved.

\begin{table}[htb]
    \centering
    \renewcommand{\arraystretch}{1.4}
    \begin{tabular}{|c|c|c|c|c|c|}
        \hline
        Wavelength (nm) & Energy (eV) & $A_1$ & {$A_2$} & $\tau_1$ (ns) & $\tau_2$ (ns) \\ \hline
        1132 & 1.095\ & 0.649  & -0.801 & 7.2 & 221.4 \\  \hline
        1632 & 0.760 & 0.841  & -1.00     & 7.1 & 218.9 \\ \hline
        1684 & 0.736\ & 0.467  & -0.636 & 6.5 & 221   \\ \hline
        1708 & 0.726 & 4.105  & -4.831 & 7   & 244   \\
        \hline
    \end{tabular}
    \caption{The parameters of fitting \dTT\ data at various wavelengths with a bi-exponential function with the following equation: $A_1 \exp (-t/\tau_1)+A_2 \exp (-t/\tau_2)$.}
    \label{tab:V1_ISC_fit}
\end{table}

The ISC process can be confirmed by measuring the GSB signal that is measured in the ground spin-quartet state, \gsq. The GSB signal is fitted with a bi-exponential function, with time constants of 8.8 and 205\,ns as illustrated in Fig.\,\ref{fig:chap6_v1_all}-d. The extracted lifetimes are in close agreement with the lifetimes reported in Table\,\ref{tab:V1_ISC_fit}. This outcome validates the hypothesis that the lower spin-doublet state decays back to the ground spin-quartet state within its lifetime.

\subsection{Photoexcitation dynamics of k-site vacancy \vsik}
\label{SM_vsik}

\begin{figure}[hbpt]
    \centering
\includegraphics[width=0.5 \columnwidth]{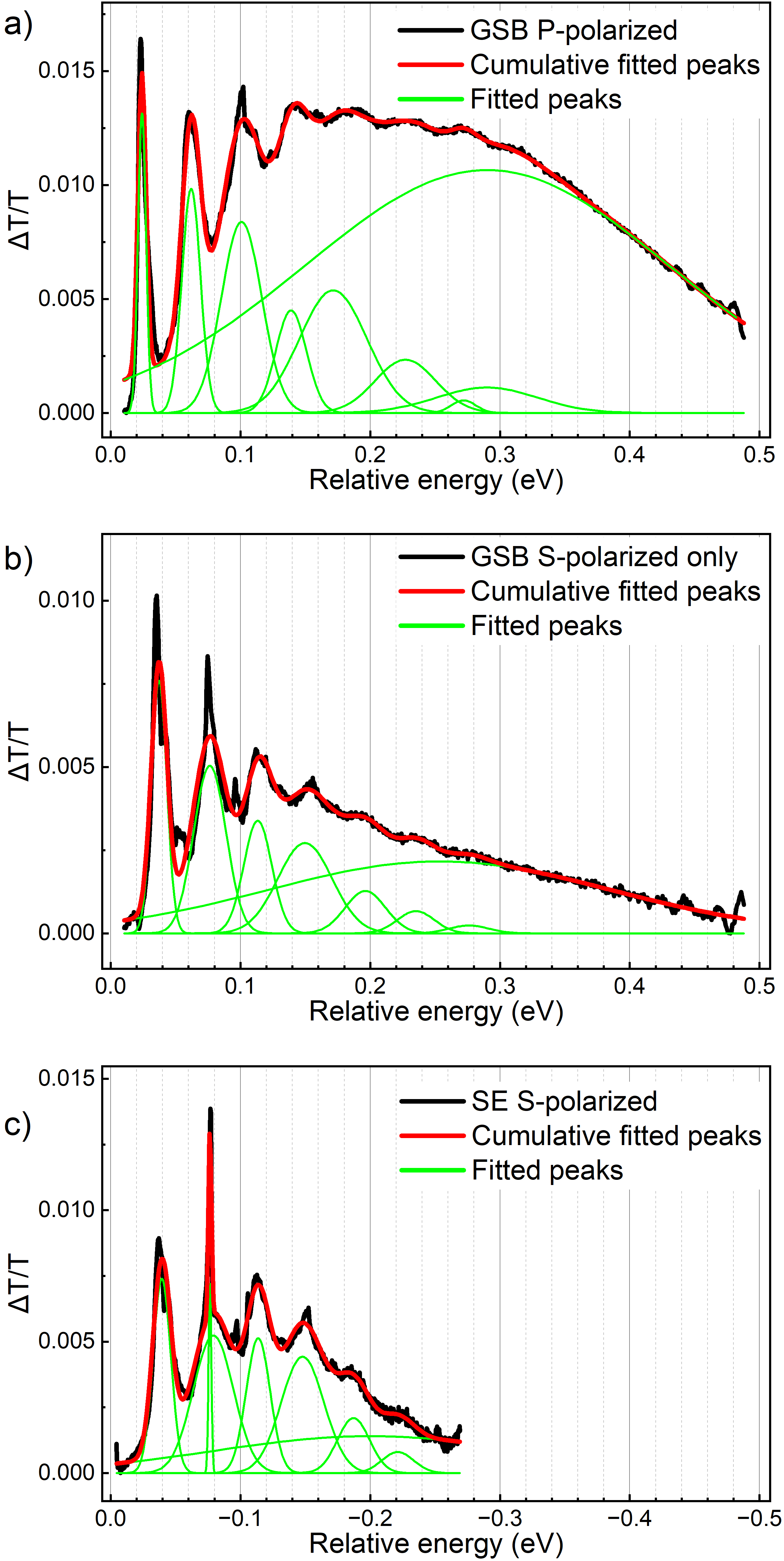}
    \caption{Fitting the PSB (GSB and SE) of \vsik for various conditions: a) GSB with 90°-polarized probe; b) GSB with 0°-polarized only probe (subtracted signal as explained in the text); and c) SE for 0°-polarized probe.}
    \label{fig:chap6_v2_PSB_fit}
\end{figure}

The following spectra are obtained from the top panel of Fig.~\ref{fig:chap6_2D_SiC}b (0°-polarized probe) and indicated with "\vsik". The PSB of \vsik on both the GSB and SE sides are shown in Fig.\,\ref{fig:chap6_v2_PSB_fit}. The shape of the PSB of \vsik is observed to depend on the polarization of the probe beam. As the polarization of the probe beam is rotated from 0° to 90°, a significant decrease in the SE is observed (consistent with the drop in V2 ZPL absorption signal), accompanied by a change in the shape of the GSB and the disappearance of some side peaks. The 90°-polarized data is then taken as a single contribution, and subtracted from the 0°-polarized data with proper rescaling. The rescaling factor was determined by eye to ensure that the subtracted data excluded the peaks present in the 90°-polarized spectrum.
The result of this subtraction is plotted in  Fig.\,\ref{fig2}-c. Hereafter, the term "0°-polarized only" data will be used to denote the subtraction result. The GSB feature for both 90°- and 0°- polarized only are fitted with Gaussian functions and plotted in Fig.\,\ref{fig:chap6_v2_PSB_fit}-a and b, respectively. The SE for 0°-polarized is also fitted with Gaussian peaks and shown in Fig.\,\ref{fig:chap6_v2_PSB_fit}-c. The SE for 90°-polarized is not analyzed because of its low intensity. The parameters of the fittings are listed in Table\,\ref{tab:V2_GSB_SE_param}. The ZPL of V2 (\vsik) at 1.353\,eV is fitted with a Lorentzian function when the sample is pumped at the sideband (880\,nm), and the parameters are listed in the table. In this calculation, the signal is rescaled to have a similar area to the ZPL-pumped case (Fig.\,\ref{fig:chap6_v2_PSB_fit}). However, due to complications of different polarization and different contributions between the GSB and SE, the ZPL area is not considered in the area ratio value. 

\begin{table}[htb]
\centering
\renewcommand{\arraystretch}{1.3}
\setlength{\tabcolsep}{6pt}
\small
\begin{tabular}{|c|c|c|c|c|c|c|}
\hline
 & \textbf{Peak Index} & \textbf{Intg. area $\times 10000$} & \textbf{FWHM (meV)} & \textbf{Max height $\times 1000$} & \textbf{Center (meV)} & \textbf{Area ratio (\%)} \\
\hline
\multirow{9}{*}{\textbf{GSB-90°}} 
& 1 & 1.02 & 7.3 & 13.1 & 25 & 2.2 \\ \cline{2-7}
& 2 & 1.77 & 17 & 9.8 & 62 & 3.8 \\ \cline{2-7}
& 3 & 3.28 & 37 & 8.4 & 101 & 7.0 \\ \cline{2-7}
& 4 & 1.33 & 28 & 4.5 & 139 & 2.8 \\ \cline{2-7}
& 5 & 3.51 & 61 & 5.4 & 171 & 7.4 \\ \cline{2-7}
& 6 & 1.37 & 55 & 201.0 & 227 & 2.5 \\ \cline{2-7}
& 7 & 1.32 & 22 & 0.56 & 272 & 0.3 \\ \cline{2-7}
& 8 & 1.09 & 91 & 1.1 & 290 & 2.3 \\ \cline{2-7}
& 9 & 34.0 & 331 & 10.7 & 290 & 71.4 \\ 
\hline
\multirow{8}{*}{\textbf{GSB-0°}} 
& 1 & 1.31 & 14 & 7.6 & 37 & 9.0 \\ \cline{2-7}
& 2 & 1.57 & 29 & 5.0 & 76 & 12.6 \\ \cline{2-7}
& 3 & 0.92 & 26 & 3.4 & 113& 7.4 \\ \cline{2-7}
& 4 & 1.43 & 49 & 2.7 & 150  & 11.4 \\ \cline{2-7}
& 5 & 0.50 & 37 & 1.3 & 196 & 4.0 \\ \cline{2-7}
& 6 & 0.24 & 34 & 0.67 & 235 & 1.9 \\ \cline{2-7}
& 7 & 6.63 & 310 & 2.2 & 253 & 53.4 \\ \cline{2-7}
& 8 & 0.24 & 36 & 0.24 & 276 & 0.7 \\ 
\hline
\multirow{8}{*}{\textbf{SE-0°}} 
& 1 & 1.27 & 16 & 7.4 & -39 & 12.5 \\ \cline{2-7}
& 2 & 0.20 & 2.6 & 7.2 & -77 & 1.9 \\ \cline{2-7}
& 3 & 2.03 & 37 & 5.2 & -79 & 20.0 \\ \cline{2-7}
& 4 & 1.19 & 22 & 5.1 & -114 & 11.6 \\ \cline{2-7}
& 5 & 1.83 & 39 & 4.4 & -148 & 17.9 \\ \cline{2-7}
& 6 & 0.65 & 29 & 2.1 & -187 & 6.3 \\ \cline{2-7}
& 7 & 2.80 & 281 & 1.4 & -200 & 27.4 \\ \cline{2-7}
& 8 & 0.25 & 29 & 0.8 & -221 & 2.4 \\ 
\hline
\textbf{ZPL-0°} & 0 & 0.81 & 0.7 & 87.0 & 0.0 & -- \\
\hline
\end{tabular}
\caption{Fitting parameters of GSB (for both 90°-polarized and 0°-polarized only, labeled as GSB-90° and GSB-0°, respectively), and SE (for 0°-polarized probe) of \vsik. The ZPL for 0°-polarized probe (ZPL-0°) is also fitted while the sample is pumped at the sideband and rescaled accordingly. Intg. area is the area under each fitted function. The area percentage is the ratio of each peak to the entire sideband excluding ZPL area.}
\label{tab:V2_GSB_SE_param}
\end{table}

\begin{figure}[htb]
    \centering
\includegraphics[width=0.8\columnwidth]{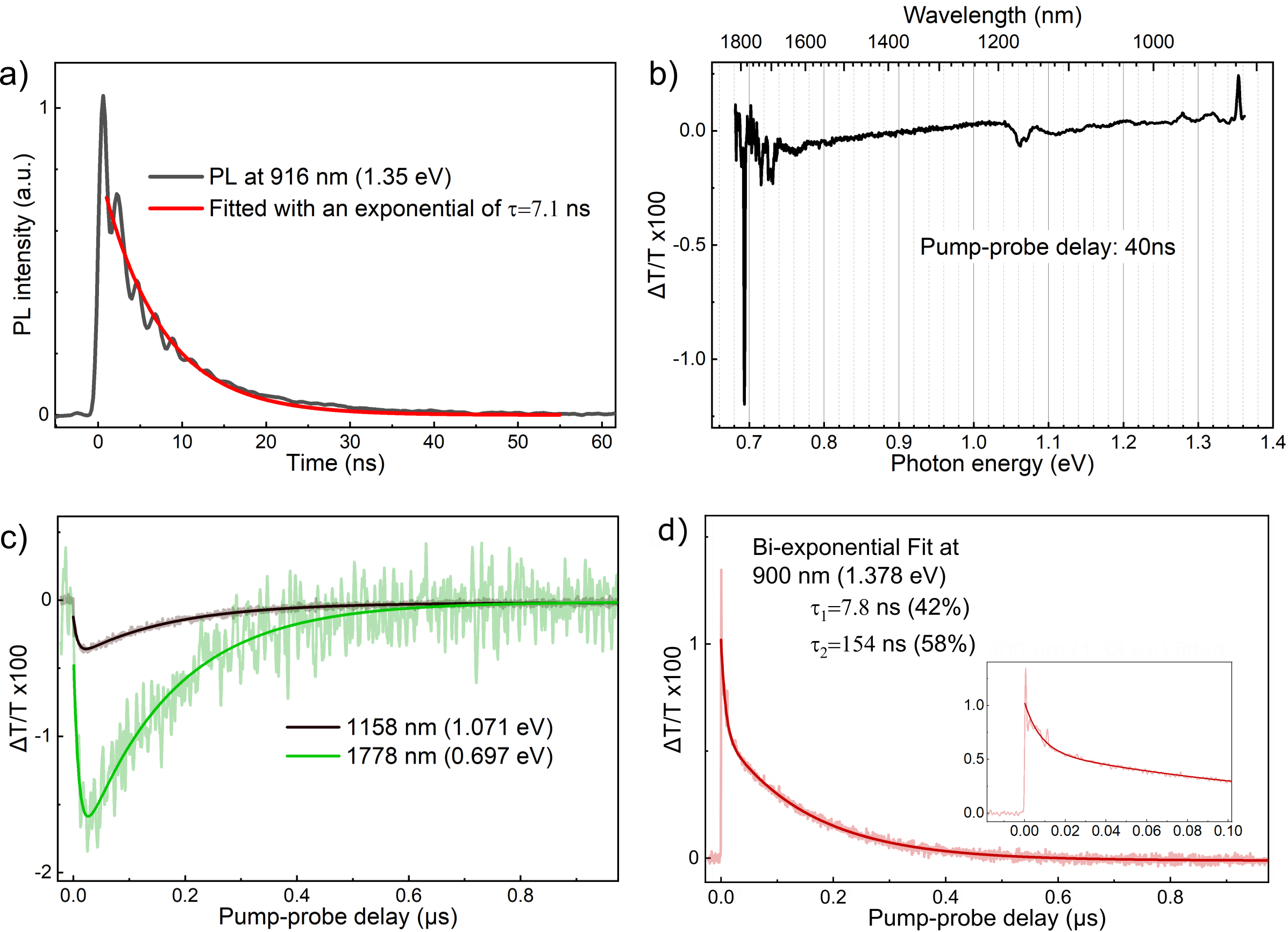}
    \caption{a) The PL decay of \vsik fitted with an exponential function with a lifetime of 7.1\,ns. b) The absorptive features from the lower spin-doublet states to the higher ones of \vsik measured at 40\,ns pump-probe delay. c) Dynamics of the absorption from the spin-doublet states, representing the ISC processes fitted with bi-exponential functions. d) The GSB dynamic of \vsik fitted with a bi-exponential function comprising lifetimes of 7.8 and 154\,ns. The percentages indicate the relative amplitude of each exponential component. Inset highlights the signal artifact within the first $\sim$10\,ns.}
    \label{fig:chap6_v2_all}
\end{figure}

The photoexcitation dynamics of \vsik are analogous to those of \vsih, wherein the pump excites the electrons from the ground to the excited spin-quartet states (\gsq\,$\rightarrow$\,\gsq). 
The electrons in the excited state either decay back to the ground state through either PL or a non-radiative channel via ISC to the spin-doublet states. 
The excited state lifetime, as determined through PL signal fitting, is estimated to be approximately 7.1\,ns, as illustrated in Fig.\,\ref{fig:chap6_v2_all}-a. 
Similar to \vsih, absorptive features corresponding to the ISC and spin-doublet states absorption are observed. They are considered to be the absorption from the lower to higher spin-doublet states. The \dTT\ spectrum showing this feature is illustrated in Fig.\,\ref{fig:chap6_v2_all}-b. 
These features look almost identical to \vsih with shifted energies. 
The graph reveals two distinct peak sets: one at approximately 1778\,nm (0.694\,eV) with a narrow linewidth, suggesting a ZPL of a transition accompanied by its PSB with higher energies from the lowest populated spin-doublet state to the intermediate spin-doublet state; and another at about 1160\,nm (1.065\,eV) with a broader linewidth than the previous one. 
 Additionally, there are several lower-energy peaks, which are presumed to be the PSB of this transition. In contrast to the \vsih case, the latter absorptive feature exhibits two peaks separated by 10\,meV (peaks at 1.071 and 1.061 eV). The dynamics of the absorptive features are plotted in Fig.\,\ref{fig:chap6_v2_all}-c. 

The dynamics of two absorptive peaks at 1158\,nm (1.071\,eV) and 1778\,nm (0.697\,eV), along with their fittings with bi-exponential functions are plotted in Fig.\,\ref{fig:chap6_v2_all}-c. The fitting parameters are listed in Table\,\ref{tab:V2_ISC_fit}. The absorption builds up with a lifetime similar to the excited spin-quartet state lifetime, around 8\,ns, representing the populating of the ground doublet state via ISC from the excited spin-quartet state. The absorption lasts around 165\,ns, which is the lifetime of the lower spin-doublet state, which decays back to the ground spin-quartet state via ISC. It is noteworthy that the lifetimes observed here are of the same order as the \vsih ISC lifetimes.

\begin{table}[htb]
    \centering
    \renewcommand{\arraystretch}{1.4}
    \begin{tabular}{|c|c|c|c|c|c|}
        \hline
        Wavelength (nm) & Energy (eV) & $A_1$ & {$A_2$} & $\tau_1$ (ns) & $\tau_2$ (ns) \\ \hline
        1158 & 1.071\ & 0.299  & -0.407 & 7.8 & 168.1 \\  \hline
        1778 & 0.697 & 1.646  & -1.972  & 9.4 & 160.2 \\ \hline
    \end{tabular}
    \caption{The parameters of fitting \dTT\ data at various wavelengths with a bi-exponential function with the following equation: $A_1 \exp (-t/\tau_1)+A_2 \exp (-t/\tau_2)$.}
    \label{tab:V2_ISC_fit}
\end{table}

As illustrated in Fig.\,\ref{fig:chap6_v2_all}-d, the GSB of \vsik is fitted with a bi-exponential function, comprising two lifetimes: 7.8 and 154\,ns. 
The fast component corresponds to the lifetime of the spin-quartet excited state, as substantiated by the PL lifetime. Similarly to the \vsih case, the difference between the measured PL and GSB lifetimes are due to the instrumental error near time-zero by the large signal artifact.
The longer component is the relaxation time of the spin-doublet states to the ground spin-quartet state via ISC. 

\subsection{V2' PSB characterization and analysis}
\label{SM_V2PSB}

\begin{figure}[b]
    \centering
    \includegraphics[width=0.45\linewidth]{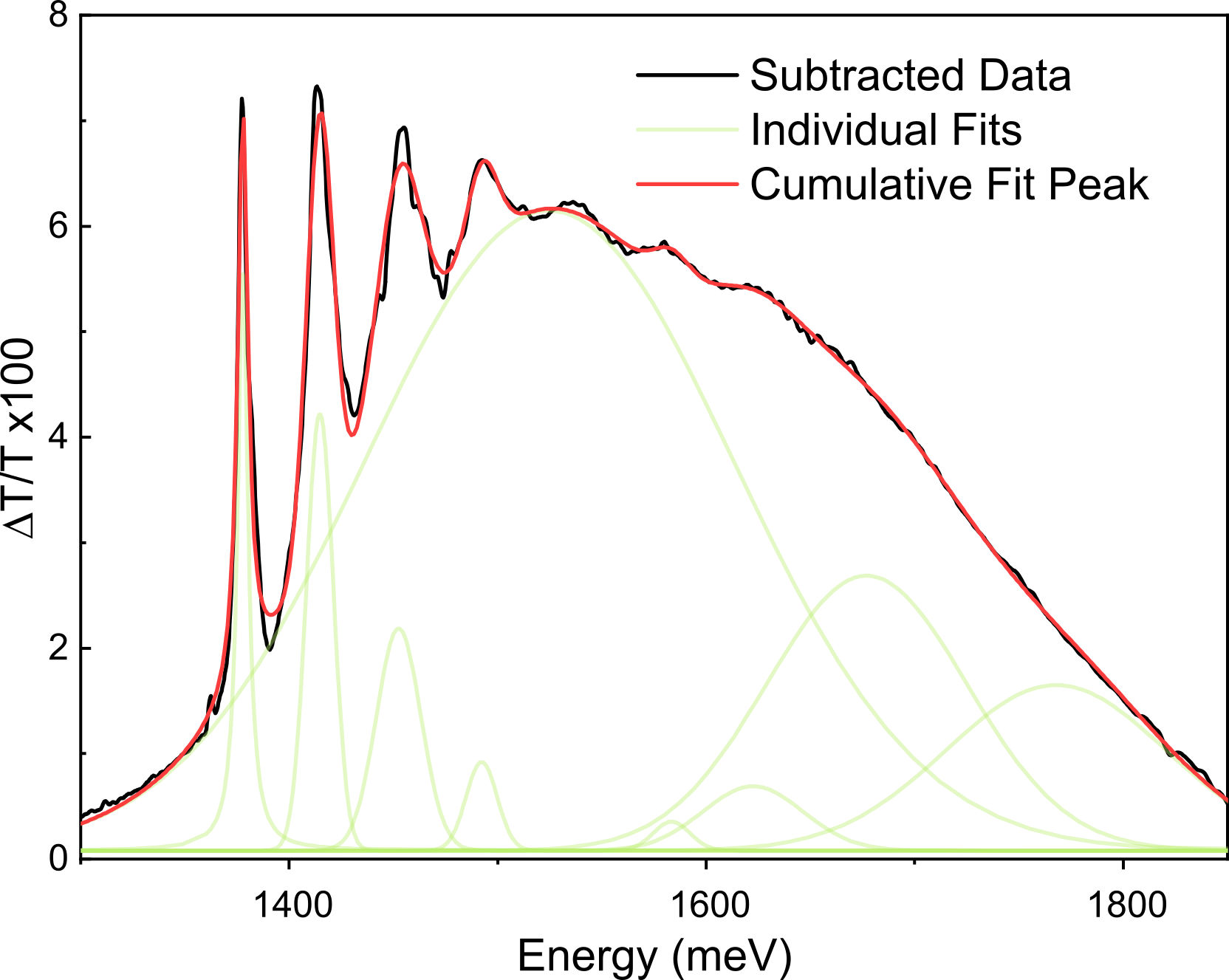}
    \caption{V2' GSB spectrum obtained from selective pumping at normal incidence. Parameters detailed in Table~\ref{table:v2prime_psb}.}
    \label{fig:v2prime_psb}
\end{figure}

In the oblique incidence case, exciting the V2' will also excite the V2 transition (would require polarization control of the pump pulse otherwise). To obtain a clean absorption spectrum of V2', we perform a TA measurement pumping selectively at the V2' ZPL at 1.38\,eV with the probe beam entering with the normal incidence case. Here, the probe only interacts with transitions allowed by $\mu_x$ and $\mu_y$, thereby eliminating the response of V1 and V2. Furthermore, pumping below the ZPL of V1' (1.43\,eV) eliminates the contribution of V1' to the spectrum, hence only the response of V2' will be observed with the probe. 

Unfortunately, the TA spectrum is susceptible to the pump scattering at the excitation wavelength. Thus, at the ZPL wavelength (and any spectral position of the pump), the spectrum has an artifact that cannot be filtered out. In order to obtain a spectrum of the V2' with both the ZPL and PSB, we take a spectrum pumped at the ZPL, then another spectrum pumped at the 1-phonon line. The resultant spectrum uses the data at the ZPL in the 1-phonon pump case, and the data at 1-phonon peak comes from the ZPL pump case. The rest of the spectrum is the averaged spectrum of the two cases. It is noted that the absorption cross-sections of the V2' ZPL and its 1-phonon peak are comparable from the SETA measurements (Fig.~\ref{fig:chap6_2D_SiC}). A FFT lowpass filter was also applied to the spectrum to remove interference artifacts due to the thin sample.
\begin{table}[t]
\centering
\begin{tblr}{
  cells = {c},
  hlines,
  vlines,
}
Peak Index & Peak Type & {Area Intg\\x10000} & {FWHM\\(meV)} & Center Energy (eV) & Area (\%) \\
1          & Lorentz   & 5.17                & 6.08          & 1.378         & 2.46   \\
2          & Gaussian  & 6.70                & 15.2          & 1.415         & 3.19   \\
3          & Gaussian  & 5.64                & 25.2          & 1.452         & 2.68   \\
4          & Gaussian  & 1.57                & 17.6          & 1.492         & 0.750   \\
5          & Gaussian  & 135                 & 209           & 1.525         & 64.2  \\
6          & Gaussian  & 0.60                & 20.7          & 1.583         & 0.287   \\
7          & Gaussian  & 3.67                & 56.6          & 1.622         & 1.75  \\
8          & Gaussian  & 31.3                & 113           & 1.677         & 14.9  \\
9          & Gaussian  & 20.6                & 124           & 1.768         & 9.78   
\end{tblr}
\label{table:v2prime_psb}
\caption{Fitting parameters of the V2' GSB spectrum as shown in Fig.~\ref{fig:v2prime_psb}.}
\end{table}

A 9-peak model was used, where the first peak is a Lorentzian fit to the expected ZPL, while the rest of the peaks are Gaussian fits to model the response of vibronic modes. The overall fit result is shown in Fig.~\ref{fig:v2prime_psb}. The response shows a rather intricate structure, with a rather large red onset of the ZPL response and a sharp response at 121\,meV, indicating a local vibrational mode (LVM) at higher energy than the LO phonon mode of 4H-SiC at 119\,meV. The red onset is modeled by a broad peak, serving similarly to a baseline for the first 4 peaks within the first replica regime. The peaks at higher energy than this broad peak models the second replica regime of the phonon modes which are much more broadened than their first replica counterparts.

\newpage
\subsection{Polarization dependence of the spin-doublet transitions}
\label{SM_pol}

In this Section, we provide an analysis of the polarization-dependent spectra of the spin-doublet transitions, which were shown in Fig.~\ref{fig4} of the main manuscript. For either h- or k-site, we model the transition from the lower spin-doublet $^2T_1$ to the splitting of the upper spin-doublet $^2E$ and $^2A_1$ (from $^2T_2$) with two Lorentzian peaks and fit the model to the experimental data. To condition for the fitting, the background signal not contributing to the peaks are subtracted from the data.

\begin{table}[h]
\centering
\begin{tblr}{
  cells = {c},
  cell{1}{1} = {r=2}{},
  cell{1}{2} = {c=2}{},
  cell{1}{4} = {c=2}{},
  cell{1}{6} = {r=2}{},
  cell{1}{7} = {c=2}{},
  vlines,
  hline{1,3,5} = {-}{},
  hline{2} = {2-5,7-8}{},
  hline{4} = {1,6}{},
}
$V_h$  & {Integrated Area\\ x10000 (a.u.)} &     & {FWHM\\  (meV)} &     & {Center Energy\\ (meV)} & Area Ratio (\%) &     \\
       & 0°                                & 90° & 0°              & 90° &                         & 0°              & 90° \\
Peak 1 & 2.8                               & 1.9 & 7.2             & 4.6 & 1091                    & 63              & 26  \\
Peak 2 & 1.6                               & 5.3 & 6.4             & 6.4 & 1095                    & 37              & 74  
\end{tblr}

\caption{Fitting parameters of the D$^2$ transitions of the \vsih\ center with 0°-polarized and 90°-polarized probe light. The curves are shown in Fig.~\ref{fig:doublet_fits}a.}
\label{table:doublet_vh}
\end{table}

\begin{table}[h]
\centering
\begin{tblr}{
  cells = {c},
  cell{1}{1} = {r=2}{},
  cell{1}{2} = {c=2}{},
  cell{1}{4} = {c=2}{},
  cell{1}{6} = {r=2}{},
  cell{1}{7} = {c=2}{},
  vlines,
  hline{1,3,5} = {-}{},
  hline{2} = {2-5,7-8}{},
  hline{4} = {1,6}{},
}
$V_k$  & {Integrated Area\\ x10000~\\(a.u.)} &     & {FWHM\\  (meV)} &     & {Center Energy\\ (meV) } & {Area Ratio~\\(\%)} &     \\
       & 0°                                  & 90° & 0°              & 90° &                          & 0°                  & 90° \\
Peak 1 & 5.7                                 & 1.9 & 9.6             & 9.7 & 1061                     & 86                  & 49  \\
Peak 2 & 0.9                                 & 1.9 & 5.8             & 6.6 & 1071                     & 14                  & 51  
\end{tblr}

\caption{Fitting parameters of the D$^2$ transitions of the \vsik\ center with 0°-polarized and 90°-polarized light. The curves are shown in Fig.~\ref{fig:doublet_fits}b.}
\label{table:doublet_vk}
\end{table}

\begin{figure}[b]
    \centering
    \includegraphics[width=1\linewidth]{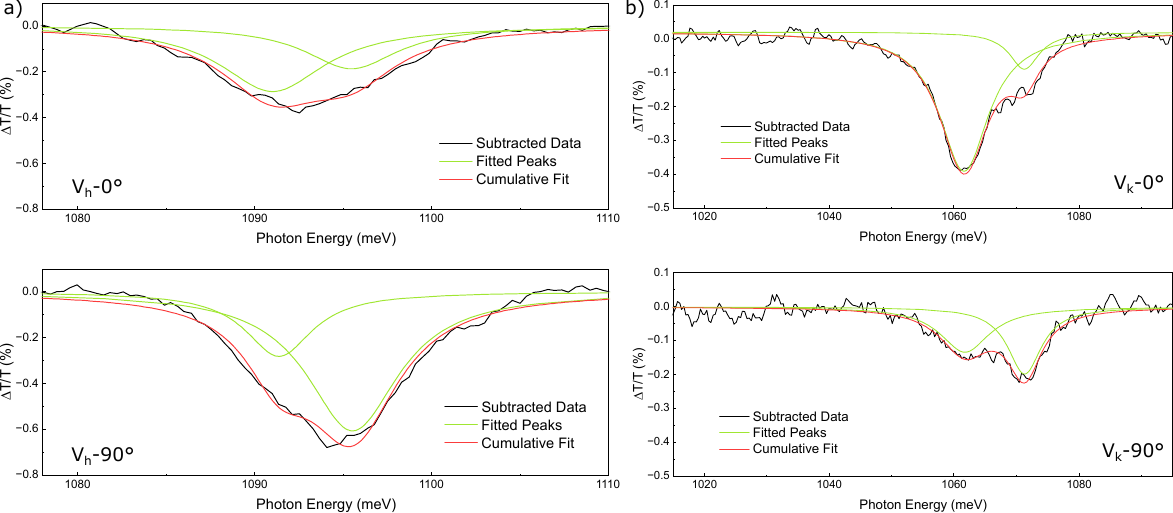}
    \caption{Lorentzian fitting of the spin-doublet transitions in either 0°-polarization or 90°-polarization cases: a) \vsih site; b) \vsik site.}
    \label{fig:doublet_fits}
\end{figure}

The fit results are shown in Fig.~\ref{fig:doublet_fits} for both \vsih and \vsik cases and summarized numerically in Table~\ref{table:doublet_vh} for \vsih and Table~\ref{table:doublet_vk} for \vsik. While the two peaks are rather well-separated spectrally in the \vsik case, the \vsih case is not as clear. This is largely because the linewidth (in FWHM) of the doublet transitions are in the 6-8\,meV range even at 5\,K and therefore a resolution of two peaks separated by less than $\sim$5\,meV is less reliable. However, the two-peak model for \vsih can still yield the relative strength of two transitions and approximate the $^2T_2$ splitting. We notice a similarity in the behavior of the relative peak strength in either \vsih or \vsik shifting from the lower energy peak to the higher energy peak when the polarization is changed from 0° to 90°. We interpret this observation as the lower-energy peak (Peak 1) representing the C transitions with the $\mu_z$ component and the higher-energy peak (Peak 2) representing the B transitions with $\mu_x$ and $\mu_y$ components. As the polarization is switched to 90°, the probe no longer probes the $\mu_z$ component and instead couples to $\mu_y$, shifting the signal towards Peak 2. From this, we conclude that ordering of the upper orbital-triplet splitting ($^2T_2$) as illustrated in Fig.~\ref{fig4}c in the main text from the assignment of the B and C transitions in both \vsih and \vsik.

\subsection{Temperature dependence of spin-doublet transitions}

\label{SM_temp}

\begin{figure}[b]
    \centering
    \includegraphics[width=1\linewidth]{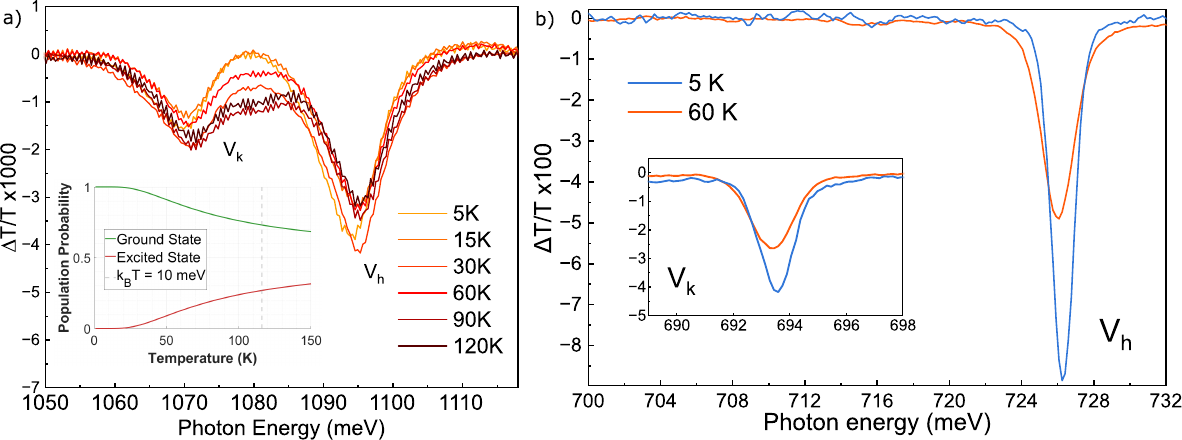}
    \caption{a) Spectra of the A transition in either h- or k-sites in the normal incidence case with temperature between 5\,K and 120\,K. Inset shows the population probability of a two-level system separated by 10\,meV in energy against temperature using a Boltzmann distribution model. b) Spectral comparison of the E transitions at 5\,K and 60\,K. Obtained with oblique incidence configuration with mixed 0°- and 90°-polarized light.}
    \label{fig:tdep_doublets}
\end{figure}

As shown in Fig.~\ref{fig:tdep_doublets}, a temperature dependence measurement was carried out in the normal incidence configuration primarily to determine the splitting energy of the lower doublet $^2T_1$, which we will call $\kappa$. The normal incidence configuration eliminates the spectral contribution of the C transition, and thus only A and B transitions are expected (transition names shown in Fig.~\ref{fig4}). 
We observed transitions at 1071\,meV and 1095\,meV (Fig.~\ref{fig:tdep_doublets}a), which are attributed to either A or B transitions in h- and k-sites, respectively. These transitions do not exhibit any substantial spectral shift or clear new spectral feature with increasing temperature, only slight red shifts of these transitions. This is further made difficult by the linewidth of these transitions, which is about 6-8\,meV (FWHM) and increases with temperature. While it is possible to analyze the slight peak shifts by considering the spectral shift of the area under curve to determine $\kappa$, this is complicated by the dependence of the relative oscillator strength between the A and B transitions. Other temperature-related effects such as thermal expansion of bulk SiC and electron-phonon coupling will also contribute to the spectral shift (as observed in the NV center in diamond \cite{luu2024nitrogen,Doherty_2014}). 

Despite such complications, we can still make a few observations about the splitting of the $^2T_1$ states. The insensitivity to temperature dependence implies two possible scenarios: 
\begin{itemize}
    \item $\kappa$ is small (sub-1\,meV): population in the $^2A_2$ and $^2E$ states are already approximately equal at 5\,K, thus increasing temperature does not change the population in either states. 
    \item $\kappa$ is large (> 10\,meV): population in the upper state cannot be thermally occupied significantly at 5\,K and elevated temperatures. From the polarization dependence measurements (Fig.~\ref{fig4}) in the main text, this case is only valid if $^2E$ is the lower energy state due to the sensitivity to the $\mu_z$ transition dipole at 5\,K.  
\end{itemize}

From the temperature dependence results here and the results observed in the polarization dependence measurements, the transitions at 1071\,meV and 1095\,meV are attributed tentatively to the B transitions in h- or k-sites, as they signify the population from the $^2\!E$, sharing the same lowest state as the C transitions. 

While a $\kappa$ $<$ 1\,meV splitting cannot be verified experimentally due to the linewidth of these transitions, we can explore the possibility that the splitting is $>$ 10\,meV by spectrally identifying transition A and F. From Fig.~\ref{fig4}c, the energy for the transition A would be \(E_A = E_B - (\kappa+\eta)\), or a spectral red-shift from the B transitions. Therefore, the expected spectral shift would be $>\sim$14\,meV and $>\sim$20\,meV for \vsih and \vsik, respectively. Transition F red-shift would simply be the $\kappa$ splitting. In the case of \vsih that means:
\begin{itemize}
    \item Transition A: In Fig.~\ref{fig:tdep_doublets}, there is a region in between 1075-1085\,meV that has a rise in signal as the temperature is increased. One may interpret this rise in signal as thermal populating of the $^2\!A_2$, thereby suggesting that the $\kappa$ splitting is $\sim$10\,meV. A simple Boltzmann factor model of a two-level system separated by 10\,meV (see the inset of Fig.~\ref{fig:tdep_doublets}a) also predicts a similar pattern to the signal rise: the signal onset is around 30\,K and saturates at around 100\,K. Unfortunately, the signal observed here could also simply arise from the broadening of the B transitions. Without precise parameters regarding the oscillator strength of the A transitions, it is not possible to discern the two possibilities in this case.
    \item Transition F: Following the observations for transition A, one would expect a resonance to appear at about 716\,meV (10\,meV red-shifted). However, there is a complete lack of any other spectral feature at elevated temperature (Fig.~\ref{fig:tdep_doublets}b) in this region. It is also possible that while the $^2\!A_2$ can be thermally populated, the F transition may be too weak to be observed.
\end{itemize}

In the case of \vsik, there is no observable red-shifted spectral feature until the spectral cutoff of our measurement at 690\,meV. This can once again point to a splitting too large to be thermally populated.
While there are some arguments supporting $\kappa>$10\,meV (at least in \vsih) from the A transition, the lack of evidence in the F transition renders this hypothesis inconclusive. A splitting much larger than this value is also not supported by the crystal field splittings observed in other states. Nevertheless, we could not rule out the possibility that the $\kappa$ is in the 10-30\,meV range for either \vsih or \vsik from the data obtained here and would likely require a different spectroscopy technique to determine this splitting.



\subsection{ZPL characteristics of the spin-doublet and spin-quartet transitions}
\label{SM_ZPL}
\begin{wraptable}{R}{6.5cm}
\centering
\renewcommand{\arraystretch}{1.25}

\begin{tabular}{|c|c|c|c|} 
\hline
\begin{tabular}[c]{@{}c@{}}Transition\\Name\end{tabular} & \begin{tabular}[c]{@{}c@{}}Center\\Energy\\(meV)\end{tabular} & \begin{tabular}[c]{@{}c@{}}FWHM\\(meV)\end{tabular} & Fit Type  \\ 
\hline
V1          & 1439         & 0.9~        & Lorentz     \\ 
\hline
V1'         & 1444         & 0.8         & Lorentz     \\ 
\hline
V2$_0$      & 1353         & 0.7        & Lorentz    \\ 
\hline
V2$_{-1}$   & 1344         & 0.7         & Lorentz   \\ 
\hline
V2$_{+1}$   & 1361         & 0.7          & Lorentz   \\ 
\hline
V2'         & 1378         & 6.1         & Lorentz   \\ 
\hline
D$^1_h$     & 726          & 1.5         & Gaussian  \\ 
\hline
D$^1_k$     & 694          & 1.4         & Gaussian  \\ 
\hline
D$^2_h$     & 1095        & 6.2 $\pm$ 0.5  & Lorentz   \\ 
\hline
D$^1_k$     & 1065        & 7.9 $\pm$ 1.8  & Lorentz   \\
\hline
\end{tabular}
\caption{Summary of the linewidth of the ZPLs for all observed transitions presented in Fig.~\ref{fig:linewidth_zpl_all}.}
\label{table:zpl_linewidth}
\end{wraptable}

\begin{figure}[b]
    \centering
    \includegraphics[width=0.9\linewidth]{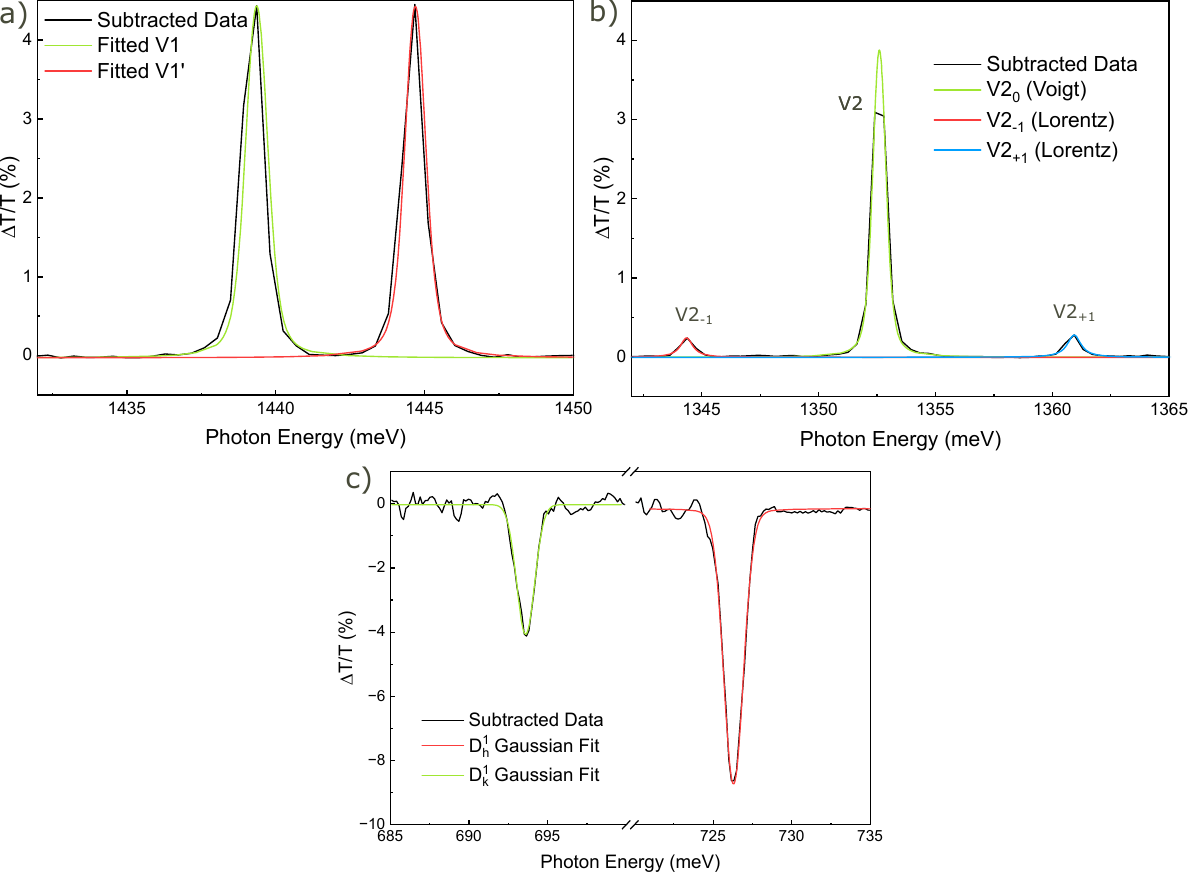}
    \caption{Peak Fitting of the ZPLs: a) V1 and V1'; b) V2 (center) and the two unknown splittings. c) D$^1$ transitions.}
    \label{fig:quartet_fits}
\end{figure}

\begin{figure}[h]
    \centering
    \includegraphics[width=0.5\linewidth]{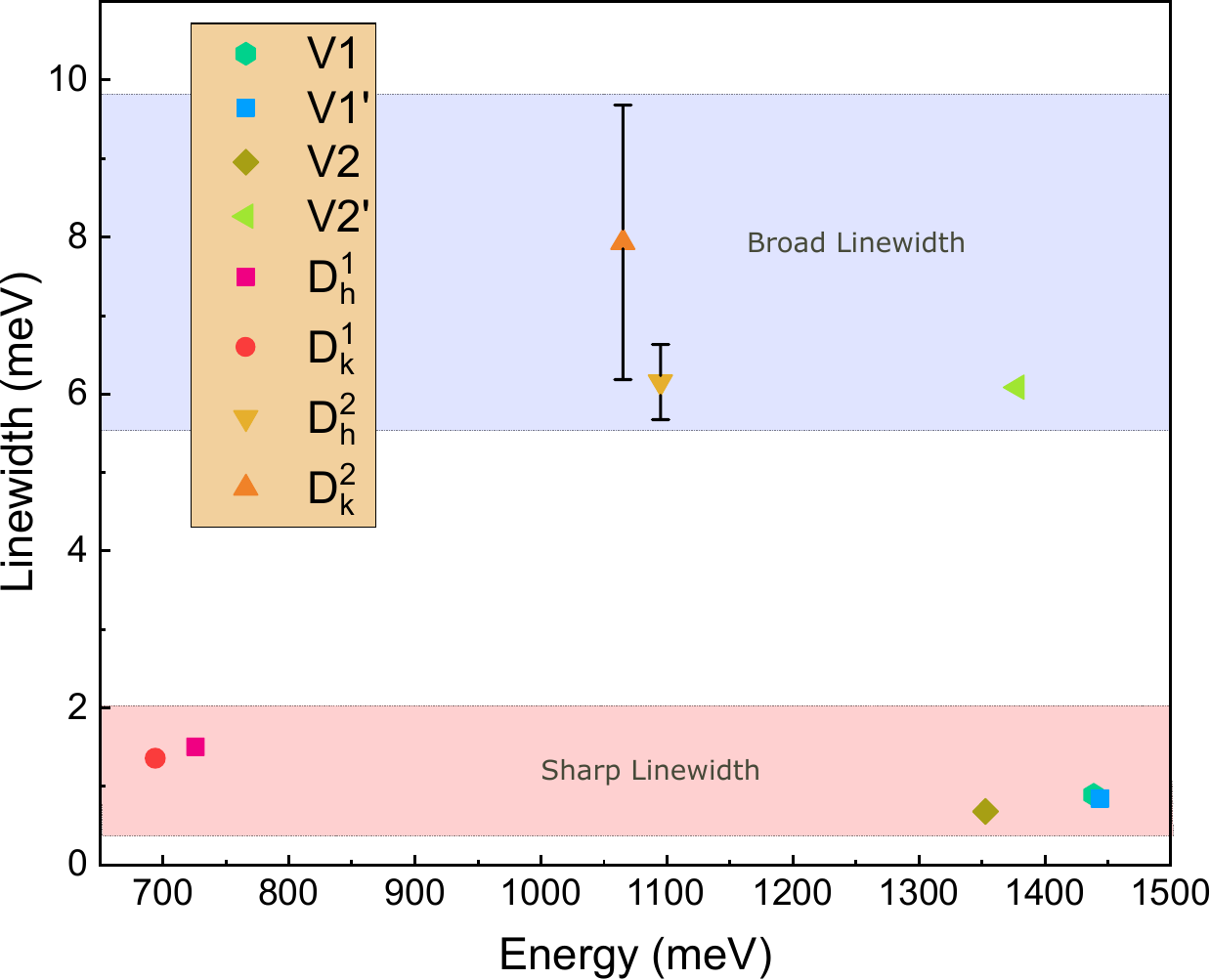}
    \caption{Linewidths of the ZPLs of all transitions observed in the spin-quartet and spin-doublet transitions. The error bars represent the FWHM distribution of D$^2$ transitions from all measurements instead of peak fitting error.}
    \label{fig:linewidth_zpl_all}
\end{figure}

The fitting and analysis of the V1, V1', V2 (and its splittings) and the D$^1$ transitions are presented in Fig.\ref{fig:quartet_fits}. The spectra for the spin-quartet transitions are taken from Fig.~\ref{fig:chap6_2D_SiC}b taken at non-resonant pumping wavelength (840\,nm/1.476\,eV for V1/V1' and 880\,nm for V2); The spectra for D$^1$ transitions are taken from the data presented in Fig.~\ref{fig4}d and e. Combined with the data already analyzed and presented in previous sections, Table~\ref{table:zpl_linewidth} summarizes the linewidth properties and lineshape functions used to determine the profile of the ZPLs. The values for the V2' line is taken from Table~\ref{table:v2prime_psb} and the D$^2$ lines from Tables~\ref{table:doublet_vh} and \ref{table:doublet_vk}. The uncertainties given for the D$^2$ lines reflect the multiple transitions arising from the splitting of the $^2T_2$ (and potentially $^2T_1$ states). Figure~\ref{fig:linewidth_zpl_all} is a graphical representation of the linewidth values shown in Table~\ref{table:zpl_linewidth}, highlighting the clear broad ($\sim$5.5-10\,meV) and sharp ($\sim$0.7-1.5\,meV) linewidth regimes of these ZPLs. 

Ideally, ZPL profiles are within the homogeneous broadening limit at cryogenic temperature (5\,K). Deviation from this model gives information on whether inhomogeneous broadening processes are involved, or whether that is simply caused by instrumental limitations. In our case, the limitation is mainly the spectral resolution of the monochromator in the ns-$\mu$s setup. This method can also be used as a tool to determine whether the optical transition observed is a ZPL (Lorentzian profile) or a vibronic mode of the PSB (Gaussian profile) - see Section~\ref{SM_ep} for a theoretical description. This was employed to determine the nature of the V2' line as a ZPL instead of a LVM peak, as demonstrated in Section~\ref{SM_V2PSB}, since Gaussian fit could not produce a solution to fit the ZPL of the V2' line.

In Table~\ref{table:zpl_linewidth}, we list the center energies, linewidths and lineshapes of all ZPLs in both quartet and doublet channels. We attribute the Gaussian lineshapes of the D$^1$ transitions to the spectral resolution of the monochromator that is used in the ns-ms setup. In this case, the "broadening" is most likely due to instrumental response rather than an inhomogeneous broadening process. The broader lineshapes of D$^2$ transitions, however, all have Lorentzian characteristics, suggesting being dominated by homogeneous broadening.

\vspace{2cm}

\clearpage

\subsection{Calculated emission and absorption spectra of \vsim} 

\label{SM_HR}

\begin{figure}[h]
  \includegraphics[width=0.90\textwidth]{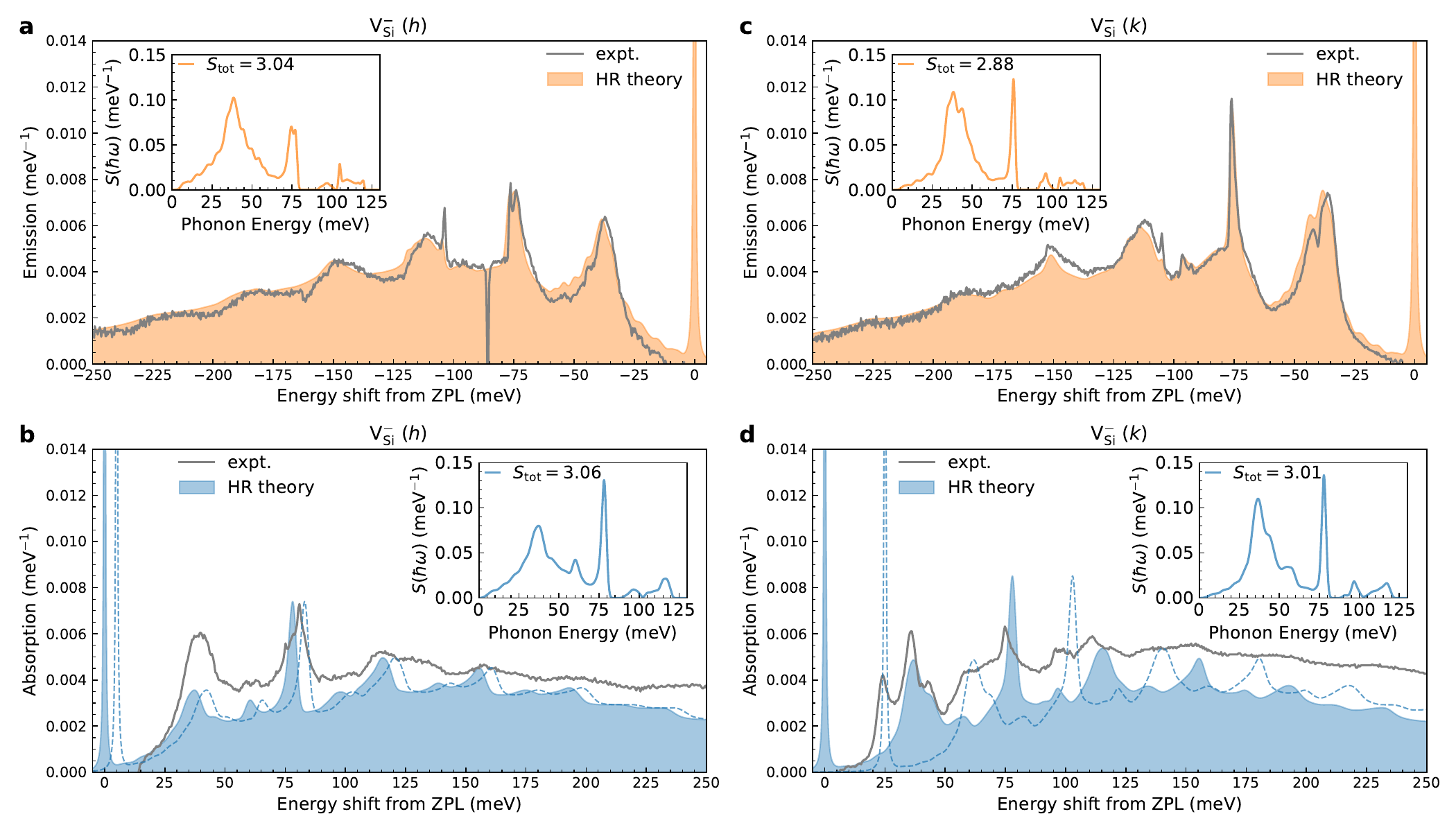}
  \caption{Calculated PSB of V1 and V2 in emission (orange-shaded) and absorption (blue-shaded) lineshapes for unstrained ($6\times6\times2$) size supercells with V$_\mathrm{Si}^{-}(h)$ and V$_\mathrm{Si}^{-}(k)$ defects. The inset figures contain the calculated spectral functions of electron–phonon coupling. The intensities of the experimental lineshapes have been scaled to match the peaks of the computed lineshapes. Lineshapes were computed with r$^{2}$SCAN functional using HR theory and extrapolated to the dilute limit, approximated by a $(25\times25\times8)$ supercell with 40\,000 atomic sites.}
  \label{fig:lum_abs_no_strain}
\end{figure}

Figure~\ref{fig:lum_abs_no_strain} shows the calculated normalized emission (${^{4}A'_{2}} \rightarrow {^{4}A_{2}}$) and absorption (${^{4}A_{2}} \rightarrow {^{4}A'_{2}}$) lineshapes for V$_\mathrm{Si}^{-}(h)$ and V$_\mathrm{Si}^{-}(k)$ defects, with the horizontal axis shifted relative to the respective V1 and V2 ZPL values. The inset figures contain the calculated spectral functions of electron–phonon coupling $S(\hbar\omega)$ together with the total electron--phonon coupling strength $S_{\mathrm{tot}}$. Lineshapes were computed with the r$^{2}$SCAN meta-GGA functional using HR theory (described in Sec.~\ref{SM_ep}) excluding the vibronic coupling between different components of the excited state and then extrapolated to the dilute limit using embedding methodology for size-effects (described in Supplementary Note~\ref{SM_emb}).

The calculated emission lineshapes are in strong agreement with the experiment, capturing the fine features in the PSB. Specifically, the double-peak feature associated with quasi-local vibrational modes for $\mathrm{V_{Si}^{-}}(h)$, observed at $(74.0,\,76.4)$~meV below the V1 ZPL, and the double-peak feature arising from bulk-like vibrational modes for $\mathrm{V_{Si}^{-}}(k)$ at $(36.2,\,42.0)$~meV below V2 ZPL are reproduced due to the large number of vibrational modes obtained using embedding methodology for size-effects.

The ${^{4}A_{2}} \rightarrow {^{4}A'_{2}}$ absorption lineshape which was calculated for $\mathrm{V_{Si}^{-}}(h)$ configuration reproduces nearly all experimentally observed vibronic features (blue shaded area in Fig.~\ref{fig:lum_abs_no_strain}b). The remaining spectral broadening can be attributed to the presence of a second ZPL located approximately 5~meV higher in energy. This effect is illustrated by an additional absorption lineshape obtained by rigidly shifting the calculated ${^{4}A_{2}} \rightarrow {^{4}A'_{2}}$ spectrum by 5~meV to the V1$^{\prime}$ ZPL energy (dashed blue line in Fig.~\ref{fig:lum_abs_no_strain}b). The reproduction of nearly all observed PSB features by adiabatic HR theory implies that non-adiabatic Jahn--Teller coupling within the $\esqE$ manifold and between the $\esqA$ and $\esqE$ states is not dominant.

For the $\mathrm{V_{Si}^{-}}(k)$ configuration the calculated ${^{4}A_{2}} \rightarrow {^{4}A'_{2}}$ absorption lineshape (blue shaded area in Fig.~\ref{fig:lum_abs_no_strain}d) does not capture several PSB features seen in the experiment, especially the feature 25~meV above the V2 ZPL. This feature is theoretically predicted to be V2$^{\prime}$ ZPL~\cite{udvarhelyi2020vibronic}. An additional absorption contribution is included by rigidly shifting the calculated V2 absorption lineshape by 25~meV (dashed blue line in Fig~\ref{fig:lum_abs_no_strain}d). The resulting superposition reproduces several additional PSB features, suggesting that overlapping absorption from two transitions may contribute to the observed spectrum. The anomalous linewidth of the V2$^{\prime}$ ZPL can be understood as arising from resonant coupling between the nominal zero-phonon state of the $\esqE$ manifold and a dense manifold of vibronic states associated with the $\esqA$ configuration. The V2$^{\prime}$ ZPL lies within an energy range that overlaps with the energy range of the first vacancy-related vibrational resonance (the first PSB peak at 35~meV). Consequently, the oscillator strength is redistributed over many closely spaced vibronic states, leading to an effective broadening of the ZPL.

\clearpage 

\subsection{Quantum embedding results}
\label{SM_QE}

\begin{wraptable}{R}{5cm}
    \centering
    \caption{Lattice parameters for the considered computational unit cells.}
    \begin{tabular}{lll}
          & $a$ ({\AA}) & $b$ ({\AA}) \\ \hline
       PBE    & 3.09    & 10.13       \\
       HSE06  & 3.07    & 10.05       \\ \hline
    \end{tabular}
    \label{tab:lat-par}
\end{wraptable}

First, spin polarized DFT was used to relax the geometry until the maximal force (acting on any ion) was smaller than 10~meV/{\AA}. 

We tested both regular semi-local DFT using the Perdew-Bruke-Ernzerhof (PBE) functional \cite{Perdew1996} exchange correlation functional as well as the the hybrid Heyd-Scuseria-Ernzerhof HSE06 functional \cite{Heyd2003}. The plane wave cutoff was 520~eV and the Brillouin zone was sampled with the zone center only. The lattice parameters are given in Tab.~\ref{tab:lat-par}. The computations were performed on a 128-atom supercell.

\def\gls#1{#1}
\paragraph{Electronic structure at the semi-local DFT level (PBE):}
The ground state of V$_{\mathrm{Si},k}^{-1}$ exhibits spin $3/2$. The $\alpha$ spin-channel contains 3 additional electrons in comparison with the $\beta$ channel (\autoref{fig:ks}a). These states are localized as given by the inverse participation ratio (IPR) (\autoref{fig:ks}b). In the $\beta$ spin-channel there are 3 unoccupied states that are localized.

\begin{figure}[h]
    \centering
    \includegraphics{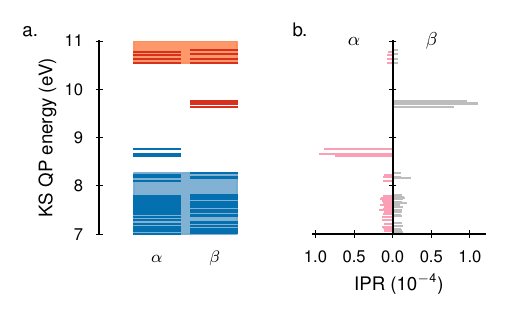}
    \caption{(a) Kohn-Sham (KS) quasi-particle energies and (b) corresponding inverse participation ratio (IPR) at the level of semi-local \gls{DFT}.}
    \label{fig:ks}
\end{figure}

\begin{figure}[h]
    \centering
    \includegraphics{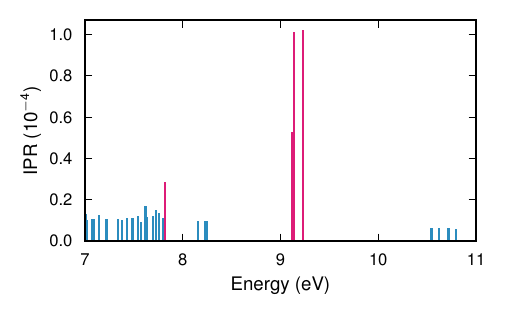}
    \caption{Inverse participation ratio (IPR) for the spin paired (spin 0) calculation. The active space states are indicated with magenta and the environment with blue color.}
    \label{fig:ks_loc_sp}
\end{figure}

The initial state was obtained by spin pairing the system. The spin-pairing forces the occupied $\alpha$-channel states up in energy and the corresponding empty $\beta$-channel states down in energy (\autoref{fig:ks_loc_sp}). There are 3 electrons that populate these quasi-degenerate states with unequal occupation factors. With very low electronic smearing (0.005~eV), the population of the in-gap states is such that the lowest populated in-gap state at 9.12~eV contains 2 electrons and the state at 9.14~eV contains a single electron and the state at 9.23~eV is empty at the PBE level. Together with a low lying fully occupied state, these states make up the active space (\autoref{fig:ks_loc_sp}).

\begin{figure}[h]
    \centering
    \includegraphics[]{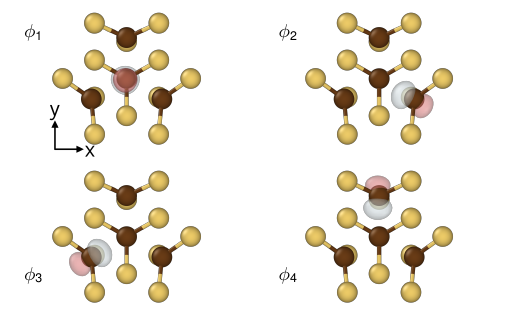}
    \caption{The single particle basis $\{\phi\}$ for the embedding.}
    \label{fig:active-space}
\end{figure}

The active space is chosen such that states with IPR above 0.2 are included. There are 4 spatial states in total, and Wannier functions corresponding to these states are constructed using $sp^3$ projectors on the neighbouring C atoms. The spread of the Wannier functions ranges between 4.0~{\AA} and 4.2~{\AA}. The resulting Wannier functions are shown in \autoref{fig:active-space}. The imaginary part is negligible.

\begin{wraptable}{R}{5.5cm}
    \centering
    \caption{Hubbard interaction parameters from cRPA.}
    \label{tab:crpa-pbe}
    \begin{tabular}{ccc}
        & bare & screened \\ \hline
      $\langle U \rangle$ &      8.70 & 1.76         \\ 
      $\langle u \rangle$ &      4.24 & 0.56        \\ 
      $\langle J \rangle$ &         0.03   & 0.01    \\ \hline
    \end{tabular}

\end{wraptable}

For the constrained random phase approximation, we consider 1120 bands in total in the cRPA calculation for the 128 atom supercell, which is 32~eV above the conduction band minimum. 
The resulting interaction parameters are tabulated in \autoref{tab:crpa-pbe}.
At the level of semi-local \gls{DFT} on the negatively charged silicon vacancy in the $k$ site, the many-body states are tabulated in \autoref{tab:mbspectra_pbe_k} and shown in \autoref{fig:mbspectra_pbe_k}. 

The current data (PBE, 128 atom supercell, FCI+cRPA) would give possible transition energies in the doublet manifold (from the ground state) of 0.86, 0.75, 0.72, 0.52, and 0.46~eV. 

\begin{figure}
    \centering
    \includegraphics[width=0.5\linewidth]{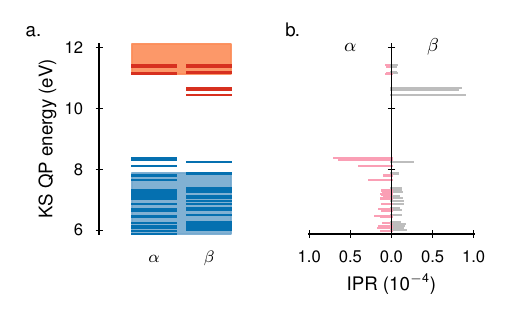}
    \caption{Kohn-Sham eigenstates using the HSE06 exchange functional.}
    \label{fig:ks_hse_128_k_p}
\end{figure}

\begin{figure}[h]
    \centering
    \includegraphics{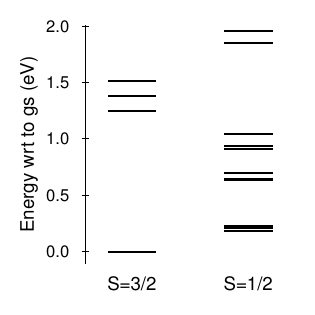}
    \caption{Many-body spectra for the $k$ site negatively charged silicon vacancy as computed with FCI+cRPA on a 128-atom supercell and using PBE for the underlying DFT calculations.}
    \label{fig:mbspectra_pbe_k}

\end{figure}

\begin{figure}[h]
    \centering    \includegraphics{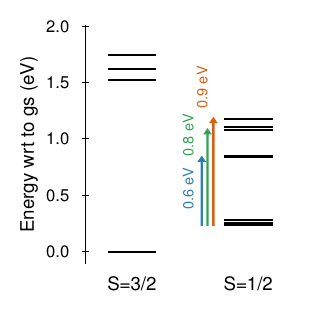}
    \caption{Many-body spectra for the $k$ site negatively charged silicon vacancy as computed with FCI+cRPA on a 128-atom supercell and using HSE06 for the underlying DFT calculations.}
    \label{fig:mbspectra_hse_k}

\end{figure}

Using the HSE06 functional, the overarching picture remains the same, but the eigenenergies change (Fig.~\ref{fig:ks_hse_128_k_p}). The resulting electronic structure using cRPA+FCI shows better agreement with experiment for V$_{\mathrm{Si},k}^{-1}$ than PBE-level calculations (Fig.~\ref{fig:mbspectra_hse_k}).

\vspace{1cm}

\begin{table}[h]
    \centering
    \caption{Many-body states of the negatively charged silicon vacancy at $k$ site as computed with FCI+cRPA on the 128 atom supercell using PBE single particle basis consisting of 4 orbitals (\autoref{fig:active-space}).}
    \label{tab:mbspectra_pbe_k}
    \begin{tabular}{cc}
     Energy (eV)    & Spin \\ \hline
     0.00 & 3/2 \\ 
     0.18 & 1/2 \\ 
     0.21 & 1/2 \\ 
     0.23 & 1/2 \\ 
     0.64 & 1/2 \\ 
     0.70 & 1/2 \\ 
     0.90 & 1/2 \\ 
     0.93 & 1/2 \\ 
     1.04 & 1/2 \\ 
     1.24 & 3/2 \\ 
     1.38 & 3/2 \\ 
     1.51 & 3/2 \\ \hline
    \end{tabular}

\end{table}

\end{document}